\documentclass[12pt]{article}

\usepackage{epsfig,amsmath,latexsym,amsfonts}
\def\ra{\rightarrow}

\def\gs{g_{\rm s}}
\def\ap{{\alpha^\prime}}

\def\N{{\cal N}}
\def\none{$\N=1$}
\def\F{{\cal F}}
\def\ntwo{$\N=2$ }
\def\s{\sigma}

\input epsf

\begin{document}
{\titlepage{
\font\cmss=cmss10 \font\cmsss=cmss10 at 7pt

\hfill IC/2003/11

\vskip .1in \hfill CPHT-RR-007-0203
\vskip .1in \hfill Bicocca-FT-03-3
\vskip .1in \hfill hep-th/0303191

\hfill

\vspace{20pt}

\begin{center}
{\Large \textbf{Supergravity duals of 
supersymmetric four dimensional gauge theories}}
\end{center}

\vspace{6pt}

\begin{center}

\textsl{F. Bigazzi $^{a}$, A. L. Cotrone $^{b,c}$, M. Petrini $^{b}$ and
A. Zaffaroni $^{d}$}  

\vspace{20pt}

\textit{$^a$ The Abdus Salam ICTP, Strada Costiera, 11; I-34014 Trieste, Italy.}

\textit{$^b$ Centre de Physique Th{\'e}orique, {\`E}cole Polytechnique, 48 Route de Saclay; F-91128 Palaiseau Cedex, France.}

\textit{$^c$ INFN, Piazza dei Caprettari, 70; I-00186 Roma, Italy. }

\textit{$^d$ Universit{\`a} di Milano-Bicocca and INFN\\ Piazza della 
Scienza 3, I-20126 Milano, Italy.}

\vspace{10pt}

\end{center}

\vspace{12pt}

\begin{center}

\textbf{Abstract}\end{center}
This article contains an overview of some recent attempts at understanding
supergravity and string duals of four dimensional gauge theories using 
the $AdS/CFT$ correspondence. We discuss the general philosophy 
underlying the various ways to realize Super Yang-Mills theories in terms of systems of branes.
We then review some of the existing duals for
${\cal N}=2$ and ${\cal N}=1$ theories. We also discuss differences and
similarities with realistic theories.

\vfill

\vskip 5.mm
\hrule width 5.cm

\vskip 2.mm

\noindent e-mail: bigazzif@ictp.trieste.it, Cotrone@cpht.polytechnique.fr,\\ 
Michela.Petrini@cpht.polytechnique.fr, alberto.zaffaroni@mib.infn.it
}}



\tableofcontents

\pagenumbering{arabic}

In recent years, a great deal of attention has been attracted by a new
kind of duality between gauge theories and string theories, known as 
the $AdS/CFT$ correspondence. According to it, certain superconformal gauge
theories have a dual description in terms of critical string
backgrounds. 
This provides the first explicit realization of the old idea that the 
strongly coupled
dynamics of a gauge theory has a description in terms of an effective
theory of strings. The correspondence also naturally implements the 't Hooft large $N$
expansion, thus providing a verification of many ideas about gauge theories at large $N$. 
In addition to these qualitative successes, 
$AdS/CFT$ also provides quantitative tools for understanding gauge theories.
For example, correlation functions of the conformal gauge theory in the 
strongly coupled regime at large $N$, which cannot be computed in perturbative quantum field theory, 
can be reduced to a classical computation in supergravity.
Originally formulated as a duality 
between $\N=4$ super Yang-Mills (SYM) and Type IIB string theory on
$AdS_5\times S^5$ \cite{malda,gkp,witten1}, the correspondence 
can also be extended to conformal gauge theories 
with less supersymmetry and in different dimensions and, nowadays, 
there are few doubts about its correctness.
It is somewhat ironic that the first successful example of a 
stringy description of gauge theories 
deals with conformal theories and not with confining ones,
where the string would be naturally 
identified with the color flux tubes of
confinement, or, briefly, the QCD strings.  

In this review, we will give an overview of some recent attempts to
extend the $AdS/CFT$ ideas to non conformal theories. Considered the
huge literature on the subject, we decided to discuss
only four-dimensional gauge theories with unitary groups. 
We will focus, in particular, on two
specific ways of generalizing  the correspondence to other pairs of
gauge/gravity duals. The first one 
consists in deforming a conformal theory for which we possess a well defined
supergravity dual. The gauge theory obtained in this way is non conformal 
at energies below the scale set by the deformation. 
The second method uses wrapped and fractional branes engineering
theories that are non-conformal at all scales.

The extension of the $AdS/CFT$ ideas to non-conformal theories
is not straightforward. 
From a technical point of view, it is difficult to avoid singularities 
in the solutions. No regular solutions dual to $\N=2$ gauge theories are indeed known. 
The $\N=1$ case is more successful: two completely regular supergravity solutions describing
$\N=1$ gauge theories have been found \cite{ks,mn2}. The road to realistic
theories, 
like QCD,
is still long. Classical supergravity solutions give a quite accurate description
of theories that are not pure YM theories, but contain infinite additional fields. It is a general expectation that classical supergravity alone cannot describe
realistic gauge theories, which contain higher spin glueballs. The dual of pure QCD is
therefore expected to be a strongly coupled string model.
The $AdS$-inspired solutions that we will describe are nevertheless interesting. 
Firstly, the possibility of re-summing all string world-sheet corrections
for a string background  is not unconceivable. The inclusion
of these corrections would give a good description of pure gauge theories
in the large $N$ limit. The computation of world-sheet corrections, which is
relatively easy in flat space-time, here is complicated by the presence of
RR-fields, but some progress in this direction has been recently made.
Secondly, the supergravity duals provide many exactly solvable models   
exhibiting confinement and other phenomena typical of the pure gauge theory.
Thus, even if not quantitatively relevant for QCD, they provide a good laboratory 
for studying the mechanism of confinement and
the qualitative properties of QCD.

The purpose of this article is to provide a general overview of the literature,
to describe the main features of the various
methods to realize interesting gauge theories and to
give  a unifying picture for various models. We 
do not plan to be exhaustive. Since there exist many good
reviews in the literature covering  some of the  models we will discuss,
we will sometimes refer to them for the details
and the proof of specific results. 
Inevitably, many methods to extend the
correspondence are not covered here, including 
some that were  largely discussed in the literature
and came first historically. Two basic subjects are not discussed 
here at all: 
the introduction 
of finite temperature and Type 0 theories. Even in the context of 
deformations and fractional/wrapped branes we will make several omissions.

The review is organized as follows.
In Section 1 we briefly review the $AdS/CFT$ correspondence for the $\N=4$ and $\N<4$ cases. 
In Section 2 
we discuss the general aspects of the deformation method while in Section 3 
we discuss fractional and wrapped branes. In Section 4 we give an overview of
the known supergravity solutions with $\N=2$ supersymmetry. In Section 5,
we discuss supergravity solutions with $\N=1$ supersymmetry. The case of softly
broken theories is discussed in the last part of Section 5. 
In each Section, we chose to cover in more detail the 
case of wrapped branes, which therefore forms the backbone  of this review.

\section{Basic dictionary of the $AdS/CFT$ correspondence}\label{uno}
A throughful introduction to the $AdS/CFT$ correspondence
\cite{malda,gkp,witten1} would 
itself require a whole review. This Section has been inserted for
completeness, to recall 
the basic facts we will use and generalize
in the following Sections. 
Therefore we suggest the reader who already has a certain knowledge of the 
correspondence to start with Section 2 and come back to Section 1
when necessary. 
On the contrary, for a more complete discussion of $AdS/CFT$ we refer the
reader to the  very good reviews in the literature \cite{magoo,petersen,divecchia}.
Here we will first focus on the best known
example of the $AdS/CFT$ correspondence, which involves $\N=4$
Super Yang-Mills theory in four dimensions. Then
we will examine some extensions to 
less supersymmetric 
models and six-dimensional theories.

\subsection{The $AdS/CFT$ correspondence: Motivations}\label{ads}

The $AdS/CFT$ correspondence 
\cite{malda} derives
from the observation that systems of D-branes in Type II string theory
(or systems of M-branes in M theory) admit a complementary
description in terms of gauge theories on their world-volume on one side and
curved supergravity backgrounds on the other side.
Consider the simplest system of branes that 
realize on the world-volume a four-dimensional gauge theory:  
a stack of $N$ parallel D3-branes in Type IIB. Since the D-branes preserve
half of the 
space-time
supersymmetry, this configuration has $\N=4$ conformal supersymmetry in four dimensions.
The massless fields on the branes form a $\N=4$ multiplet containing a $U(N)$
gauge field 
$A_{\mu}$, four Weyl
fermions $\lambda_a$ and six scalars $\phi_i$, all transforming in the adjoint 
representation of $U(N)$.
The $SO(6)\sim SU(4)$ symmetry of the space transverse to the branes is realized  as the
field theory 
R-symmetry, under which the fermions transform in the representation
 $\bf{4}$ and the scalars in the representation $\bf{6}$. The theory has a 
$6N$ dimensional moduli
space of vacua labeled by the Cartan values of the adjoint scalar VEVs. 
In a generic vacuum, the gauge group is broken to its maximal abelian subgroup $U(1)^N$.
Being BPS objects, the D-branes can be separated with no cost in energy.
The generic vacuum of the gauge theory is then represented by a string configuration
where the branes have arbitrary positions in the transverse space $\mathbb{R}^6$. 
Notice that a typical 
massive $W$-boson in a generic vacuum is represented by an open string connecting two branes
and its mass is given by $m=\Delta r/\alpha^\prime$,
where $\Delta r$ is the brane separation.

At low energies, the system is conveniently described by  
the $\N=4$ massless fields on the branes coupled to the massless fields of Type IIB
supergravity in the bulk. The low energy Lagrangian for the coupled brane/bulk system reads
\begin{equation}
-\frac{1}{8\pi g_s}\int d^4x \sqrt{g} \, Tr(F^2)+\frac{1}{(2\pi)^7\alpha^{\prime 4}g_s^2}
\int d^{10}x \sqrt{g}R + \cdot\cdot\cdot\label{e15}
\end{equation}

In this expression, we integrated out all the open and closed string oscillator modes.
In the low energy limit $E\ll 1/\sqrt{\alpha^\prime}$ 
the gauge theory on the branes decouples from the bulk
and we recover $4d$ $\N=4$ SYM theory with gauge coupling $g_{YM}$ determined by the string coupling: $g_{YM}^2=4\pi g_s$\footnote{In this review we
use the conventions (see for example \cite{peskin})
${\cal L}= - \frac{1}{4g_{YM}^2}F^a_{\mu\nu} F^{a \mu\nu}  + \frac{\theta_{YM}}{32\pi^2}
F^a_{\mu\nu}{\tilde F}^{a\mu\nu}= -\frac{1}{2g_{YM}^2}Tr(F_{\mu\nu}F^{\mu\nu}) + \frac{\theta_{YM}}{32\pi^2}
 Tr F_{\mu\nu}{\tilde F}^{\mu\nu}$. The complex coupling is $\tau=\frac{\theta_{YM}}{2\pi}+i\frac{4\pi}{g_{YM}^2}$.}.
Since the mass of the generic gauge excitation in a broken vacuum is 
of order $m=\Delta r/\alpha^\prime$, we can still detect the existence of
a moduli space by focusing on the region very close to the branes, or equivalently
by rescaling the distances $\Delta r$.

The stack of $N$ D3-branes has an equivalent
description in terms of a 3-brane extremal solution in IIB supergravity. 
This solution 
contains a constant dilaton, a RR four-form potential and a metric given (in
the string frame) by 
\begin{eqnarray}
ds^2&=&Z(r)^{-1/2}dx_{\mu}dx^{\mu}+Z(r)^{1/2}(dr^2+r^2d\Omega_5^2),\nonumber\\
C_{(4)}&=& Z(r)^{-1}  dx^0 \wedge dx^1 \wedge dx^2 \wedge dx^3
\wedge dx^4, \nonumber\\
Z(r)&=&1+\frac{4\pi g_sN\alpha^{\prime 2}}{r^4}.
\label{e17}
\end{eqnarray}
In this description the decoupling  limit can be
realized by sending $\alpha^\prime\rightarrow 0$ while keeping 
the parameters of the gauge theory fixed. As we saw, we should also rescale distances
to preserve the existence of a moduli space.
We then send $\alpha^{\prime}\rightarrow 0$ keeping $g_s$ and
$r/\alpha^{\prime}\equiv U$ fixed.  
In the limit we have just described, we can discard the $1$ in the 
expression~(\ref{e17}) for $Z$.
This is equivalent to focusing on the near-horizon geometry
\begin{equation}
ds^2=\alpha^{\prime}\left\{ R^2\frac{dU^2}{U^2}+\frac{U^2}{R^2}dx_{\mu} dx^{\mu}
+R^2d\Omega_5^2\right\}.
\label{e18}
\end{equation}
The metric  is the direct product of two spaces of constant 
curvature, $AdS_5\times S^5$, with the same radius $R^2=\sqrt{g_{YM}^2N}\alpha^\prime$. 

This is the observation that led Maldacena \cite{malda} to conjecture that four-dimensional 
$\N=4$ $SU(N)$ SYM in $3+1$ dimensions is equivalent to  Type IIB string theory on $AdS_5\times S^5$. 
This is the content of the so-called $AdS/CFT$ correspondence. The matching of
the parameters on the two sides of the correspondence reads
\begin{eqnarray}\label{par}
4\pi g_s&=& g_{YM}^2=\frac{x}{N},\\ \nonumber
\frac{R^2}{\alpha^\prime}&=& \sqrt{g_{YM}^2N}=\sqrt{x},
\end{eqnarray}
where $x=g_{YM}^2N$ is the 't Hooft coupling. 
The string theory is weakly coupled if we first send $N\rightarrow\infty$ 
at fixed $x$ (thus suppressing  string loops), and then we take
$x\gg 1$ (thus suppressing world-sheet corrections). 
From eq. (\ref{par}) we see that 
the latter condition means that 
the large-$N$ gauge theory is strongly coupled. 
We can therefore think in terms of a duality: strongly
coupled phenomena in the large $N$ limit of a gauge theory are
described by a dual weakly coupled  string background. 
The double perturbative expansion of string theory,
in powers of $g_s$ (string loop) and $\alpha^\prime$ (higher derivative
terms) is associated respectively with the $1/N$ expansion (at fixed $x$)
and the $1/x$ expansion at each order in $N$. 
The old proposal that gauge theories at large $N$
have a dual description in terms of a string theory is explicitly
realized.

It is also instructive to compare the symmetries of the two theories.
$\N=4$ SYM is invariant under the conformal group $SO(4,2)$,
has $\N=4$ supersymmetry that is doubled with the  addition of  the
superconformal generators, and a $SO(6)$ R-symmetry. 
In the dual theory  
$SO(4,2)$ is the isometry group of $AdS_5$, the $\N=8$ supersymmetries
are those of Type IIB supergravity compactified on $AdS_5 \times S^5$, and
$SO(6)$ is the isometry group of $S^5$.
In a word, the symmetries on both sides form the superconformal 
group $SU(2,2|4)$.

\subsection{Precise definition of the $AdS/CFT$ correspondence}\label{definition}

The $AdS/CFT$ correspondence, in a little more general form than the one introduced 
in the previous Section, relates a $4d$ $CFT$ to a 
critical string in $10d$ on $AdS_5\times H$.
If $H$ is compact, the string theory is effectively five-dimensional.
The $AdS_5$ factor guarantees that the dual theory is conformal, since its isometry
group  $SO(4,2)$ is the same as the group of conformal 
transformations of a four-dimensional quantum field theory.

To define the correspondence, we need a map between the observables in the two
theories and a prescription for comparing physical quantities and amplitudes.  
The correspondence is via holography \cite{gkp,witten1}. 
Let us start by writing the $AdS$ metric as
\begin{equation}
ds^2=dy^2+e^{2 y /R}dx_\mu dx^\mu,
\label{AdSmetric}
\end{equation}
where the radial coordinate $y$ is related to that in eq. (\ref{e18}) by $y= R\log (U/R)$. 
We see that the metric has a conformal boundary
at $y=\infty$ isomorphic to Minkowski space-time and this will play an important role in the following.
The $CFT$ is specified by a complete set of conformal operators. In a gauge theory at large $N$,
a distinguished role will be played by single-trace operators\footnote{Multiple trace operators are usually
associated with multi-particles states in $AdS$.}. The fields in $AdS$, on the other hand,
are the excitations of the string background.
They certainly contain the metric and many other fields. We may assume that, when a semi-classical 
description is applicable, their interaction is described by an effective action
$S_{AdS_5}(g_{\mu\nu},A_{\mu},\phi,...)$.
Suppose that we have a map between observables in the two theories.
We can formulate a prescription to 
relate correlation functions in the $CFT$ with scattering amplitudes
in $AdS_5$.
In $CFT$ we can define the functional generator $W(h)$ for
the connected Green functions for a given operator $O$.
 $h(x)$ is
a source, depending on 4 coordinates, which is coupled to the operator $O$ through
\begin{equation}
L_{CFT}+\int d^4x h O.
\label{e20}
\end{equation}
$O$ is associated with a scalar field $\hat h$ in $AdS$, which, for simplicity, we assume to
be a canonically normalized scalar: 
$S_{AdS}=\int d^4x dy \sqrt{g} [(\partial \hat h)^2-m^2\hat h^2+...]$.
The solution of the equation of motion of  $\hat h(x,y)$ for large $y$ is
\begin{equation}
\hat h(x,y)\rightarrow e^{(4-\Delta)y/R} \hat h_{\infty}(x),
\label{prescr}
\end{equation}
where
\begin{equation}
m^2=\frac{\Delta (\Delta -4)}{R^2}.
\label{mass}
\end{equation}
Since we expect that the large $y$ behavior 
of $\hat h$ reflects the conformal scaling of the field we identify $\Delta$ 
with the dimension of the dual operator $O$.
The prescription for identifying correlation functions with scattering
amplitudes is the following: given a solution $\hat h$ of the equations of motion  
derived from $S_{AdS}$ that reduces to $\hat h_{\infty}(x)\equiv h(x)$ 
at the boundary, we claim that 
\cite{gkp,witten1}\footnote{The equations of motion in $AdS$ are second order 
equations, but the
extension of the boundary value inside the space is unique. What we implicitly
impose is regularity in the interior of $AdS$.}
\begin{equation}
e^{W(h)}=\left\langle e^{\int h O}\right\rangle=e^{-S_{AdS_5}(\hat h)}.
\label{n1}
\end{equation}
This prescription is valid in the low energy limit where
supergravity is valid. 
In full string theory, the right-hand side of the last equation
should be replaced by some S-matrix element for the state $\hat h$.
Notice that we used equations of motion in $AdS$: an off-shell theory in four dimensions 
corresponds to an on-shell theory in $5d$. 
This is a generic feature of all the $AdS$-inspired correspondences.
The previous prescription allows to compute Green functions for a
strongly coupled gauge theory at large $N$ using classical supergravity.
In all the computations done up to now, there is an amazing agreement
between the $CFT$ and the supergravity predictions, whenever a comparison
can be made. This leaves very few doubts about the validity of the $AdS/CFT$ correspondence. For more details on the subject, the reader is referred to \cite{magoo}.

The map between $CFT$ operators and $AdS$ fields should be worked out
case by case. For specific operators the dual field can be found using symmetries.
For example, the natural couplings
\begin{equation}
L_{CFT}+\int d^4x\sqrt{g}(g_{\mu\nu}T_{\mu\nu}+A_{\mu}J_{\mu}+
\phi F_{\mu\nu}^2+\cdot\cdot\cdot )
\label{e21} 
\end{equation}
suggest that the operator associated with the graviton is the stress-energy
tensor and the operator associated to a gauge fields in $AdS$
is a $CFT$ global current.
In general, global symmetries in $CFT$ correspond to gauge 
symmetries in $AdS$.
In the previous formula, we also included a coupling that is very natural
in string theory. Since $g_s\sim g_{YM}^2$,  
the operator associated to the dilaton is the derivative
of the classical Lagrangian with respect to $1/g_{YM}^2$. 

We are mainly interested in the limit where string theory is weakly coupled, and reduces to
Type IIB supergravity. Since $H$ is compact,
the bosonic massless modes in $10d$ can be expanded in a set of Kaluza-Klein (KK) modes on $H$
with masses of order $1/R^2$.
We see that all operators with finite dimension 
for $x \rightarrow\infty$ (supergravity limit) should
correspond to Kaluza-Klein (KK) modes on $AdS\times H$. We can 
explicitly describe the relation for $\N=4$ SYM, where the large
amount of supersymmetry allows for a complete classification. 
All KK modes on $AdS_5\times S^5$ were computed in the eighties
\cite{kim}.
They are organized in $\N=8$ multiplets \cite{gunaydin}. 
The difference in spin in a generic $\N=8$ multiplet 
may reach four units. The KK multiplets, with maximum spin
2, should correspond to short (and therefore protected) multiplets.
$\N=8$ short multiplets $A_k$ are labeled by integers $k\ge 2$, and their lowest state 
is a scalar in the $k$-fold symmetric representation of $SO(6)$ with mass $m^2=k(k-4)/R^2$.
The KK spectrum contains each $A_k$ for $k\ge 2$ exactly once.
The corresponding multiplets on the $CFT$ side are obtained by applying the
supersymmetry charges to the operator \cite{witten1,sergio} 
\begin{equation}
Tr\phi_{\{ i_1}\cdot\cdot\cdot\phi_{i_k\}}-{\rm traces}
\label{new2}
\end{equation}
of dimension $k$. One can prove that these multiplets are short and therefore
have protected dimensions. 
There is a complete correspondence with the KK spectrum. It is believed
that the previously defined $CFT$ multiplets  exhaust 
the (single trace) short multiplets of $\N=4$ SYM.
A special role is played by $A_2$, which is the supergravity massless multiplet (in five-dimensional sense)
containing the graviton and the 15 gauge fields of $SU(4)$. It corresponds to the
supermultiplet of currents in the $CFT$ side. 

For further reference, we list the lowest fields/operators
appearing in the KK spectrum:
\begin{center}
\begin{tabular}{llll}
SU(4) rep.&operator&multiplet/dim.&mass\\
$\bf{20}$& Tr$\phi_{\{ i}\phi_{j\}}-{\rm traces}$&$A_2\,\,\, \Delta =2$\,\,\, & $m^2=-4$\\
$\bf{50}$& Tr$\phi_{\{ i}\phi_{j}\phi_{k\}}-{\rm traces}$&$A_3\,\,\, \Delta =3$\,\,\, & $m^2=-3$\\
$\bf{10_c}$& Tr$\lambda_a\lambda_b+\phi^3$&$A_2\,\,\, \Delta =3$\,\,\, & $m^2=-3$\\
$\bf{105}$& Tr$\phi_{\{ i}\phi_{j}\phi_{k}\phi_{p\}}-{\rm traces}$&$A_4\,\,\, \Delta =4$\,\,\, & $m^2=0$\\
$\bf{45_c}$&Tr$\lambda_a\lambda_b\phi_i+\phi^4$ &$A_4\,\,\, \Delta =4$\,\,\, & $m^2=0$\\
$\bf{1_c}$&${\rm on-shell\, Lagrangian}$&$A_2\,\,\, \Delta =4$\,\,\, & $m^2=0$\\
\end{tabular}
\end{center}
Some of the masses of $AdS$ fields are negative but this does not represent an instability.  
Due to the negative curvature,
a mode is stable if $m^2R^2\ge -4$ (Breitenlohner-Freedman bound \cite{BF}).
Using formula~(\ref{mass}), we see that a scalar field has negative, null or positive mass
when it corresponds to a relevant, marginal or irrelevant $CFT$ operator, respectively.
In the previous table we listed all the relevant scalar operators
 appearing in the KK spectrum.

Let us also briefly consider the stringy states. In the supergravity limit, 
all stringy states are very massive and should decouple. In the $CFT$ these states correspond to operators with large 
anomalous dimension\footnote{Using formula~(\ref{mass})  with 
$m^2={\rm integer}/\alpha^\prime$ we predict $\Delta\sim(x)^{1/4}$.}, which, 
for consistency, decouple from all the OPEs and Green functions.
It is then a prediction of the $AdS/CFT$ correspondence that, at large $N$ and at large 't Hooft coupling
the only single trace $\N=4$ SYM operators with finite dimensions are the
protected ones, which have been classified above. 
The simplest example of an operator dual to a string state is 
$Tr(\phi_i\phi_i)$ (missing in the previous classification). 
Some progress in the
understanding of a certain class of stringy states has been made in \cite{ppwave,gkp2}.

We finish with one particularly important comment. 
Notice that the theory realized on the world-volume of D3-branes in Type IIB is $\N=4$ SYM
with gauge group $U(N)$. It is believed that the $U(1)$ 
factor is not described by the
correspondence. Some evidence for the disappearing of the $U(1)$ factor comes from the
analysis of the spectrum of $AdS_5\times S^5$. Indeed, 
the KK multiplet $A_1$, which would be associated with the $CFT$ multiplet
with lowest component ${\rm Tr}\phi_i$, is not present in the supergravity spectrum.
This is a strong evidence that the SYM gauge group is $SU(N)$ and not $U(N)$, as one could
naively expect. Another strong evidence comes from the existence of 
a baryonic vertex 
(obtained as a wrapped brane in the bulk \cite{witten3}), which can only exist in a $SU(N)$ theory. The gauge $U(1)$ factor on the D3-brane
theory is frozen in the holographic dual, meaning that it reduces to a global symmetry.
In this particular case, the $U(1)$ is completely decoupled.

\subsection{The correspondence for NS5-branes}\label{LST}

A generalization of the $AdS/CFT$ correspondence that we will need
in the following deals with 5-branes in Type II.
Consider a stack of $N>1$ 
coincident Type IIB NS5-branes in flat space-time. 
The background they generate is
\begin{eqnarray}
ds^2&=& dx_{\mu}dx^{\mu}+Z(r)
\left(dr^2 + r^2d\Omega_3^2\right),\nonumber\\
e^{2\Phi}&=&g_s^2Z =g_s^2\left(1+
\frac{\alpha^\prime N}{r^2}\right),\nonumber\\
B_{ (6)} &=& (Z^{-1} -1)g_s 
dx^0 \wedge \cdots \wedge dx^5.
\label{nsfive}
\end{eqnarray}

Via S-duality, one can write an analogous solution for a set of D5-branes.
Both NS5 and D5-branes are BPS objects, so they preserve $(1,1)$ supersymmetry 
on the six dimensional world-volume. At very low energies the world volume 
theory is a 6$d$ $(1,1)$ supersymmetric gauge theory with coupling
($m_s^2=1/\alpha'$)
\begin{equation}
\frac{1}{g_{D}^2}\sim \frac{m_s^2}{g_s},\qquad\qquad \frac{1}{g_{NS}^2}\sim m_s^2.
\end{equation}

This result is easily deduced from the Born-Infeld action for the 
D5-branes. An S-duality gives then the result for the NS5-branes.
Now, in the limit $g_s\rightarrow 0$ with $m_s$ fixed
the bulk modes which interact
with a NS5 brane through the string coupling $g_s$ would decouple.
We are thus left with a six dimensional, non gravitational theory with 
sixteen supercharges and a mass scale $m_s$ \cite{little0}.  
Since $m_s\ne 0$, the theory still contains strings (they emerge, 
for example, as instantons of the low energy gauge theory), 
from which the name ``Little String Theory'' (LST). It is a non-local 
theory of strings, exhibiting a form of 
T-duality. We refer the interested reader to 
\cite{kutasov} and references therein for a comprehensive 
review of the subject.
At very low energy, the LST reduces to $(1,1)$ SYM in six dimensions.
The massless fields of the 6$d$ theory are: one gauge  vector field, 
4 scalars parameterizing  the directions transverse 
to the branes and  two symplectic Majorana fermions, 
all in the adjoint of the
 gauge group. The scalars and the spinors transform as the 
$(\bf{1},\bf{4})$ and $(\bf{4}_{+}, \bf{2}_{+})+(\bf{4}_{-}, \bf{2}_{-})$ of 
$SO(1,5)\times SO(4)$, respectively, where 
the isometries of the transverse $\mathbb{R}^4$ become the $SO(4)$ 
R-symmetry group of the world volume theory. 

What is important for the gauge/gravity correspondence is that LST has 
a holographic dual \cite{little}.
Send $r\rightarrow 0$ at the same rate as $g_s$ in 
the background (\ref{nsfive}).
Defining $r=g_se^\rho$, we find 
\begin{equation}
ds^2=dx_\mu dx^\mu + N\alpha'(d\rho^2 + d\Omega_3^2), \qquad \Phi=-\rho+const.\label{lindil}
\end{equation}

This is the so-called  ``linear dilaton background''.
String theory on this space-time has an exact conformal field theory description in terms of six free coordinates, a Liouville field for the radius and an $SU(2)$ WZW model at level $N$.
We will be mostly satisfied with the supergravity approximation.
This is reliable in the large $N$ limit, as usual, and far away from the 
branes, where the string coupling is vanishing as can be seen 
from (\ref{lindil}).
As one approaches the branes, 
the string coupling diverges and one has to go to the S-dual D5-brane description.

A discussion of the observable mapping can be found in \cite{kutasov}. 
Obviously, the non-locality of the theory makes  
the mapping difficult. 
However in the low energy limit, where the LST reduces to SYM, the operators 
become local and we can still make some natural identifications. In particular we can use the $SO(4)$ global symmetry to classify the operators.
On the supergravity side,  
we will perform a dimensional reduction on the $S^3$ transverse 
to the branes, obtaining a tower of KK states. As we will see, 
the massless multiplet
will be described by an $SO(4)$ gauged supergravity in seven dimensions, where
we will be able to identify the dual operators. 
In the KK spectrum, 
in analogy with $\N=4$ SYM,  we expect to find scalars dual to the operators
${\rm Tr} X_{\{i_1}...X_{i_k\}}$-traces, where $X_i$ are the
four massless scalars in the (1,1) SYM theory.

\subsection{Conformal field theories with $\N <4$: Orbifolds}\label{orbifold}

In general, in the gauge/gravity correspondence the amount of supersymmetry can
be reduced by placing the branes
in curved geometries. Since the $AdS/CFT$ correspondence involves 
the near-brane region and every smooth manifold is 
locally flat, we may expect to find new models only when
the branes are placed at a
singular point of the transverse
space \cite{kacsil,kehag,kw1,morpless}. There is no general
method for determining the gauge theory living on the world-volume of
branes placed at generic singularities. For orbifold
singularities, however, such a method exists and we will start
reviewing it.

Consider $N$ D3-branes sitting at the singularity of 
the orbifold $\mathbb{R}^6/\Gamma$, where $\Gamma$ is a
discrete group $\Gamma\subset SO(6)\sim SU(4)$. 
The supergravity solution in this case reads
\begin{equation}
ds^2=Z^{-1/2}(r)dx_{\mu} dx^{\mu}+Z^{1/2}(r)(dr^2+r^2 ds^2_{S^5/\Gamma}),
\label{st2}
\end{equation}
where $Z$ was given in eq.~(\ref{e17}).
We also used the fact that 
the radial coordinate $r^2=\sum_{i=1}^6x_i^2$ is unaffected by the projection since $\Gamma\subset SO(6)$.
We see from eq.~(\ref{st2}) that the near-horizon geometry is $AdS_5\times(S^5/\Gamma )$.
The presence of an $AdS$ factor predicts that the field theory 
on the world-volume of the D3-branes is conformal,
at least at large $N$ \cite{kacsil}.

The supersymmetries preserved by the orbifold projection
are determined by considering the action of $\Gamma$ on 
the holonomy group of the transverse space and are summarized as follows:
\begin{itemize}
\item{$\Gamma$ subgroup of $SU(2)$: $SU(4)\rightarrow SU(2)_R\times U(1)_R$
$\rightarrow$ ${\cal N}=2$ supersymmetry.}
\item{$\Gamma$ subgroup of $SU(3)$: $SU(4)\rightarrow U(1)_R$ $\rightarrow$
${\cal N}=1$ supersymmetry.}
\item{$\Gamma$ subgroup of $SU(4)$ $\rightarrow$
${\cal N}=0$ supersymmetry.}
\end{itemize}
In the previous list we also reported the subgroup of $SU(4)\sim SO(6)$
that survives the projection and appears as the R-symmetry of the brane world-volume theory.

Similarly, to determine the gauge theory living on branes on
 $\mathbb{R}^6/\Gamma$ \cite{dm}, we have to study the action of the orbifold 
projection on the world volume fields. 
For simplicity, we consider only abelian groups $\mathbb{Z}_k$. 
In the covering space
$\mathbb{R}^6$,  a D3-brane has $k-1$ images under $\mathbb{Z}_k$. We can think of 
a collection
of $k$ D3-branes as making a physical D3-brane. $\mathbb{Z}_k$ acts on 
the set of $k$ branes by a cyclic permutation: this is called 
the regular representation
of $\Gamma$. 
Before the projection, a set of $kN$ branes realizes a $U(kN)$ gauge theory. 
Let each element $\alpha \in \Gamma$ act on the Chan-Paton factors with
a matrix $\gamma_\alpha$ in the regular representation of $\Gamma$.
The projected theory is then obtained by 
\begin{eqnarray} A_\mu&=&\gamma_\alpha A_\mu\gamma^{-1}_\alpha,\nonumber\\
\lambda_a&=&R(\alpha )_{ab}\gamma_\alpha \lambda_b\gamma^{-1}_\alpha,\nonumber\\   
\phi_i&=&R(\alpha )_{ij}\gamma_\alpha \phi_j\gamma^{-1}_\alpha,\label{rule}\end{eqnarray}
where $i,j=1,\ldots,6$ and $a,b=1,\ldots,4$. 
The matrices $R(\alpha )$ take into account that the original $\N=4$ scalars 
and fermions
transform non trivially
under $SO(6)\sim SU(4)$ (in the $\bf{6}$ and the $\bf{4}$, respectively) and
therefore under its subgroup $\Gamma$.

As an example, consider an $\N=2$ theory: the orbifold $\mathbb{R}^4/\mathbb{Z}_2\times \mathbb{R}^2$.
Representing $\mathbb{R}^6$ with three complex coordinates $z_i$ , the action of $\mathbb{Z}_2$  is given by
\begin{equation}
z_1\rightarrow -z_1,
z_2\rightarrow -z_2, z_3\rightarrow z_3.
\label{s4}
\end{equation}
There is only one non trivial matrix $\gamma_\alpha$ corresponding to the generator of $\mathbb{Z}_2$ and it can be chosen as $\gamma_{\alpha} ={\rm diag}
\{ I_N,-I_N\}$. A simple application of the previous rules shows that the
gauge group is $U(N) \times U(N)$, with adjoint ${\cal N} = 2$ vector
multiplets and two bi-fundamental hypermultiplets.

The gauge theories obtained as projections have a characteristic quiver
(or moose) form. In the $\N=2$ case, a complete classification exists
\cite{dm} based on the Mc Kay correspondence \cite{mckay}. 
The discrete subgroups of $SU(2)$ are 
in one-to-one correspondence with the simply-laced Lie algebras
$A_k,D_k$ and $E_6,E_7,E_8$. The gauge theory on $N$ physical branes
at a singularity $\mathbb{R}^4/\Gamma\times \mathbb{R}^2$ is
associated with the affine Dynkin diagram of the Lie algebra corresponding
to $\Gamma$. A $U(n_iN)$ vector multiplet is associated with each node with Dynkin 
label $n_i$ and a bi-fundamental hypermultiplet is associated with each
link connecting two different nodes. The $\N=1$ case is considerably more
complicated. We refer to \cite{he} for a detailed discussion.
We just notice that, in the $\N=1$ case, the quiver theory inherits
a superpotential from the projection of the $\N=4$ one.

The Green functions for $\Gamma$-invariant operators
are also obtained by projection 
from $\N=4$ SYM: they are identical to those of the parent theory  in
 the large $N$ limit \cite{kacsil,LNV}. 
Notice, however, that the orbifold may have extra fields
and operators that are not invariant under $\Gamma$. 
In string theory, they come from twisted 
sectors. Their Green functions are obviously 
not determined by those of $\N=4$.

Finally, the $AdS/CFT$ correspondence predicts that all the orbifold 
theories constructed as above are $CFT$ at large $N$.  
It is easy to check that the  one-loop beta function is 
zero in all these theories \cite{kacsil,LNV} .

\subsection{Conformal field theories with $\N <4$: Conifolds}
\label{conifolds}

Another efficient way of obtaining $CFT$'s makes use of conifold singularities.
We place branes at the singularity of a six-dimensional Ricci-flat manifold 
$C_6$ whose metric has the conical form
\begin{equation}
ds_{C_6}^2=dr^2 + r^2ds_{H_5}^2.
\end{equation}
The supergravity solution for $N$ branes is of the form
\begin{equation}
ds^2=Z^{-1/2}(r)dx_{\mu} dx^{\mu}+Z^{1/2}(r)(dr^2+r^2 ds^2_{H_5}),
\label{st2222}
\end{equation}
with $Z$ given in eq.~(\ref{e17}).
One can prove that $C_6$ is a Calabi-Yau if $H_5$ is a  five-dimensional Einstein manifold 
\cite{kehag,kw1,morpless}.
The $AdS/CFT$ correspondence then applies for the background 
$AdS_5\times H_5$, which is the near horizon limit of the previous metric. 

Useful and simple Einstein manifolds are the cosets $G/K$, where $G$ 
and $K$ are Lie groups. 
There are only two supersymmetric examples in five dimensions: 
$S^5=SO(6)/SO(5)$ with $\N=8$ supersymmetry, 
corresponding to $\N=4$ SYM, and
$T^{1,1}=(SU(2)\times SU(2))/U(1)$ with $\N=2$ supersymmetry.
In this Section we discuss the solution corresponding to 
$AdS_5\times T^{1,1}$ \cite{kw1}.

The manifold $C_6$ relevant for this example  
can be written as a singular quadric in $\mathbb{C}^4$, $\, \sum_{a=1}^4 w_a^2=0\,$ \cite{candelas}, or equivalently 
\begin{equation}
{\rm det}W=0, \quad  (W\equiv\sigma^aw^a,\, \sigma=(\sigma^i,\, i{\mathbf1})), 
\label{quadric}
\end{equation}
$\sigma^i$ being Pauli matrices. This equation is invariant 
under $SO(4)\times U(1)_R\sim SU(2)\times SU(2)\times U(1)_R$.
The constraint~(\ref{quadric}) can be solved 
in terms of complex doublets $A_i,B_j$ ($W_{ij}\sim A_iB_j$)
satisfying
\begin{equation}
|A_1|^2+|A_2|^2= |B_1|^2+|B_2|^2, \quad A_i\sim e^{i\alpha}A_i,\,\, B_i\sim e^{-i\alpha}B_i .
\label{s8} 
\end{equation}

$C_6$ is a cone over $T^{1,1}$. 
The base of the cone is obtained by intersecting
$C_6$ with the sphere $\sum_{a=1}^4 |w_a|^2=1$, or, equivalently,
by restricting $\sum |A_i|^2=\sum |B_i|^2=1$ in eq.~(\ref{s8}).
In this way we obtain an equation for $(S^3\times S^3)/U(1)=
(SU(2)\times SU(2))/U(1)=T^{1,1}$. 
To write a metric on $T^{1,1}$ we can introduce the following basis of
 one forms
\begin{equation}
g^1 = {{e^1 - e^3}\over\sqrt2}, \quad g^2 = {{e^2 - e^4}\over\sqrt2},\quad
g^3 = {{e^1 + e^3}\over\sqrt2}, \quad g^4 = {{e^2 + e^4}\over\sqrt2},\quad
g^5 = e^5,
\label{fbasis}
\end{equation}
with
\begin{eqnarray}
e^1= -\sin\theta_1d\phi_1, \qquad e^2= d\theta_1, \qquad e^3= \cos\psi\sin\theta_2d\phi_2 - \sin\psi d\theta_2, \nonumber \\
e^4= \sin\psi\sin\theta_2d\phi_2 + \cos\psi d\theta_2, \qquad e^5= d\psi + \cos\theta_1d\phi_1 + \cos\theta_2d\phi_2.
\end{eqnarray}

An Einstein metric on $T^{1,1}$ is
\begin{equation}
ds_{T^{1,1}}^2 = {1\over 9}(g^5)^2 +{1\over 6}\sum_{i=1}^4(g^i)^2 
={1\over9} (d\psi + \sum_{i=1}^{2}\cos\theta_{i}d\phi_{i})^2 + {1\over6}\sum_{i=1}^{2}(d\theta_{i}^2 + \sin^2\theta_{i}d\phi_{i}^2).
\label{t11metric}
\end {equation}
The angular variable $\psi$ ranges from $0$ to $4\pi$, while 
$(\theta_1, \phi_1)$ and $(\theta_2, \phi_2)$
parameterize two $S^2$'s in the  standard way. 
The expression above shows that $T^{1,1}$ is
an $S^1$ bundle over $S^2\times S^2$. The metric is invariant under
$SU(2)\times SU(2)\times U(1)_R$, where the $SU(2)$ factors act on
the two $S^2$ and $U(1)_R$ shifts the angle $\psi$.
By forgetting an $SU(2)$, $T^{1,1}$ can be also written as an $S^3$
bundle over $S^2$. It can be proved that such bundle is topologically
trivial (see for instance \cite{kw1}), so that $T^{1,1}$ is isomorphic
to $S^3\times S^2$. In particular, $T^{1,1}$ has non-trivial
two and three cycles where we could wrap D-branes. In Sections 3.3
and 5.3 we will need to wrap D5-branes on a two cycle; for the
metric~(\ref{t11metric}) a minimal
volume $S^2$  is parameterized by
$\theta_1=\theta_2, \phi_1=-\phi_2$.

It is difficult, in general, to determine the world-volume theory of
branes sitting at singularities different from the orbifold ones.
A powerful hint in this direction is provided by the observation that
the space transverse to the branes should describe the moduli space 
of the gauge theory. 
In our case, equations~(\ref{s8}) can be viewed as the D-terms of an
$\N=1$ abelian gauge theory \cite{kw1} $U(1)\times U(1)$ with two sets 
of chiral multiplets 
$A_i$ and $B_i$ with charges $(\mathbf{1},\mathbf{-1})$ and $(\mathbf{-1},\mathbf{1})$, respectively.
Here the diagonal $U(1)$ factor is decoupled while the other linearly 
independent combination of the $U(1)$'s
acts as in eq.~(\ref{s8}). 
We  identify this theory  with that living on the world-volume of a 
brane placed at the conifold singularity. The moduli space of vacua of 
such abelian 
$\N=1$ theory is in fact identical to $C_6$. 
When we consider a stack of $N$ parallel D3-branes at the singularity, we have to extrapolate this result to 
the non-abelian case. We then consider a $U(N)\times U(N)$ theory with two sets of
chiral fields $A_i,B_i$ transforming in the representations
({\boldmath $N$},{\boldmath $\overline{N}$}) 
and ({\boldmath $\overline{N}$},{\boldmath $N$}). We must also  add to the theory the superpotential
\begin{equation}
W=h \, \epsilon_{ij} \, \epsilon_{pq}Tr(A_iB_pA_jB_q).
\label{s10}
\end{equation}
Such superpotential respects all the symmetries of the model and
is crucial for avoiding a proliferation of geometrically-redundant
non-abelian modes \cite{kw1}. 
The global symmetry of the $CFT$ is $SU(2)\times SU(2)\times U(1)_R$,
which corresponds to the isometry of $T^{1,1}$.

There are various strong checks that the identification is correct.
First of all, the theory has to be conformal. Using the results of \cite{kw1}
it can be rigorously proved that this non-abelian gauge theory flows
at low energies to an interacting conformal field theory. 
Indeed, even though the theory  depends on various parameters, 
the couplings $g_{YM,i}$ and $h$, the conditions for conformal 
invariance \cite{LS} impose a single relation among them \cite{kw1}.  
For both groups, the vanishing of
the exact NSVZ beta functions \cite{NSVZ}\footnote{In $\N=1$ gauge theories,
 if we use a holomorphic scheme,
the beta function is completely determined  at 1-loop. From this result one can then
deduce the following beta function for the 1PI coupling
$\mu\frac{d}{d\mu}\frac{8\pi^2}{g_{YM}^2}= f(g_{YM})(3 C_2(G) -\sum T(R_a)(3-2\Delta_a))$
where $C_2$ is the second Casimir of the group $G$,
$T(R_a)$ are the dimensions of the representations
$R_a$ of the matter fields, and $f(g_{YM})$ is a positive scheme dependent function of the coupling.
With a Pauli-Villars regularization $f(g_{YM})= 1/(1-Ng_{YM}^2/8\pi^2)$.
The knowledge of $f(g_{YM})$ is not necessary when imposing
the scheme independent condition $\beta(g_{YM})=0$.} gives
the relation 
\begin{equation}
\mu\frac{d}{d\mu}\frac{8\pi^2}{g_{YM,i}^2}\sim 3 C_2(G) -\sum T(R_a)(3-2\Delta_a)= N (2(\Delta_A+\Delta_B)-3)=0,
\label{betaex}
\end{equation}
where $\Delta_{A,B}(g_{YM,i},h)$ are the dimensions of the fields 
$A_i$ and $B_j$. These dimensions 
do not depend on the indices  $i,j$ due to the $SU(2)\times SU(2)$ 
invariance.
When~(\ref{betaex}) is satisfied, the last condition, which requires 
that the superpotential has scaling dimension three \cite{LS},
is automatically satisfied. We are thus left with
 a manifold of fixed points, defined implicitly by the requirement
that the dimension of the gauge invariant operator  $Tr(AB)$ is 3/2. 
As a further check, the complete KK spectrum of Type IIB compactified on $T^{1,1}$ has been 
computed \cite{ferrara}, 
finding a complete agreement with $CFT$ expectations. Let us recall that all $U(1)$ factors are not
described by the $AdS$ dual. 
In this case one of them is decoupled while the other reduces to a global baryonic symmetry. The
existence in this model 
of solitonic objects dual to baryons (obtained as D3-branes wrapped on
$S^3$) \cite{gk} is the best evidence that the $U(1)$'s 
are not dynamical.

\section{Breaking conformal invariance I}\label{nonconfI}

There are various ways to
construct string duals of non-conformal gauge theories. 
Since the conformal group is equivalent on the supergravity side of
the correspondence to the isometry group of $AdS_5$, one 
can for instance consider small deformations of the $AdS$ background.
In this case the background still asymptotes to 
$AdS_5 \times S^5$ and  we can still easily apply the $AdS/CFT$
dictionary.
Alternatively, one can consider completely different geometries
generated by fractional and wrapped branes in singular spaces.
Another  possibility, 
which we will not discuss here, is to
consider theories at finite temperature. Historically this was the first
example of non-conformal gauge/gravity
duals. We refer the reader  to \cite{witten2,magoo} 
for a discussion of the subject. 
We will describe the systems of fractional and wrapped branes in the next
Section, while here we will focus on the  deformations of Anti-de Sitter.

Notice that in all such constructions we will be eventually able 
to perform predictive calculations only in the limit 
where the supergravity approximation is valid. It is then difficult 
to study the complete dynamics of ``realistic'' theories such as 
pure Yang-Mills in this context.
To understand this point, consider a specific example. We can
obtain  pure YM 
 by adding a mass deformation $M$ to a $CFT$ that 
possesses a holographic dual, 
 for example $\N=4$ SYM. 
The mass parameter induces a dimensionful scale $\Lambda\sim Me^{-1/Ng^2_{YM}}$.
The limit where the low energy theory decouples from the $CFT$ 
is $M\rightarrow\infty$, $x=Ng_{YM}^2\rightarrow 0$, with $\Lambda$ fixed. 
However, we can trust supergravity in the opposite limit $x\gg 1$.
Thus the description of the low energy pure YM theory 
requires the knowledge of the full string theory. 
Similar arguments apply to all the non-conformal models constructed so far. 
The expectation
that the spectrum of bound states in any realistic model should contain
higher spin glue-balls suggests that more than supergravity
is required to describe the pure YM theory. 
In the previous example, it would be sufficient to re-sum all 
world-sheet $\alpha^\prime$ corrections
in the string background to correctly describe pure YM in the large $N$ limit. 
World-sheet corrections
are, in principle, more tractable than loop corrections. In flat space, for example, all the
$\alpha^\prime$ corrections are computable. In the $AdS$ case, the analogous computation is made
difficult by the presence of RR-fields. 
In this review, we will mostly remain in the supergravity regime. 
We may take various attitudes towards the solutions we will find.
In the previous example, we may consider the supergravity solution as 
a description of pure YM 
with a finite cut-off $\Lambda\sim M$. The situation is similar, 
in spirit, to
a lattice computation at strong coupling. In general, in all the models 
discussed so far,
the supergravity solution describes a YM theory with many non-decoupled 
massive modes. 
These theories can be considered as cousins of pure YM, and they have 
often the same qualitative
behavior. At present we have many examples of theories that are, 
in a certain sense,
generalization of pure glue theories. They are interesting as exactly 
solvable toy models.
Moreover, it is interesting to investigate which properties of pure YM, 
that  are not
consequences of symmetries, are also realized in these generalized models.

\subsection{The radius/energy relation}

A crucial ingredient in all the models obtained by the 
$AdS/CFT$ correspondence
is the identification of the radial coordinate in the supergravity 
solution with an energy scale in the dual field theory.

Let us first consider a conformal field theory and its $AdS$ dual.
The identification between radius and energy
follows from the form~(\ref{AdSmetric}) of the $AdS$ metric. 
A dilatation $x_{\mu}\rightarrow \lambda x_{\mu}$ in the boundary $CFT$
corresponds in $AdS$ to the $SO(4,2)$ isometry 
\begin{equation}
x_\mu\rightarrow \lambda x_\mu,\qquad
y\rightarrow y-R\log\lambda.
\end{equation}
We see that we can roughly identify $e^{y/R}$ with an
energy scale $\mu$. The boundary region of $AdS$ ($y\gg 1$)
is associated with the UV regime in the $CFT$, while the horizon region ($y\ll 1$) is
associated with the IR. 
This is more than a formal identification: holographic 
calculations of Green functions or Wilson loops associated
with a specific reference scale $\mu$ are dominated by bulk contributions from the region
$y=R\log\mu$. Examples and further references can be found in \cite{magoo}. 

Obviously, a change of scale in a $CFT$ has little physical
meaning. In a non conformal theory, however, the quantum field theory couplings
run with the scale. This suggests we can interpret the running couplings 
in terms of a specific radial dependence of the fields in the supergravity 
solution.
Moreover, we are also lead to interpret solutions interpolating between different
backgrounds  as an holographic realization of the 
Renormalization Group (RG) flow between the dual QFTs.
This interpretation works very well at the qualitative level and
we will see many explicit examples in this review.
As in the $AdS$ case, the region with large (small) radius will be associated
with the UV (IR) dynamics of the gauge theory.
However, the quantitative identification of the radius with the scale 
can be difficult to find. For non-conformal theories the precise form of the relation 
depends on the physical process we use to determine it \cite{pp}. 
The radius/energy relation can be found for instance by considering the
warp factor multiplying the flat four-dimensional part of the metric $ds^2=Z(y)dx_\mu dx^{\mu}+ \dots$,
since $Z(y)$ is a redshift factor connecting the
energies of observers at different points
in the bulk: $Z(y^\prime)^{-1/2} E^\prime = Z(y)^{-1/2} E$. Alternatively, 
we can compute  a Wilson loop in supergravity \cite{magoo}:
the energy of a string
stretched between the boundary and a fixed IR reference radius
represents in the gauge theory the self-energy of a quark.   
Finally, one can
also extract the radius/energy relation by analyzing the equation of motion of
a supergravity mode with fixed four-dimensional momentum. 
While for conformal theories  all the different methods 
give the same result, 
this is no longer true for gravity duals
of non-conformal theories. In particular, it is known \cite{pp} that for the non-conformal
six-dimensional theories living on D5-branes, the radius/energy identification can be ambiguous.
This will make the extension of the $AdS/CFT$ 
dictionary to systems with wrapped branes somehow less clear. Also in the relatively well
understood case of the Klebanov-Strassler solution, the different prescriptions give different results
\cite{HKO}.
We will be more fortunate in the case of $\N=2$ theories, where supersymmetry and
the existence of a moduli space will give a natural method for determining the
radius/energy relation.

\subsection{Deformations of ${\cal N}=4$ Super Yang-Mills}

To break the conformal invariance of  ${\cal N}=4$ Super Yang-Mills 
we introduce a scale in the theory. This can be done either by 
deforming the action with gauge invariant operators
\begin{equation}
S\rightarrow S+\int d^4x\, h_i\,O_i(x),
\end{equation}
or by considering the theory for non-zero VEV of some operators 
$<O_i(x)>$.
For energies lower than the deformation
scale, the coupling will run and we expect a Renormalization Group flow to the
IR. Depending on the deformation the theory could flow to an IR fixed point, 
or develop a non trivial IR non-conformal dynamic, like confinement.
The choice of the deformation also determines the amount of preserved
supersymmetry. 

Typically one considers relevant or marginal deformations, i.e. operators with
classical conformal dimension $\Delta \le 4$. This is because we want the
deformation to affect the IR dynamic of the theory, being negligible in the
UV. As we saw in Section 1.2, almost all the mass terms for scalars and fermions
have duals in the KK tower and can be described in the supergravity
approximation. The only exception is a diagonal mass term for the scalars,
$Tr\phi_i\phi_i$, whose dual operator is a genuine string state.
A mass term generically breaks all
supersymmetries, but we can also easily consider supersymmetric mass deformations.
The classical examples are 
\begin{equation}
\delta {\cal L} = \int d\theta^2 \left( \sum_{i,j=1}^3 m_{ij} 
Tr(\Phi_i \Phi_j) + h.c \right) = \sum_{i,j=1}^3
\left(m_{ij} \lambda_i \lambda_j + m_{ik}m_{kj}^* \phi_i^{\dagger}  
\phi_j\right),
\end{equation}
where $ \Phi_i$ are the three chiral multiplets of ${\cal N}=4$ in 
${\cal N}=1$ notation. The IR theory is generically non-conformal:
$m_{ij}=  \delta_{ij}$  breaks to ${\cal N}=1$ SYM, 
$m_{11}=m_{22}=m$ and  $m_{33}=0$ gives ${\cal N}=2$ SYM, 
and finally 
$m_{11}=m_{22}=m$ and  $m_{33}=M$ with $M \le m$ gives the soft breaking
from  ${\cal N}=2$ to  ${\cal N}=1$.
Since we will always consider those theories in regimes where the
massive modes are not decoupled, we will denote these theories with a star, for example: ${\cal N}=2^*$.
In some case we can get an IR fixed point. It can be shown
that the deformation $m_{11}=m_{22}=0$ and  $m_{33}=m$ 
flows to the conformal Leigh and Strassler fixed point \cite{LS}.

Similarly, a simple 
example of spontaneous symmetry breaking by non-zero VEVs is
provided by the Coulomb branch of ${\cal N}=4$ SYM, where the operators
$Tr\Phi^k$ acquire a VEV.

\subsection{The dual supergravity solutions}\label{thedual}

The construction of  the supergravity duals of deformed ${\cal N}=4$
SYM relies on a simple application of the $AdS/CFT$ dictionary \cite{gppz1,dz,freed1}, 
using the map between gauge invariant operators and supergravity states, and the radius/energy relation.
The idea is to look for IIB solutions with a non trivial
radial dependence and interpret them as RG flows in the dual gauge theory.
The  candidate backgrounds will be of the form
\begin{eqnarray}
ds^2 &=& F \left(dy^2 + e^{2 Y(y)} dx^{\mu} dx_{\mu}\right) + G ds^2_H,
\nonumber\\
\varphi&=& \varphi(y),
\end{eqnarray}
where $H$ is the internal 5$d$ manifold, $F,G$ are generic warp factors and 
$\varphi$ is the
supergravity field dual to the operator $O(x)$. 
Notice that the 5$d$ space-time part of the metric is not any longer $AdS$, consistently with the fact that the field theory is not conformal;
the ansatz is dictated by the requirement of Poincar{\'e}
invariance of the dual field theory, which only leaves undetermined a 
single function, 
the 5$d$ warp factor $Y$.
For large values of $y$, interpreted as the UV region, the solutions 
are asymptotic to 
$AdS_5 \times S^5$ with the field dual to the gauge theory deformation turned
on. This translates into boundary conditions for the 4$d$
dimensional warp factor $Y$ and the field $\varphi$: $Y \rightarrow y/R$ and
$\varphi(y) \rightarrow 0$ for $y\rightarrow\infty$.
For small values of $y$, corresponding to the IR region, 
the geometry of the solution can be completely
different. If the dual gauge theory has an IR fixed point, we expect the
background to be of the form $AdS_5 \times H_{IR}$, where the $AdS$ factor
reflects the restoration of conformal invariance at the fixed point. Usually the IR
$AdS_5$ has a different cosmological constant (and a different radius $R$) 
from the UV one, corresponding 
to a different number of degrees of freedom in the dual gauge theory.
Alternatively, a non conformal gauge theory should correspond to a geometry with
a horizon or a singularity.
All along the flow  the isometries of the internal part of the metric
determine the global symmetries of the dual gauge theory.

An important point in the identification of the gauge and gravity sides is the
fact that supergravity
solutions can represent both deformations of a $CFT$ and different vacua of the
same theory \cite{bala,kw2}. 
The asymptotic UV behavior of the solutions discriminates between the 
two options. To this extent it is enough to look at the 5$d$ space-time part of
the solution. 
In the asymptotic $AdS$ region, we just need a linearized analysis.
The fluctuation $\varphi(y)$ for a minimally coupled scalar field with mass $m$ 
in the asymptotically $AdS$ background satisfies 
\begin{equation}
\varphi'' +\frac{4}{R}\varphi'=m^2 \varphi,\label{I2}
\end{equation}
where primes denote derivatives with respect to $y$.
The previous equation has a solution depending on two arbitrary
parameters
\begin{equation}
\varphi(y)=Ae^{-(4-\Delta )\frac{y}{R}}+Be^{-\Delta \frac{y}{R}},\label{I3}
\end{equation}
where $\Delta$ (see also Section 1.2) is the dimension of the dual operator,
$m^2=\Delta (\Delta -4)/R^2$ \cite{gkp,witten1}.
We are interested in the case of relevant operators, where $\Delta\le 4$.
We associate
solutions behaving as $e^{-(4-\Delta )\frac{y}{R}}$
with deformations of the ${\cal N}=4$ theory
with the operator $O$. On the other hand, solutions asymptotic
to $e^{-\Delta \frac{y}{R}}$ (the subset with $A=0$) are associated with a different
vacuum of the UV theory, where the operator $O$ has a non-zero
VEV\footnote{We are not careful about 
subtleties for particular values of $\Delta$ \cite{kw2}.} \cite{bala,kw2}.

Solutions of Type IIB equations of motion with the above properties are
difficult to find, even at the perturbative level. However for many of 
the cases at hand, it is sufficient to consider a lower dimensional truncation
of the theory, namely 5$d$   ${\cal N}=8$ gauged supergravity, with gauge
group $SO(6)$ \cite{warn}.
This  is the low energy effective
theory for the ``massless'' modes of the compactification of Type IIB on
$AdS_5 \times S^5$. It is believed to be a consistent
truncation of Type IIB on $S^5$ in the sense that every
solution of the 5$d$ theory can be lifted to a consistent 10$d$ Type IIB
solution.
5$d$ ${\cal N}=8$ gauged supergravity has 42 scalars, which
transform as the 
$\bf{1_c}$, $\bf{20}$, and $\bf{10_c}$ of $SO(6)$ (the ${\cal N}=4$ SYM 
R-symmetry $SU(4)$).
The singlet is associated with the marginal deformation
corresponding to a shift in the complex 
coupling constant of the $\N=4$ theory.
The mode in the $\bf{20}$ has mass square $m^2=-4$ and is
associated with a symmetric traceless mass term for the scalars
$Tr(\phi_i\phi_j)$, ($i,j=1,...,6$) with $\Delta = 2$.
The $\bf{10_c}$ has mass square $m^2=-3$ and corresponds to 
the fermion mass term $Tr(\lambda_a\lambda_b)$, ($a,b=1,...,4$) of
dimension 3.
Thus the scalar sector of ${\cal N}=8$ gauged supergravity is enough to discuss
at least all mass deformations
that have a supergravity description.

The five-dimensional Lagrangian for the scalars of ${\cal N}=8$ gauged
supergravity  \cite{grw2}
\begin{equation}
{\cal L} = \sqrt{-g}\left[-{{\cal R}\over 4}
- {1\over 24} Tr (U^{-1}\partial U)^2 + V(U)
\right]\label{m32}
\end{equation}
is written in terms of a
$27\times 27$ matrix $U$, transforming in the fundamental
representation of
$E_6$ and parameterizing the coset $E_6/USp(8)$.
In a unitary gauge, $U$
can be written as $U=e^X, X=\sum_A \varphi_A T_A$,
where $T_A$ are the
generators of $E_6$ that do not belong to $USp(8)$.
This matrix has exactly 42 real independent parameters, which are the
scalars of the supergravity theory.
Typically the solutions we are looking for only involve a small subset
of the 42 scalars, those dual to the gauge theory deformation.  Thus
by a suitable truncation and parameterization of the coset element
$U$, eq. (\ref{m32}) can be reduced to the Lagrangian 
for some scalars minimally coupled to gravity. 
The non trivial scalar  potential $V$ is typical of gauged
supergravities and has only isolated minima  (apart from one flat
direction, corresponding to the dilaton). 
There is a central critical point with
$SO(6)$ symmetry and with all the 
scalars $\varphi$ vanishing: it corresponds to the unperturbed ${\cal N}=4$ 
SYM theory.
Non-zero VEVs of some of the scalars characterize minima
where part of the gauge group is spontaneously broken. Those 
other minima should correspond to IR conformal field theories.

With a metric of the form $ds^2= dy^2 + e^{2 Y(y)}dx_\mu dx^\mu$,
a standard computation shows that the Einstein and scalar equations of
motion following 
from eq.~(\ref{m32}) can be deduced from the
effective Lagrangian
\begin{equation}
{\cal L} = e ^{4 Y} \left[ 3 \left( \frac{d Y}{dy}\right)^2 
- {1\over 2} G_{ab} 
\frac{d \varphi^i}{d y} \frac{d \varphi^j}{d y}
  - V(\varphi)\right ],
\label{5dlagr}
\end{equation}
supported by the zero energy constraint $3(Y')^2 - \frac{1}{2} G_{ab} 
(\varphi^i)'(\varphi^j)' + V(\varphi)=0$.
The independent
equations of motion and constraints read
\begin{eqnarray}
\frac{d}{d y}\left(G_{ij} \frac{d\varphi^j}{d y}\right) 
+4 G_{ij} \frac{d  Y}{d y} \frac{d \varphi^j}{d y}=
{\partial V\over\partial\varphi^i},
\nonumber\\
6\left( \frac{d Y}{d y} \right)^2=G_{ij} \frac{d \varphi^i}{d y}
\frac{d \varphi^j}{d y} -2V.\label{I7}
\end{eqnarray}  
Thus the problem of finding interpolating solutions of IIB
supergravity reduces to finding solutions of the above equations 
that for large values of $y$ tend to the maximally symmetric vacuum
(on the gauge theory side the UV theory is ${\cal N}=4$ SYM). 
However, the presence of the potential $V$, which generally is an
exponential in the
scalar fields, makes such solutions not very easy to find. For most of 
the flows interpolating between two fixed points, the best one can do is to
prove that such solutions exist.
Things are simpler when some supersymmetry is preserved. In these cases, one can look
for solutions for which the fermionic shifts
vanish, thus reducing the second order equations to first order ones.

In refs.~\cite{freed1,townsk} 
the conditions for a supersymmetric flow were found\footnote{Analogous 
BPS domain wall solutions were originally found for four-dimensional
supergravity (see \cite{cveticrey}).}.
For a supersymmetric solution, the potential $V$
can be written in terms of a superpotential $W$ as
\begin{equation}
V = \frac{1}{8} G^{ij} 
\frac{\partial W}{\partial \varphi^i} 
\frac{\partial W}{\partial \varphi^j} 
- \frac{1}{3} \left|W \right|^2.\label{I9}
\end{equation}
The equations of motion reduce to
\begin{eqnarray}
\frac{d \varphi^i}{d y}&=&\frac{1}{2} G^{ij}\frac{\partial W}{\partial \varphi^j},\nonumber\\
\frac{d Y}{d y}&=& - \frac{1}{3} W.\label{I10}
\end{eqnarray}
It is easy to check that a solution of eqs. (\ref{I10}) satisfies also the
second order equations ~(\ref{I7}).
Supersymmetry also helps in  unambiguously identifying the UV
behavior of the solutions. 
Close to the boundary, we can 
always find a basis where the scalar fields are canonically normalized
and the superpotential $W$ has the expansion (this is actually possible
around any minimum of the potential) 
\begin{equation}
W = -\frac{3}{R} + \frac{1}{2} \left. \frac{\partial^2 W}{\partial \varphi^i
  \partial \varphi^j }\right|_{\varphi_i=0} \varphi^i \varphi^j + \ldots
\end{equation}
Notice that the value of the superpotential for zero VEV is related to
the cosmological constant of $AdS_5$. From the mass matrix 
$W_{ij}=\frac{\partial^2 W}{\partial \varphi^i \partial\varphi^j }$
we can read the UV asymptotics of the fields dual to the
gauge theory deformations. 
More precisely, for a diagonal $W_{ij}= \frac{2 w_i}{R} \delta_{ij}$\footnote{For the non diagonal
case the same reasoning applies after
  diagonalization of the mass matrix $W_{ij}$.},
we have
\begin{equation}
w_i= \left\{ \begin{array}{c@{~\Rightarrow~}l} 
- \Delta & {\rm VEV,} \\
\Delta -4 & {\rm deformation.} 
\end{array}
\right. 
\end{equation}
The last step in the construction of the supergravity solutions is the lift 
to ten dimensions. This is necessary for a correct holographic 
interpretation of the flows, since the 5$d$ solutions encode in a very 
complicated way the gauge theory information.
A typical example is the identification of the gauge coupling constant 
with the dilaton, which is correct only in 10$d$, since the 5$d$ dilaton is
always constant in these solutions.
The knowledge of the  ten-dimensional solutions is also needed
to address another common problem of such
supergravity solutions, namely
the presence of a naked IR singularity. Usually the 10$d$ geometries are 
still singular but the singularities are milder and may have
a physical interpretation as distributions of D-branes or other extended objects 
\cite{freed2, bpp, michela,ps}.

The procedure for the lift to ten dimension is known in principle
\cite{pw,pwN1}. The 10$d$ metric is expected to be
\begin{equation}
ds^2 = \Omega^{-2/3} ds_5^2 + ds_{\tilde{S}^5}^2,
\end{equation}
where $ds_5^2$ and $ds_{\tilde{S}^5}^2$ are the metric of the 5$d$ solution
and the deformed five-sphere, respectively. The warp factor $\Omega$ is a 
function of the deformed five-sphere metric and it is usually responsible 
for the mildening of the IR singularity. 
The ansatz for the dilaton and the metric of the deformed $\tilde S^5$ 
are given in full generality in terms of the scalar coset element $U$ 
\cite{pw,pwN1}.
Then the only difficulty relies in the explicit computation which can
be quite awkward depending on the scalars
involved in the solution.
More complicated is the ansatz for the RR forms, which has to be
guessed for every solution on the basis of the symmetries of the
problem \cite{pw,pwN1}.

We end this Section  with a short list of the known five-dimensional
solutions and their lifts. 
For deformations flowing to IR fixed points, the following $CFT$ theories 
can be obtained: 
\begin{itemize}
\item{Three ${\cal N}=0$ theories with symmetry $SU(3)\times U(1)$, 
$SO(5)$ and $SU(2)\times U(1)^2$ \cite{pilch,gppz1,dz}. 
All these theories are unstable
and correspond to non-unitary $CFT$s. 
} 
\item{A stable ${\cal N}=1$ theory with symmetry $SU(2) \times U(1)$. 
It corresponds to the $\N=4$ theory deformed with a mass for one of the three
${\cal N}=1$ chiral superfields. The results and the supergravity description
\cite{lust,freed1} are almost identical to the $T^{1,1}$ case discussed in Section 1.5,
which is a sort of $\mathbb{Z}_2$ projection of this example. The 10$d$ lift can be found 
in \cite{pw}.}
\end{itemize}
The solutions dual to non-conformal  gauge theories are:
\begin{itemize}
\item{An ${\cal N}=1$ solution with residual symmetry $SU(3)$ \cite{gppz3}.
It is dual to the flow from ${\cal N}=4$ to ${\cal N}=1$, after
soft breaking with a mass term for the 
chiral multiplets. It has mass gap and gaugino
condensates, and is one of the few solutions
 known analytically. The 10$d$ solution
has still a mild singularity \cite{pwN1}.}
\item{Solutions corresponding to the Coulomb branch of $\N=4$ or $\N=2$
theories. The solutions for the Coulomb branch of $\N=4$ SYM \cite{freed2}
have various residual symmetries. 
The 10$d$ lifts correspond to distributions of branes. The family 
of  ${\cal N}=2$ solutions \cite{naked,pw,sfetsos} has residual symmetry 
$SU(2) \times U(1)$ and corresponds to points on
the moduli space of ${\cal N}=4$ broken to ${\cal N}=2$ by 
a mass term for two chiral multiplets. In this case the lift is completely
known \cite{pw} and presents an enhan\c con type of 
singularity \cite{jpp,bpp, michela}. 
It will be discussed in Section \ref{ndue}.}
\item{Solutions describing other patterns of supersymmetry breaking,
including a subsequent breaking
 ${\cal N}=4 \rightarrow {\cal N}=2 \rightarrow {\cal N}=1$, by giving
equal masses to two chiral multiplets and a smaller one to the third \cite{ep},
and examples of ${\cal N}=0$ solutions \cite{evans}.}
\end{itemize}
In  \cite{ps} another  ${\cal N}=1$ solution has been
constructed directly in 10$d$ using configurations of polarized D3-branes.
We will give a brief description of such a solution in Section \ref{N1star}.

\subsection{Holographic RG flow and the $c$-function}

The identification of the radial coordinate of $AdS_5$ with the energy
of the dual gauge theory motivates the interpretation of the
supergravity solutions as Renormalization Group flows. The radial 
profile 
of the scalars can be  associated 
with the running of the coupling constants in
the gauge theory. However, as already mentioned, 
the precise identification of the couplings in the five-dimensional solutions 
is often ambiguous and only in the ten-dimensional solution the dictionary can be reliably applied. 
Here we want to stress that in spite of the above limits, it is possible to 
extract interesting  results already in the
five-dimensional approach. 
One general result about
these classes of solutions is the existence of a $c$-theorem.
For the class of field theories that have a supergravity dual one can
define a $c$-function. In a $CFT$, the central charge $c$ is defined via the OPE of two
stress-energy tensors. On the supergravity side, it corresponds to the
cosmological constant at the critical points of the potential \cite{hs,g}. 
In fact, from eq.~(\ref{m32}), we can see by a simple scaling that, at a 
fixed point, 
\begin{equation}
\langle T(x) T(0)\rangle = {c\over |x|^8}\,\,\Rightarrow c\sim R^3\sim (\Lambda )^{-3/2}.
\label{scaling}
\end{equation}
More interestingly, all along the flow it is possible to define a
$c$-function that is monotonically decreasing \cite{gppz1,freed1} 
$c(y)\sim (Y')^{-3}$ and reduces to the previous result at the fixed points.
The monotonicity of $c$ can be easily checked from the equations of motion~(\ref{I7})
and the boundary conditions of the flow \cite{gppz1}. It can be also related to the weak
positive energy condition \cite{freed1} 
that is expected to hold in all physically relevant supergravity solutions.
Let us stress that the value of $c$ is well defined only at a fixed point, where
it represents a central charge. In QFT, the value of $c$ along the flow  
is scheme dependent. Similarly, in supergravity there are several possible 
definitions
of monotonic functions interpolating between the central charges at the
 fixed points \cite{gppz1,freed1,verlinde,ans}. 

To strengthen the holographic RG flow interpretation, 
some attempts to identify more precisely the five-dimensional 
equations of motion in supergravity with 
the renormalization group equations have been made in \cite{porr,verlinde}. 
Also, 
correlation functions along some of the supersymmetric flows have 
been explicitly
computed using the $AdS/CFT$ prescription \cite{bianchi}.

\section{Breaking conformal invariance II}\label{nonconfII}

In this Section we will discuss wrapped and fractional branes.
We will only consider
the engineering of the gauge theory in terms of systems of branes. 
The holographic duals will be discussed in the
next Sections. Here and in Section \ref{ndue}
we will mainly work in units $2\pi\alpha^\prime=1$.

\subsection{Fractional and wrapped branes: General observations}\label{genobs}

Most of the recently proposed duals of non-conformal theories are based on 
wrapped and fractional branes. 
The philosophy may be exemplified in the $4d$ case as follows.
Consider a geometry with a non-trivial two-cycle $S^2$ 
on which we wrap a D5-brane. The world-volume of the brane is thus of the 
form ${\mathbb{R}}^4\times S^2$, and at energies lower than the inverse 
radius of $S^2$ the theory living on the world-volume is effectively four 
dimensional. String theory has many moduli, some geometrical
in nature and some related to the bundles of antisymmetric forms which are 
always present in string theory. 
For simplicity, we focus on two specific moduli associated with $S^2$: 
the volume of $S^2$ and the integral of the $B$-field over the cycle. Only the first 
modulus has a geometrical meaning.
These moduli appear in the Born-Infeld action for the D-brane\footnote{Our conventions
for the BI action for a Dp-brane are: $\frac{-1}{\alpha^{\prime (p+1)/2}(2\pi)^p} \int d x^{p+1}
e^{-\Phi}\sqrt{G+(2\pi \alpha^\prime F +B)}$, $G+(2\pi \alpha^\prime F +B)\equiv -det(G_{ab}+(2\pi \alpha^\prime F_{ab} +B_{ab}))$.}
\begin{eqnarray}
& &-\frac{1}{(2 \pi)^2} \int dx^6 \, e^{-\Phi}\sqrt{G+F+B} = \qquad \qquad \nonumber\\
& & =  -\frac{1}{(2 \pi)^2}  \int dx^4\left[ e^{-\Phi}\int_{S^2}d\Omega_2 \sqrt{(G+B)_{S^2}}\right]\sqrt{(G+F+B)_{\mathbb{R}^4}}.
\end{eqnarray}
We see, by expanding the last square root, that the four dimensional gauge theory has an effective coupling which reads
\begin{equation}
{1\over g^2}\sim e^{-\Phi}\int_{S^2} d\Omega_2\sqrt{(G+B)_{S^2}}.\label{costanti}
\end{equation}
Whenever the quantity on the r.h.s. of this equation runs, also the coupling does, and the resulting theory is non-conformal.
We can then have two basic different models:
\begin{itemize}
\item{Wrapped branes: configurations of D5-branes wrapped in a supersymmetric fashion on a non-vanishing two-cycle $Vol(S^2)\neq0$.
There is no need to introduce a $B$-field.}
\item{Fractional branes: configurations of D5-branes wrapped
on collapsed cycles. If $\int_{S^2}B\neq0$, the corresponding four-dimensional theory has still a non-vanishing well-defined coupling constant. 
Manifolds with collapsed cycles are singular, and fractional branes must 
live at the singularity. We discussed some examples of singular manifolds
in Section \ref{uno}.}
\end{itemize}
The amount of supersymmetry preserved in these kinds of model depends on how the $S^2$ is embedded in the background geometry.

\subsection{Wrapped branes}\label{wrapped}

Wrapping a brane on a generic cycle breaks supersymmetry.
It turns out that the conditions for a cycle to be supersymmetric
are equivalent to the partial twist of the brane theory \cite{vafa}.
To understand what this means consider the case of $N$ IIB D5-branes 
wrapped on a two-sphere. Using S-duality, we can equivalently think in terms of NS5-branes.
A 4$d$ supersymmetry is preserved if and only if there 
exists a covariantly constant spinor on the sphere
\begin{equation}
(\partial_\mu +\omega_\mu)\epsilon=0,
\end{equation}
where $\omega_\mu$ is the spin connection.
It is well known that the sphere admit no covariantly constant spinors.
However, the theory contains other fields:
for example, the external gauge fields $A_\mu$,  
which couple to the $SO(4)$ R-symmetry currents. 
One can then redefine the covariant derivative as to 
include a gauge connection $A_\mu$ in a $U(1)$ 
subgroup of the R-symmetry group 
\begin{equation}
D_\mu=\partial_\mu +\omega_\mu + A_\mu.
\end{equation}
This operation is called a $twist$ of the original theory.
The new theory obtained this way can be shown to be topological.
In our case the twist is made only in the directions tangent to the sphere, so
that the remaining flat four dimensions will support an ordinary field theory.

We can preserve supersymmetry by taking the gauge connection 
to be opposite to the spin connection \cite{mn1}, so that
\begin{equation}\label{twistc}
(\partial_\mu +\omega_\mu + A_\mu)\epsilon=\partial_\mu\epsilon
\end{equation}
admits now solutions, the constant spinors.
The number of surviving supersymmetries depends on the way the $U(1)$ gauge
connection is embedded in $SO(4)$. 
The $6d$ theory on the 5-brane world-volume has ${\cal N}=(1,1)$ supersymmetry
generated by two symplectic Majorana
fermions, $\eta_+$ and $\eta_-$, transforming as $(\bf{4}_{+}, \bf{2}_{+})$ 
and $(\bf{4}_{-},
\bf{2}_{-})$ of the unbroken $SO(1,5)\times SO(4)$ subgroup of $SO(1,9)$. 
Wrapping the NS5-branes on $S^2$ further breaks the isometries of the 
world-volume
as $SO(1,5)\ra SO(1,3)\times SO(2)$.
Then, imposing the chirality and symplectic conditions, one finds that each 
6$d$ supersymmetry generator 
contributes two Weyl fermions in four dimensions with the following 
$SO(2) \times U(1)^+ \times U(1)^-$  charges
\begin{eqnarray}
\eta_+ &\rightarrow& (\mathbf{1},\mathbf{1},\mathbf{0})_{+}  + (\mathbf{1},\mathbf{-1},\mathbf{0})_{+} \equiv p+q,\nonumber\\
\eta_- &\rightarrow& (\mathbf{1},\mathbf{0},\mathbf{1})_{-} + (\mathbf{1},\mathbf{0},\mathbf{-1})_{-}\equiv \tilde p +\tilde q, 
\end{eqnarray}
where the subscripts $\pm$ indicate the 4$d$ chirality, 
$SO(2)$ is the connection on $S^2$ and $U(1)^+$, $U(1)^-$ are the abelian
factors in $SO(4) \sim  SU(2)^+ \times SU(2)^-$.
Thus, if we want to preserve 8 supercharges we have to identify 
the gauge connection with (the opposite of) the diagonal of the two abelian
subgroups $U(1)_D=\frac{1}{2}(U(1)^++U(1)^-)$. Similarly, 
${\cal N}=1$ supersymmetry is
obtained with the choice of the gauge connection in (the opposite of) 
$U(1)^+$.

In all our examples, the sphere will be a non-trivial two-cycle in a
Calabi-Yau. The R-symmetry group is the structure group of 
the bundle normal to the branes.
Thus, the twist condition is the requirement
that the tangent space group of the two-cycle is identified with
a $U(1)$ subgroup of the structure group of the normal bundle.
If the  six dimensional manifold is a (non-compact) CY$_3$ the gauge theory
will be ${\cal N}=1$ and if it is a non-compact version of K3$\times \mathbb{R}^2$ (an ALE space
times $\mathbb{R}^2$) the gauge theory will be ${\cal N}=2$; in all other cases no supersymmetry survives.

The $SO(4)$ subgroup left unbroken by the twist provides the 
R-symmetries of the 4$d$
theory. For ${\cal N}=2$ these are  $SU(2)_R \times
U(1)_R^{{\cal N}=2}$, where  
$U(1)_R^{{\cal N}=2}$ corresponds to the untwisted 
$U(1)=\frac{1}{2}(U(1)^+ - U(1)^-)$
and the action of  the abelian subgroup $U(1)_J\subset SU(2)_R$ 
on the massless modes can be identified with
both the $SO(2)$ spin connection and $U(1)_D$.  
In the  ${\cal N}=1$  case the R-symmetry $U(1)_R^{{\cal N}=1}$ 
is the twisted one, $U(1)^+$.

The twist also determines the field content of the 4$d$ theory. In particular, 
the massless states consist of the zero modes of the
compactification on the two-sphere.
The two fermions of 6$d$ SYM  have the same decomposition as the SUSY 
generators, thus giving four Weyl fermions with the following charge
assignments

\vskip 0.5truecm
\begin{center}
{\small
\begin{tabular}{ccc|cccccccccccc}
 & & & & & & & & & &   \\[-1mm]
 & & & & $p=\lambda$ &  & $\tilde{p}=\bar{\psi}$ &  & $q$ &  & $\tilde{q}$  \\[-1mm]
 & & & & & & & & & &  \\[-1mm]
\hline
 & & & & & & & & & &   \\[-1mm]
 & $ U(1)_R^{{\cal N}=2}= \frac{1}{2}(U(1)^+ - U(1)^-)$
 & & & 1 &  & -1 &  & -1 &  & 1  \\[-1mm]
 & & & & & & & & & & & & & &  \\[-1mm]
 & $ U(1)_D = \frac{1}{2}(U(1)^+ +U(1)^-)$ & & & 1 &  & 1 &  & -1 &  & -1  \\[-1mm]
 & & & & & & & & & &  \\[-1mm]
 & $ U(1)_R^{{\cal N}=1}=  U(1)^+$ & & & 1 &  & 0 &  & -1 &  & 0  \\[-1mm]
 & & & & & & & & & &   \\
\end{tabular}}

\vskip 0.3truecm
Table 1: Charge assignment of the spinors.
\end{center}

\vskip 0.5truecm

The spinors  $p$, $q$ ($\tilde{p}$, $\tilde{q}$) have positive
(negative) chirality. 
Thus the ${\cal N}=2$ twist gives mass to the $q$ spinors, while the other two
have the right quantum
numbers to be identified with the two spinors of 4$d$ ${\cal N}=2$ SYM:
$p\sim \lambda$ and $\tilde{p}\sim \bar \psi$, where $\lambda$ is the gaugino.
On the contrary $p\sim \lambda$ is the only massless spinor in the 
${\cal N}=1$ case.
The 6$d$ theory on the brane also contains 
four scalars, $X_i$, $i=1,\ldots,4$, transforming as 
$(\bf{2},\bf{2})$ under
$SU(2)^+ \times SU(2)^-$. 
Only those neutral under the
twisted gauge field give zero modes. Since all the scalars are charged under
$U(1)^+$, none of them survives the ${\cal N}=1$ twist. On the contrary, 
in the 
${\cal N}=2$ case there are two neutral scalars.
These are the two scalars, say $X_3$ and $X_4$, parameterizing
the motion of the brane in the two flat directions transverse to the ALE
space. They combine to give the complex scalar of ${\cal N}=2$ SYM (with
charges 0 and 2 under $U(1)_J \times U(1)_R^{{\cal N}=2}$).
Finally, in both cases the gauge field has no zero modes on the two sphere.
In summary, the massless states of the ${\cal N}=2$ twisted theory form a four
dimensional ${\cal N}=2$ vector multiplet, namely a gauge vector, two Weyl
fermions and a complex scalar. Similarly, the vector and the Weyl spinor
surviving the ${\cal N}=1$ twist form an ${\cal N}=1$ vector multiplet.

It is important to notice that in the $\N=2$ case the associated $U(N)$ gauge theory
has a moduli space of vacua, since the adjoint scalar fields can acquire a VEV. 
The moduli space is labeled by the $N$ Cartan values of the scalars and it is represented
in the string construction by the possibility of placing the branes in arbitrary
positions in the two flat directions.

In this paper we will only discuss geometries with a single $2$-cycle, which
give rise to gauge theories with a single gauge factor $U(N)$.
More complicated models can be realized by considering geometries with 
several $2$-cycles\footnote{Considering geometries with multiple cycles,
we obtain gauge theories with gauge factors associated with 
the cycles and
bi-fundamental fields associated with all pairs of intersecting cycles. 
The ${\cal N}=2$ theories we can construct in this way are then very similar to the ones 
obtained by placing fractional branes at ${\cal N}=2$ orbifolds. 
}.  

The holographic duals of ${\cal N}=2$ models with wrapped branes are discussed 
in Section 4 and those
of $\N=1$ models in Section 5.
As mentioned in Section 1.3, the natural setting to study the systems 
with a single set of NS5-branes 
is seven dimensional gauged supergravity, 
which is a consistent truncation of the ten dimensional $\N=1$ sector of Type IIB supergravity. 
As usual, all the $U(1)$ factors are not described by the holographic duals.

\subsection{Fractional branes}\label{fractional}

Fractional branes exist both at orbifold and conifold singularities.
Let us consider the orbifold case first \cite{kn,bertolini,polch2}. 
 In Section \ref{orbifold} we have seen that 
projecting the $\N=4$ theory with the regular representation of
the orbifold discrete group $\Gamma$ on the Chan-Paton factors
gives a conformal theory.
We can also use a representation 
that is not the regular one. In this way we can obtain non-conformal  
theories. 
When taking a non-regular representation of the orbifold group, we 
obtain fractional branes in Type II. 
As example we consider again 
 Type IIB string theory on $\mathbb{R}^4/\mathbb{Z}_2$. 
Choose the coordinates $(x_6,x_7,x_8,x_9)$ for $\mathbb{R}^4$. String theory 
on $\mathbb{R}^4/\mathbb{Z}_2$ can be defined with
an orbifold construction and  possesses a twisted sector localized 
at $x_i=0,i=6,...,9$.
The massless fields in the twisted sector form a tensor multiplet of $(2,0)$ $6d$ supersymmetry,
containing 5 scalars $\chi_I$. $\mathbb{R}^4/\mathbb{Z}_2$ is thus the singular point of a family of regular backgrounds
parameterized by the VEVs of the five scalar moduli. Three $\chi_I$, let's say $I=1,2,3$, 
correspond to the geometrical moduli of a two-sphere replacing the singular point. The geometry
of the background with $\chi_I\ne 0,I=1,2,3$ is that of an ALE space. 
Since fractional branes are associated with singular geometries, in this Section we are
particularly interested in the other two scalars $(b,c)\equiv (\chi_4,\chi_5)$. They correspond to the
flux of the NSNS and RR $2$-form along the $2$-cycle: $2\pi b = \int_{S_2} B_{(2)}$, $2\pi c =
\int_{S_2} C_{(2)}$. One can show that these fields are periodic (in our
conventions $b,c\in [0,1)$). 
$b$ and $c$ are non vanishing and well-defined even
for singular geometries where the $2$-cycle should be thought of as
hidden in the orbifold singularity\footnote{It is known that the standard orbifold construction of string theory
is perturbative in nature and corresponds to a regular world-sheet $CFT$; the non-zero value 
of the $B$-field flux is $b=1/2$
\cite{aspinwall}. String theory develops a singularity and becomes really non-perturbative 
only when all the $\chi_I=0$, modulo periodicities. 
At these points we expect 
non-perturbative phases of the theory with tensionless strings \cite{wittencomments}.}.

Add now D3-branes with world-volume $(0123)$ at the point
$x_6=x_7=x_8=x_9=0$. There are two different RR 4-forms in the orbifold 
theory. 
One is the untwisted RR form $C_{(4)}$ and the second one, $C_{(4)}^T$, 
comes from 
the twisted sector. $C_{(4)}^T$ is a six-dimensional field localized at 
the fixed point
and it can be dualized to give a scalar field which, as one can show \cite{dm},
 we can identify with $c$. 
Consequently, there are two basic types of D3-branes in this
theory, which we call fractional and anti-fractional D3-branes\footnote{The two
types of D3-branes are mutually BPS. We use the name 
anti-fractional with an abuse of language, following the
interpretation as wrapped D5-branes.}. 
Fractional branes have charges
$(b,1/2)$ with respect to the RR forms $C_{(4)}$ and $C_{(4)}^T\sim c$,
respectively; anti-fractional branes have
charges $(1-b,-1/2)$.  With a fractional and an anti-fractional
D3-brane we can make a physical D3-brane, whose charge is
$(1,0)$. There are several complementary descriptions for fractional
branes:
\begin{description} 
\item{(i)} $\,$ Consider the perturbative construction of the orbifold. Each brane at 
$x_i^{(0)},i=6,7,8,9$ has an image in $-x_i^{(0)}$. A brane and its image 
make up a physical brane, which can be moved to an arbitrary 
point in $\mathbb{R}^4/\mathbb{Z}_2$. For $x_i^{(0)}=0$, a physical brane appears as a
composite object and can be split in the plane $(x_4,x_5)$.  
The constituents of a physical brane are the two 
types of fractional branes. It is clear that they can only live at the singular point.
The $\mathbb{Z}_2$ action on the Chan-Paton factors on $n_1$ fractional and $n_2$ anti-fractional
branes can be represented with the matrix $\gamma_{\alpha}={\rm diag} \{I_{n_1},-I_{n_2}\}$. 
That charges and tensions of these objects agree with the mentioned value
follows from a direct computation in the orbifold construction~\cite{dm} or in the boundary 
state formalism~\cite{Billo2001}.
\item{(ii)} $\,$ We can make contact with the discussion in Section 
\ref{genobs} using the following observation.
A fractional brane can be represented as a D5-brane wrapped 
on the collapsed two-cycle of $\mathbb{R}^4/\mathbb{Z}_2$~\cite{PK3,dMtheory}. This object appears as a 3-brane and,
as we will see shortly, it carries D3-charge.
Similarly an anti-fractional
brane is an anti-D5-brane with one unit of flux for the gauge field
living on it: $\int_{S_2} F=-2\pi$~\cite{PK3,dMtheory,Billo2001}.  
This representation is particularly useful when $b\ne 1/2$ and 
the perturbative description of the orbifold is not adequate. 
In this representation, $C_{(4)}^T$ is the reduction of $C_{(6)}$ on the two-cycle and
the corresponding charge is just the D5-charge.
D3-charges and tensions can be read from the action for a D5- or an anti-D5-brane
\begin{eqnarray}
&& -\frac{1}{(2\pi)^2}\Bigl[\int dx^6 e^{-\Phi}\sqrt{G+F+B} \\
&&\quad \pm \int(C_{(6)}+C_{(4)}\wedge (F+B)+\frac{1}{2}C_{(2)}\wedge (F+B)^2+
\frac{1}{6}C_{(0)}\wedge (F+B)^3)\Bigr]. \nonumber
\label{BIfrac}
\end{eqnarray} 
The induced D3-charges are $b$ and $(1-b)$, while the
tensions are proportional to $|b|$ and $|1-b|$. For $b\in [0,1)$, these values satisfy the
BPS condition.
\item{(iii)} $\,$ For readers familiar with the Hanany-Witten construction~\cite{wittenM,hw}, we mention that
the same system is T-dual to a set of 
D4-branes stretched between NS5-branes in Type IIA.
The D4-branes have world-volume in the space-time directions $(0,1,2,3,6)$.  
The direction $x_6$ is compactified on a circle of radius $L$. 
The two NS5-branes have world-volume $(0,1,2,3,4,5)$ and sit at 
$x_6=0$ and $x_6=2\pi bL$ respectively, with  $x_7=x_8=x_9=0$.  
The fractional branes can be identified 
with the D4-branes stretched from the first to the second NS5-brane,
the anti-fractional branes with the D4-branes stretched from the
second to the first. A fractional and an anti-fractional brane can
join and give a physical D4-brane, which can move away in
$(x_6,x_7,x_8,x_9)$.  
\end{description} 

\begin{figure}[h] 
\centerline{\epsfig{figure=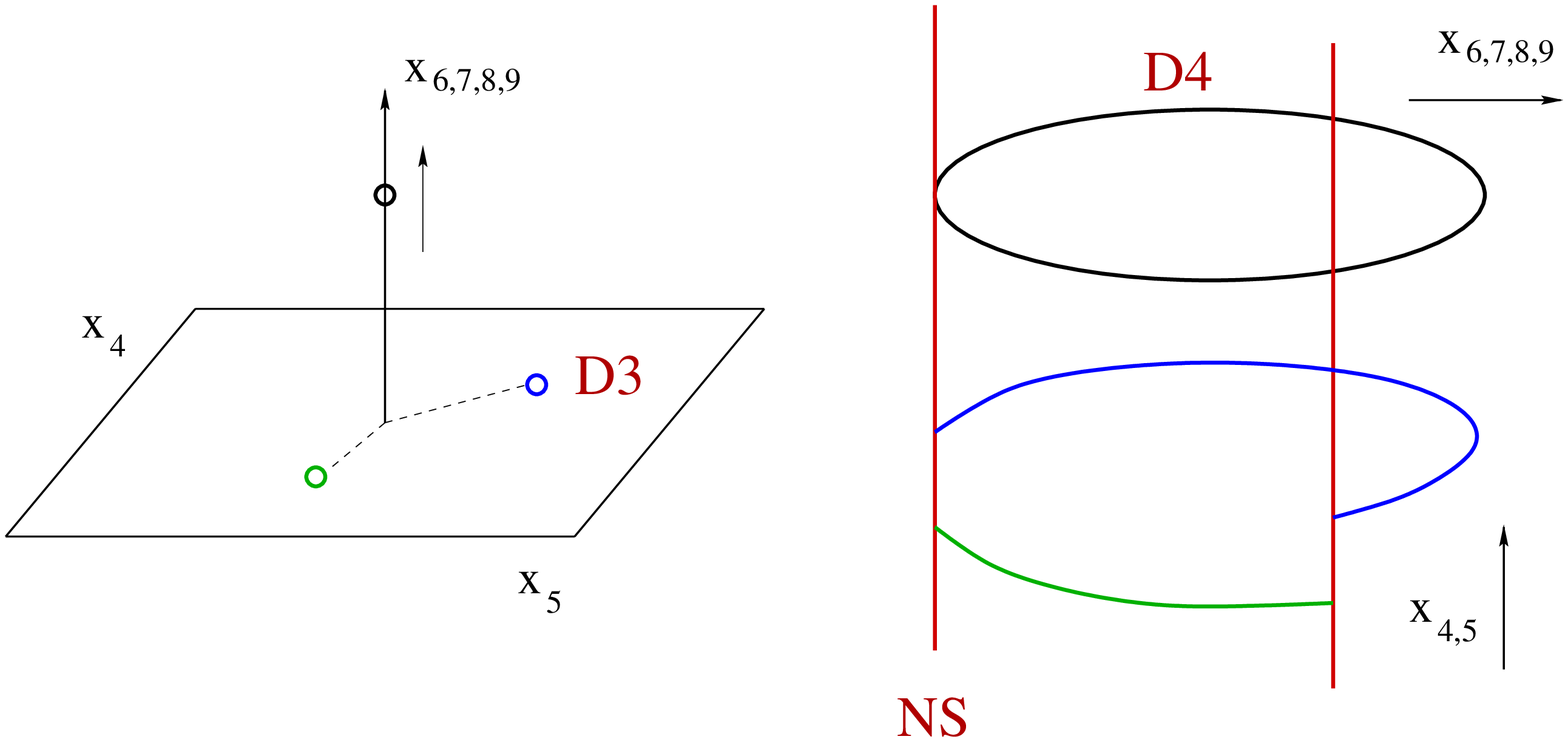,height=5 cm, width=14cm}} 
\caption{Type IIB and IIA picture for physical and fractional branes.} 
\end{figure} 

Applying the rules discussed in Section \ref{orbifold} to the representation
$\gamma_{\alpha}={\rm diag} \{I_{n_1},-I_{n_2}\}$ shows
that the gauge theory corresponding to $n_1$ fractional and $n_2$ anti-fractional
branes  
is $U(n_1)\times U(n_2)$ with two bi-fundamental hypermultiplets.
It is instructive to identify the field theory R-symmetry $SU(2)_R\times U(1)_R$ in terms
of the symmetries of the string construction. 
$SU(2)_R$ is identified with the subgroup of the $SO(4)$ rotating the coordinates $(6,7,8,9)$
that is left unbroken by the orbifold projection. $U(1)_R$ is instead identified with
the rotations in the plane $(4,5)$.  
The theory has a moduli space of vacua which consists in a Higgs and a Coulomb branch.
In this review, we will only consider
the Coulomb branch, which is labeled by the Cartan values of the adjoint
scalar fields in the
vector multiplets, consisting in $n_1+n_2$ complex VEVs. These moduli have an obvious
interpretation as the positions of the fractional branes in the plane $(4,5)$. This
is consistent with the fact that the scalars in the vector multiplets are rotated by $U(1)_R$.
For completeness, we notice that the Higgs branch is instead 
parameterized by the VEVs of the hypermultiplet scalars, rotated by $SU(2)_R$, and it
corresponds to the motion of physical branes in the directions 
$(6,7,8,9)$. 

The gauge couplings of the two groups, $\tau_1,\tau_2$,  
are determined (for $b\in [0,1)$) in terms of 
the space-time fields by equation~(\ref{BIfrac})
\begin{equation} 
\tau_1=(b \tau +c),\qquad\qquad  
\tau_2=(1-b)\tau - c, 
\label{eee2} 
\end{equation} 
where $\tau=C_{(0)} + i {\rm e}^{-\Phi}$ is the complex dilaton of Type
IIB. As we have already discussed, the case $n_1=n_2$ corresponds to a conformal field theory. 
The complex coupling constants of the two groups  
are exactly marginal parameters and the  
theory has an $AdS$ dual: $AdS$$_5\times$S$^5/\mathbb{Z}_2$~\cite{kacsil}. 
When $n_1=N +M$ and $n_2=N$, the theory is no longer conformal and the 
coupling constants run at all scales. 
One of the two gauge factors is not 
asymptotically free and it is ill-defined in the UV. We can nevertheless 
make sense of these theories by finding an $\N=2$ UV completion 
that is a $CFT$. For example, the $\N=2$
theory $U(n_1)\times U(n_2)$ can be considered as the low energy limit of a broken
phase of the $CFT$ $U(N)\times U(N)$, $N>{\rm max} (n_1,n_2)$ 
where some scalar fields developed a vacuum expectation
value. The case of a pure $SU(M)$ $\N=2$ gauge theory can be realized
by setting $n_1=M,\,n_2=0$.

In general, in orbifold theories there are as many types of fractional branes as there are nodes of the quiver diagram.
We can therefore construct non-conformal
$\N=2$ gauge theories which are products of groups with bi-fundamental hypermultiplets.
At least one gauge group is not asymptotically free in this construction, but the
theory can be safely embedded in a UV $\N=2$ $CFT$.

Fractional branes exist in all backgrounds with collapsed 2-cycles. In particular,
we may define fractional branes in the conifold geometry defined in Section~\ref{conifolds} \cite{kn,kt1,ks}.
$T^{1,1}$ has the topology of $S^2\times S^3$ and at the tip of the 
cone over $T^{1,1}$
both the $2$-cycle and the $3$-cycle are vanishing. In view of our previous discussion we are
mostly interested in the $2$-cycle where we can wrap a D5-brane of Type IIB. 
While the description (i) for fractional branes given above in the orbifold case
is no more applicable,
the description (ii) in terms of D5-branes wrapped on $2$-cycles can be
repeated almost verbatim. We obtain an $\N=1$ gauge theory of the form $U(n_1)\times U(n_2)$
with bi-fundamental chiral fields and a superpotential inherited from the conformal case.
The theory has a Higgs branch where the bi-fundamental fields acquire VEVs.
The coupling constants of the two
groups can be determined using eq.~(\ref{costanti}) and read
 \begin{equation}
{1\over g_{1}^2} + {1\over g_{2}^2}= {1\over 4\pi g_{s}},\qquad
{1\over g_{1}^2} - {1\over g_{2}^2}= {1\over 4\pi^2 g_{s}}\left(\int_{S^2}B-\pi 
\right) .
\label{pip}
\end{equation}
For $n_1=n_2$ the theory is conformal and the two coupling constants correspond to 
two exactly marginal parameters in $AdS_5\times T^{1,1}$: the dilaton and the value
of the $B$-field on $S^2$. For $n_1=N+M$ and $n_2=M$ the theory is no longer conformal. 
One of the two gauge factor is not asymptotically free in the UV. 
There is a curious UV completion of 
this theory in terms
of $U(\infty)\times U(\infty)$, based on Seiberg
duality; the details will be discussed in Section 5.

The holographic duals for $\N=2$ theories with fractional branes are discussed in Section 4
and those for $\N=1$ theories in Section 5.
Fractional branes act as sources for closed string
states. We will find holographic duals where the corresponding
fields depend on the radial coordinate. 
In particular, since the gauge theory coupling constants run
 with the scale,
we expect to find duals where the twisted fields $b$ and $c$ run with the radial coordinate
in the $\mathbb{R}^4/\mathbb{Z}_2$ case (see eq.~(\ref{eee2})) and $\int B$
runs in the conifold example (see eq.~(\ref{pip})).
Notice, however, that 
the diagonal coupling $\tau_1+\tau_2=\tau$ does not run in every background 
where the Type IIB dilaton remains constant.
As usual, all the $U(1)$ factors are not described by the holographic duals.

\section{Supergravity duals of $\N=2$ gauge theories}\label{ndue}

In the first part of the present Section we review some basic  
properties of the $\N=2$ supersymmetric gauge theories that 
will be used in the following.
Then we discuss some of the corresponding string/gravity duals 
available in literature. 
They can be obtained as mass deformations
of $\N=4$ SYM \cite{naked,pw,sfetsos,bpp,michela}, using fractional branes
at orbifold singularities \cite{kn,pg,bertolini,polch2} or five-branes wrapped on two-cycles \cite{martelli,articIII,fayya}.
As already mentioned in the introduction, we will discuss in detail
the case of wrapped branes, while for the other examples we will simply review
the results. We will see that the conjectured dual supergravity backgrounds are in general plagued by singularities that can be resolved by stringy effects, such as the so-called enhan\c con mechanism \cite{jpp}.

\subsection{Some remarks on $\N=2$ SYM}\label{SW}

We briefly review what is known about the ${\cal N}=2$ physics
of the theories we are interested in.  
For a general discussion about $\N=2$ supersymmetry, Seiberg-Witten 
theories and more information about the material  
in this Section  we refer the reader to the many good reviews in the 
literature \cite{peskin,dhokphong,alvgaume,divesw}.

As already mentioned, one important property of $\N=2$ theories is that they possess a moduli space
of vacua. The scalars in the $\N=2$ multiplets can have a vacuum
expectation value. In the moduli space we can distinguish 
a Higgs
branch, where we give VEV to hypermultiplet scalars, and a Coulomb
branch, where we give VEV to the complex scalars in the vector multiplets.
We will be mainly interested in the Coulomb branch. 
In a generic Coulomb branch vacuum, the gauge group is broken to
the maximal abelian subgroup and the only massless fields are $n$
abelian vector multiplets, where $n$ is the rank of the gauge group.
There are correspondingly $n$ massless complex scalar fields
$u_i$, whose VEVs 
parameterize the Coulomb branch.  
At low energies, we can write an effective Lagrangian for the massless
fields. It is a consequence of $\N=2$ supersymmetry that the effective
Lagrangian is completely determined in terms of a single holomorphic
function ${\cal F}$ of the $u_i$, called the prepotential,
\begin{equation}
{\cal L}\sim Im(\tau_{ij})F_{\mu\nu}^iF^{\mu\nu j}+Re(\tau_{ij})F_{\mu\nu}^i\tilde F^{\mu\nu j}+ Im \left ( \partial_\mu u_i \partial^\mu\frac{\partial {\cal F}}{\partial u_i}\right )+{\rm fermions},
\end{equation}
where $\tau_{ij}=\partial^2 {\cal F}/\partial u_i\partial u_j$.
This effective Lagrangian is a good description of the physics
except for certain
values of $u_i$, which correspond to singularities in the moduli
space associated with new physical massless particles.

The perturbative contribution to $\F$ is exhausted at 1-loop, all other
corrections being given by instantons. To fix the ideas, we discuss
the case of the simplest $\N=2$ gauge theory, with gauge group $SU(N)$ 
and no flavors. 
The 1-loop prepotential reads
\begin{equation}
2\pi i \F^{(1)} = 
-{1 \over 4} \sum _{j\not=i} (u_i-u_j)^2 \log {(u_i - u_j)^2 \over \Lambda ^2},\label{1loop}
\end{equation}
where $\{u_i\}$ are the $N$ eigenvalues of the $SU(N)$ adjoint scalar field satisfying
$\sum_i^N u_i=0$. The expressions for product groups or adjoint massive 
hypermultiplets will be written when needed.
Formula~(\ref{1loop}) fails at scales of order $\Delta u=\Lambda$,
where the instantonic contributions to the prepotential become
relevant. The apparent singularity in the 1-loop formula is resolved
by adding an infinite series of instantonic contributions.
The full prepotential can be determined using the Seiberg-Witten curve.
This is a family of Riemann surfaces $\Gamma
(u_1, \cdots , u_N)$, parameterized by the $N$ complex parameters
 $u_1, \cdots , u_N$, which label the flat directions
of the $\N=2$ vacua. 
For the derivation of 
${\cal F}$ from the SW curve the reader is referred to \cite{peskin,dhokphong,alvgaume,divesw}. Here we  only give the form of the curve since it will
 be used in Section 4.5. 
The curve for pure $SU(N)$ is given by the genus $(N-1)$ 
hyperelliptic Riemann surface
\begin{equation}
y^2=P(x)^2-\Lambda^{2N},\label{curvesun}
\end{equation}
where the polynomial $P(x)$ is expressed in terms of the moduli 
\begin{equation}
P(x) = \prod _{i=1} ^N (x-u_i).
\end{equation}
The importance of this curve is that it has a clear interpretation
in terms of branes, which will be discussed in Section 4.5.

The existence of a moduli space allows us to probe the theory. Consider just a single modulus $z$ of the pure $SU(N+1)$ theory. 
A non-zero VEV for $z$ corresponds to the breaking $SU(N+1)\rightarrow SU(N)\times U(1)$.
Under certain conditions, namely when $z$ is sufficiently large and $N$ big enough, 
we can study the physics of $SU(N)$ by looking at the effective action for the $U(1)$ factor. We then consider a point in the Coulomb branch
where the moduli read $(\bar{u}_1 - z/N,...,\bar{u}_{N} -z/N,z)$.
The $\bar{u}_i$'s indicate the point in the moduli space of $SU(N)$
that we would like to investigate.
The coupling constant associated with $z$ can de determined from 
\begin{equation}
\tau(z)=\frac {\partial^2 \F (z,\bar{u}_i)}{\partial z^2}
\label{run}
\end{equation}
at fixed $\bar{u}_i$. 
For example, the 1-loop contribution reads
\begin{equation}
\tau(z)=\frac{i}{\pi}\sum_{i=1}^{N} \log \frac{(z-\bar{u}_i)}{\Lambda}.
\label{run2}
\end{equation}
For large $N$, we may expect that the introduction of a probe
would not seriously alter the physics of $SU(N)$; we may also be tempted to 
send the probe
very close to the other moduli, to investigate the non-perturbative dynamics 
of the gauge theory.
For comparison with holographic duals we need to consider the large $N$ limit 
of the gauge theory. 
One should not make the mistake of neglecting instantonic contributions
for $N\gg 1$ due to the naive estimate $e^{-1/g_{YM}^2}=e^{-N/x}$: 
differences of VEVs, which appear in $\F$, 
can be so small to compensate this exponential factor and this typically
happens in strongly coupled vacua \cite{ds}. The effective action for a probe
 is accurate at 1-loop
for $z$ greater than the dynamically generated scale of the theory $\Lambda$. 
Instantonic corrections rise up very sharply
(at large $N$) near $\Lambda$ and dominate the IR physics.

The probe computation allows to compare directly the gauge theory 
results with a calculation in the holographic dual. When we engineer
a system of branes corresponding to an $\N=2$ gauge theory with
a Coulomb branch, we expect that the constituent branes possess a moduli
space of vacua isomorphic to that of the gauge theory. We typically have a 
set of branes
that can be arbitrarily distributed in a plane in space-time, as we 
explicitly saw 
in Sections \ref{wrapped} and \ref{fractional} for the $\N=2$ theories 
with wrapped and fractional branes.
If we have an holographic dual for our gauge theory, obtained as the near horizon geometry
of the system of branes, we may think of studying it by 
sending in a probe. The probe is represented in the string theory construction by a physical, fractional or wrapped
brane which is sent in the background of a large number $N$ of other branes.
If the theory is $\N=2$, such a brane is a BPS object which can freely (that is without feeling
any force) move in the moduli space. The effective
action on the probe can be rigorously written using the Born-Infeld action for
branes in a given
background. This result, which is greatly constrained 
by $\N=2$ supersymmetry, must
agree with the gauge theory result computed via formula~(\ref{run}).
It is important 
to stress that the ${\cal N}=2$ effective action is determined by
holomorphicity. Holomorphic (or BPS) quantities are protected
and can be often computed in the supergravity regime, despite the
presence of many un-decoupled modes. 
 The possibility
of comparing the probe action with the $\N=2$ effective
action in the dual field theory also provides an unambiguous way of 
determining the radius/energy relation in $\N=2$ solutions: the modulus 
$z$, which represents the energy scale we are probing, can
be identified with the space-time position of the probe. 

The supergravity solutions that we are going to discuss, correctly
capture the one loop contribution in field theory
 but are plagued by singularities at the
scale where, in field theory, instantons become important. This is a
situation where, as we will discuss in Section 4.5, one can learn
from field theory, specifically from the SW curve, 
how the supergravity singularity is possibly resolved.
For completeness, the SW curve associated with the $\N=2$ theories
that will be considered in this review are explicitly discussed in Appendix \ref{swcurve}.

\subsection{$\N=2$ SYM from wrapped five-branes}\label{nduewrapp}

As observed in Section 3.2, one way to realize pure $U(N)$ $\N=2$ SYM in $(1+3)$-dimensions 
is to consider the low energy theory on the world volume of $N$ NS5-branes 
wrapped on a non-trivial cycle in a geometry of the
form $\mathbb{R}^2\times$ALE \cite{martelli,articIII}.
Let us summarize the basic ingredients in this construction.
There is a massless complex scalar field $\phi$ on the world-volume of the
branes that parameterizes their motions on the $\mathbb{R}^2$ plane.
The generic vacuum in the Coulomb branch of the gauge theory is
labeled by the $N$ eigenvalues of $\phi$. They are given
by the arbitrary positions of the $N$ branes 
on the  $\mathbb{R}^2$ plane. 
Moreover, for the compactification to
preserve ${\cal N}=2$ supersymmetry, the theory has to be twisted:
the spin connection on $S^2$    
has to be identified with a background $U(1)$ field    
in the $SO(4)$  R-symmetry group \cite{vafa,mn2}, which 
corresponds to the diagonal subgroup $U(1)_D=\frac{1}{2}(U(1)^-+U(1)^+)$ in 
the decomposition  $SO(4)\rightarrow SU(2)^+\times SU(2)^-$.
The $U(1)_R$ symmetry of the gauge theory corresponds instead 
to a rotation in the plane $\mathbb{R}^2$.

To construct the dual supergravity solutions, 
consider first a set of flat NS5-branes. Such a configuration  
admits a holographic description in terms
of the  linear dilaton background \cite{little}, with a ten dimensional 
metric of the form $\mathbb{R}^{5,1}\times \mathbb{R}\times S^3$ 
(see eq.~(\ref{lindil})).
It is then natural to associate the
 solutions for wrapped branes to deformations of the
linear dilaton background where the flat six-dimensional part
of the metric is replaced by a metric of the form 
$\mathbb{R}^{3,1}\times S^2$, the $S^3$ transverse geometry is possibly
deformed and a background abelian field in $SO(4)$ is turned on.
Since the ultraviolet gauge theory is six dimensional, for large $\rho$ 
the solutions must
asymptote  the linear dilaton background.

The actual computation of the solution using the ten-dimensional Type IIB
equations is usually quite awkward. As in Section 2, we can try to
consider compactifications to lower dimensions, find solutions in the lower 
dimensional supergravity and then lift them 
to get the full 10$d$ solutions. 
In the present case, where we deal with  deformations of an 
$\mathbb{R}^{5,1}\times \mathbb{R}\times S^3$ metric, the theory we need is
provided by the seven-dimensional $SO(4)$    
gauged supergravity, corresponding to the truncation    
of the $\N=1$ sector of Type IIB on the 3-sphere transverse to the NS5-branes.  This is a consistent choice    
since the NS5-branes only couple to the NS sector of Type IIB   
supergravity \cite{mn2}. 

The bosonic sector of seven dimensional $SO(4)$ gauged
supergravity \cite{sz} consists of the metric, $SO(4)$ gauge fields, a
three-form 
and ten scalar fields. The Lagrangian for these fields (which can be obtained as a suitable
singular limit \cite{cve1,cve2} of the maximally supersymmetric $SO(5)$ 
Lagrangian \cite{pernici}) reads (we use the conventions of \cite{minasian})
\begin{equation}
2\kappa^2e^{-1}{\cal L} = R + {1\over 2} m^2 (T^2 - 2T_{ij}T^{ij}) -
Tr(P_{\mu} P^{\mu}) - {1\over 2} (V_I{}^i V_J{}^j F_{\mu \nu}^{IJ})^2,
\label{sug}
\end{equation}
where 
$I$, $i$ are the gauge and composite $SO(4)$
indices. $F_{\mu \nu}^{IJ}$ is the gauge field strength.   
$T_{ij}$ is a symmetric 
matrix parameterized by the ten scalar fields and is defined in terms of
the $SL(4,R)/SO(4)$ coset element 
$V_i^I$ as $T_{ij} = V_i^{-1\,I}
V_j^{-1\,J} \delta_{IJ}$,  $T=T_{ij}\delta_{ij}$.
The kinetic term for the scalars, $P_{\mu}$, is the symmetric part of $V_i^{-1\,I} {\cal D}_{\mu} V_I{}^j = \left( Q_{\mu} \right)_{[ij]} + \left( P_{\mu}
\right)_{(ij)}$, where the
covariant derivatives are defined as ${\cal D}_\mu V_I{}^j=\partial_\mu
V_I{}^j+2m A_{\mu\, I}^J V_J{}^j$ on the scalars and ${\cal D}_\mu
\psi=(\partial_{\mu}+{1 \over 4}Q_{\mu ij}\Gamma^{ij}+{1 \over
  4}\omega_{\mu}^{\nu\lambda}\gamma^{\nu\lambda})\psi$ on the spinors; $m$ is
the mass parameter (set to one in our conventions) which by supersymmetry is equal to one half of 
the gauge coupling constant.

As standard in the $AdS/CFT$ correspondence,    
the $SO(4)$ gauge fields correspond   
to the isometries of the 3-sphere and are dual to the R-symmetry fields. 
Because of the twist condition, we set the $U(1)_D$ gauge field equal to 
minus the
spin connection on $S^2$.
The scalar matrix $T_{ij}$ can always be brought 
to diagonal form with an $SO(4)$ gauge rotation
\begin{equation}\label{scalars}
T_{ij} = {\rm diag}( e^{2\lambda_1},  e^{2\lambda_1},
e^{2\lambda_2},  e^{2\lambda_3}).
\end{equation}
Before the twist, it is natural to associate the matrix $T_{ij}$ with 
the dual operator ${\rm Tr}X_{\{i}X_{j\}}$ constructed with the four 
scalars  living on the NS5-branes.
In the representation for the $SO(4)$ we use, the twisted $U(1)_D$
corresponds to a rotation of the first two entries $i=1,2$, 
while the R-symmetry $U(1)_R$ corresponds to a rotation of the last
two $i=3,4$. This makes it clear that the first two entries
of the previous matrix correspond to the scalars that become massive 
upon twist. Their equality is required by the ${\cal N} =2$ twist. 
The last two entries 
correspond to bilinear operators in the scalar field 
$\phi$ parameterizing the ${\cal N} =2$ moduli space. In particular,
the $U(1)_R$ charges suggest that
$\lambda_2+\lambda_3$ and  $\lambda_2-\lambda_3$ are dual to 
${\mbox Tr} \phi\bar\phi$ and ${\mbox Tr} \phi^2$ respectively.

Thus the general seven dimensional solutions we are
interested  in, 
involve a non trivial profile for the $U(1)$ gauge field and
some of the above scalars
\begin{eqnarray}\label{ansa}
ds_7^2 &=& e^{2f} (dx_4^2 + N\alpha' d\rho^2) 
           +  e^{2g} (d\theta^2 + \sin^2 \theta
d\phi^2) , \nonumber\\
A_3 &=& \frac{1}{2} \cos\theta d\phi,\\
T_{ij}& =& {\rm diag}( e^{2\lambda_1},  e^{2\lambda_1} 
e^{2\lambda_2},  e^{2\lambda_3}),
\nonumber
\end{eqnarray}
where the fields depend only on $\rho$.
The seven dimensional three-form is set to zero.

As long as we are interested in supersymmetric solutions, it is not necessary
to look at the equations of motion. It is indeed possible to reduce
the problem to the solution of a set of first order equations. This can be done
in various way. One can explicitly solve
 the fermionic shifts \cite{articIII}, as reviewed in 
Appendix \ref{fermishift}.
Alternatively, one can first write 
an effective Lagrangian for the radial dependence of the scalars,
by substituting the ansatz~(\ref{ansa}) in the Lagrangian~(\ref{sug}) and 
integrating
over $S^2$ \cite{martelli,apreda}. Imposing $f= - \lambda_1 -
(\lambda_2+\lambda_3)/2 \,$\footnote{This is allowed by the equations of motion and permits to set the warp factor for the world-volume part of the string frame metric equal to one, as in the linear dilaton background \cite{martelli}.} we obtain
\begin{eqnarray}
{\cal L}&=& {3\over16}e^{4Y}[16Y^{'2}-2h^{'2}- \frac{1}{4}(2 \lambda_1'-
\lambda_2'-\lambda_3')^2- \frac{1}{2}(\lambda_2'-\lambda_3')^{2}+ 2 e^{-2h}+ \nonumber \\
& & -\frac{1}{2}e^{-4h-2\lambda_1+\lambda_2 +
  \lambda_3}+4\cosh(\lambda_2 -
\lambda_3)-2e^{-2\lambda_1+\lambda_2+\lambda_3}
\sinh^2(\lambda_2 -\lambda_3)],
\label{Lag}
\end{eqnarray}
with $4Y= 2h +5f +\log(16/3)$, $h=g-f$  and
the prime denotes derivation with respect to $\rho$.
Then, one should find a superpotential for this system, in the sense discussed in Section \ref{thedual}. With the conventions used there, the superpotential is
\begin{equation}
W= -{3\over8} [2 e^{\lambda_1 - (\lambda_2+\lambda_3)/2} + 2
e^{-\lambda_1 +(\lambda_2+\lambda_3)/2} 
\cosh(\lambda_2 -\lambda_3) + e^{-2h -\lambda_1 + (\lambda_2+\lambda_3)/2}].
\end{equation}
There are two
 families of solutions corresponding to 
$\lambda_2=\lambda_3$ (solution $A$) \cite{martelli,articIII} and 
$\lambda_2 \neq \lambda_3$ (solution $B$), respectively \cite{articIII}.
Here we will only give the solution $B$, solution $A$ being
a particular case of the former (see also Appendix \ref{duesol})
\begin{eqnarray}
e^{2h} &=& u , \nonumber \\
e^{\frac{\lambda_2+\lambda_3}{2} -\lambda_{1}} &=& \sqrt{{{e^{4u}+b^4}\over{e^{4u}-b^4}}-{1\over2u}+ {2Ke^{2u}\over{u(e^{4u}-b^4)}}}, \nonumber \\
e^{\frac{\lambda_2+\lambda_3}{2}+\lambda_{1}} &=& \left({e^{2u}\over{e^{4u}-b^4}}\right)^{1/5} \left[{{e^{4u}+b^4}\over{e^{4u}-b^4}}-{1\over2u}+ {2Ke^{2u}\over{u(e^{4u}-b^4)}}\right]^{-{1\over10}} ,\nonumber \\
e^{\lambda_2 -\lambda_3} &=& {{e^{2u}-b^2}\over{e^{2u}+b^2}},
\label{duscasol}
\end{eqnarray}
with ${du\over d\rho}\equiv e^{(\lambda_2+\lambda_3)/2-\lambda_{1}}$.

Compared with the deformations of $AdS_5$, the lift to ten dimensions is
much simpler and it is known for a generic seven-dimensional solution
\cite{cve1,cve2}. The 10$d$ solutions contain the metric, the dilaton and 
the NSNS two-form. In the Einstein frame the solution is
\begin{eqnarray}
ds_{10}^2 &=& X^{1\over8}\left(\Delta^{1\over4}ds_{7}^2 + \Delta^{-3\over4} T_{ij}^{-1}{\cal D}\mu^i{\cal D}\mu^j\right), \nonumber \\
e^{-\Phi}\star H_{(3)} &=& -U\epsilon_7 +T_{ij}^{-1}\star{\cal D}T_{jk}\wedge(\mu^k{\cal D}\mu^i)- {1\over2}T_{ik}^{-1}T_{jl}^{-1}\star F^{ij}\wedge {\cal D}\mu^k\wedge{\cal D}\mu^l,\nonumber \\
e^{2\Phi} &=& \Delta^{-1}X^{3\over2}.
\label{cvlift}
\end{eqnarray}
where $\mu^i$ ($\mu^i\mu^i=1$) are $S^3$ angular variables, $\, \Delta=T_{ij}\mu^i\mu^j$, $\, U=2T_{ik}T_{jk}\mu^i\mu^j - \Delta T_{ii}$, and  $\, X=det(T_{ij})$. Moreover $\, {\cal D}\mu^i = d\mu^i + A^{ij}\mu^j$, $\, {\cal D}T_{ij}=dT_{ij} +A_i^{k}T_{kj} + A_j^{k}T_{ik}$. 
Applying the lift formulae to our case and passing to the string frame,
we have
\begin{eqnarray}\label{solB}
ds^2 &=& e^{(2\lambda_1+\lambda_2+\lambda_3)}(
ds_{7}^{2} + {1\over \Delta}
\left\{ e^{-2\lambda_{1}}[d\mu_1^2+d\mu_2^2
+\cos^{2}\theta(\mu_{1}^2+\mu_{2}^2)d\phi^{2} \right. \nonumber \\
& & \left. -2\cos\theta(\mu_{1}d\mu_{2}-\mu_{2}d\mu_{1})d\phi] 
+e^{-2\lambda_{2}}d\mu_3^2 + e^{-2\lambda_3}d\mu_4^2
\right\}),\nonumber\\
e^{2\Phi} &=& e^{(3\lambda_{2}+3\lambda_3+6\lambda_{1})}\Delta^{-1}
\end{eqnarray}
and $H_{(3)}$ can be deduced from (\ref{cvlift}), provided we identify 
$\Delta= e^{2\lambda_{1}}(\mu_1^2+\mu_2^2) +
e^{2\lambda_{2}}\mu_3^2+e^{2\lambda_3}\mu_4^2$, 
$\, \mu_{1,2}=\cos\theta^\prime(\cos\phi_1,\sin\phi_1)$ and 
$\, \mu_{3,4}=\sin\theta^\prime(\cos\phi_2,\sin\phi_2)\,$ 
($0 \le \theta \le \pi$, $\, 0 \le \phi \le 2\pi$~; $\, 0 \le
\theta ^\prime \le \pi/2$, $\, 0 \le \phi_1, \phi_2 \le 2\pi$). 
For $b=0$ one recovers solution $A$  with $\lambda_{2}=\lambda_3$.

Solution $A$ has two $U(1)$ isometries corresponding to shifts in
$\phi_1$ and  $\phi_2$ which  
are easily identified with the R-symmetries. $\phi_1$ 
rotations correspond to $U(1)_D$, which coincides with $U(1)_J$ on
the massless fields, while 
shifts in $\phi_2$ are associated to $U(1)_R$. 
On the contrary, in solution $B$, the scalar $\lambda_3\neq \lambda_2$ explicitly breaks
$U(1)_R$, so that the only isometry is the other $U(1)$.

In the UV, $\rho \rightarrow \infty$, the two solutions are asymptotic 
to the linear dilaton  background with the radius of $S^2$ going to infinity.
These are exactly the boundary conditions the two solutions have to satisfy.

For both solutions the metric is singular 
and the nature of the singularity depends on
the value of the integration constant $K$ in (\ref{duscasol}). 
For $K\le (1-b^4)/4$, $u\in [u_0,\infty)$, where $u_0\geq 0$ is 
determined by $e^{-\lambda_1+(\lambda_2+\lambda_3)/2}=0$,
 and the solutions are  singular for $u \rightarrow u_0$ and $\theta^\prime=\pi/2$ ($u_0=0$ for $K=(1-b^4)/4$). For $K> (1-b^4)/4$ the singularity is at
$u=0$ and it  seems to be of the bad type according to the criterium in
\cite{mn1}\footnote{This states that the (Einstein frame) $g_{00}$ component of a metric conjectured to be dual, in some region containing the singularity, to the low energy regime of a field theory, cannot increase while approaching the singularity. This is because fixed proper energy excitations should correspond to low energy ones as measured by an observer at infinity.}. Therefore we will not discuss the $K>(1-b^4)/4$ solutions in the
following.

Close to the singularity, the dilaton becomes very large and, to avoid
string corrections, we have to pass to the S-dual solution for the D5-branes. 
The two solutions are related by the standard S-duality 
transformation (in string frame)
\begin{eqnarray}
\Phi_{D} &=& -\Phi, \nonumber \\
ds^{2}_{D} &=& e^{\Phi_{D}}ds^{2}_{NS}, \nonumber \\
dC_{(6)} &=& \star F_{(3)}= e^{-2\Phi}\star_{NS}H_{(3)}.
\label{eq2}
\end{eqnarray}

The presence of a (naked) singularity seems to be a common 
feature of all the supergravity solutions describing 
${\cal N}=2$ gauge theories. The problem is then to understand 
whether the
singular behavior is an artifact of the supergravity approximation, 
which can
be resolved in the full string theory, or it signals a 
pathological behavior of the solution.
A standard technique consists in studying the low energy effective 
action of a single brane probe in the geometry.
As already mentioned, the result has a two-fold interpretation: on the
supergravity side it describes the effective geometry seen by the probe, thus 
shading light on the
nature of the the singularity, while on the field theory side it helps
in identifying the vacuum of the gauge theory.

It is clear from Section 3.1 that, for the D5 solution,
our probe will be a D5 brane wrapped on $S^2$, 
whose low energy effective action is 
\begin{equation}
(2\pi)^2 S=-\int d^{6}\xi e^{-\Phi_{D}}\sqrt{G+F} + \int C_{(6)} +{1\over2} \int C_{(2)}\wedge F\wedge F ,
\label{azione}
\end{equation}
where 
\begin{equation}
G_{\alpha\beta}=\partial_{\alpha}x^{M}\partial_{\beta}x^{N}g_{MN}
\end{equation}
is the induced metric on the probe world-volume 
($\alpha, \beta=0,1,...,5$ label 
the world-volume coordinates, while $M,N=0,1,...,9$ are space-time indices), 
and $F$ is the gauge 
field strength on the brane (the $B$ field is zero for the D5 solution).

We are interested in the low energy action describing the slow motion of the
probe in the directions transverse to the background branes. To this
purpose we can expand the action (\ref{azione}) up to quadratic order in
the derivatives of the transverse scalars.
We work in the static gauge 
($\xi^{\alpha}=x^{\alpha}$, $\alpha=0,...,5$)
and consider slow varying scalar fields in the transverse directions
$x^{m}=x^{m}(x^\mu)$, ($m=6,...,9$, $\mu=0,...,3$). 
The contribution from the DBI part
has the general expression 
\begin{equation}\label{sdbiI}
S_{DBI} \sim \int dx_4^2 \,\, d\Omega_2 \,\, e^{-\Phi_D}\sqrt{g}\left [
1+ \frac{1}{2} g^{\mu\nu} g_{mn} \partial_\mu x^m 
\partial_\nu x^n - \frac{1}{2} g^{\mu\tau}g^{\nu\rho}
F_{\mu\nu}F_{\tau\rho} \right].
\end{equation}
The $\sqrt{g}$ part in eq. (\ref{sdbiI}) 
combines with  $\int C_{(6)}$ to give the 
potential term in the low energy action, 
while the rest provides the kinetic term for the 
scalars and the gauge fields on the brane. The kinetic term 
for the scalars gives the metric on the moduli space of the gauge theory.
Finally the remaining CS term will 
give the $F\tilde{F}$ part of the SYM action.
Notice that this is a very general pattern appearing in all the various
examples of probe computations.

In our case, the potential for the probe reads
\begin{equation}
V= \int dx_4^2 \,\,  d\Omega_2 \,\,  e^{2h+2\Phi_D}\left ( 1-\sqrt{1+\frac{e^{\lambda_2+\lambda_3-2h} \cos^2\theta^\prime}{\Delta \tan^2\theta}}\right ).
\end{equation}
There is a region where the potential
term vanishes and the probe can move freely.
We will focus on solution $A$, solution $B$ being a
lengthy but straightforward generalization.
For all values of the parameter $K$, the potential vanishes for
$\theta'= \pi/2$, corresponding to  a motion of the probe in the plane
$(u, \phi_2)$. This is naturally identified with the moduli space of
the gauge theory, since $\phi_2$ generates the $U(1)$ in the
R-symmetry group. 
A bit more surprising is the fact that for $K<{1\over 4}$ the probe is BPS also
outside the $(u, \phi_2)$ plane, namely on the spherical disk defined  by
$u=u_0$, $0\le \theta' \le \pi/2$.

The next step is to look at the kinetic terms for the scalars and the
gauge fields to extract the probe tension, i.e. the 
gauge coupling $\tau$. Note however that
in order to be able to identify $\tau$ we need to recover the standard
structure of the ${\cal N}=2$ effective Lagrangian with
\begin{equation}
Im \left(\tau(z)\right) F^2 + Re\left(\tau(z)\right) F \tilde{F}+ Im \left(\tau(z)\right) \partial z \partial \bar{z}\label{f1}
\end{equation} 
for the gauge and scalar kinetic terms. It is important to notice that the same function $\tau$ appears both in the moduli space metric and in the gauge kinetic term.  
The problem is then to find the appropriate change of
coordinates that brings the effective action to the above form.

Let us first consider the solution $K=1/4$. In this case the moduli
space is given by the plane $(u, \phi_2)$, and with the coordinate
choice $z=e^{u + i \phi_2}$ the gauge coupling reads 
\begin{equation}\label{locusI}
\tau(z) = \frac{i N}{\pi} \log\frac{z}{\Lambda}. 
\end{equation}
For the solutions with $K <1/4$, the probe can move on the $(u,\phi_2)$ plane 
down to the radius $u_0$ and then it starts moving on the
spherical disk. The moduli space metric can be computed in both loci: on the
$(u,\phi_2)$  plane we obtain the same result for $\tau$ as in (\ref{locusI}),
  while on the disk $\tau$ assumes the constant value
\begin{eqnarray}\label{locusII}
\tau(z) =  \frac{iN}{\pi} \log{z_0 \over \Lambda},
\end{eqnarray} 
where $|z_0|=e^{u_0}$.
By comparing the supergravity results for $\tau$ with the gauge theory
expectation, we can determine the distribution of branes that generate the
solutions. At a generic point on the Coulomb branch of 
 ${\cal N}=2$ $SU(N)$ SYM, the 
one loop expression for $\tau$ as a function of the VEVs 
is (see (\ref{run2}))
\begin{equation}
\tau(z)={i\over\pi}\sum_i \log(z-a_i)\sim i\int da \mu (a)\log (z-a),
\label{tau2}
\end{equation}
where $a_i$ are the classical\footnote{In the SW solution
one distinguishes between classical VEVs, $a_i$, and quantum VEVs,
$u_i$. At one loop however there is no difference between the two and
we use the classical expression.} VEVs
and $\mu(a)$ is the VEV distribution in
the continuum limit. By equating it with the supergravity expression,
eqs. (\ref{locusI}), (\ref{locusII}), we find 
\begin{equation}
\mu (a)=\frac{N}{2\pi z_0}\delta (|a|-z_0),
\end{equation}
which corresponds to circular $U(1)_R$ invariant distributions of VEVs, with
radius $z_0$. This fits with the fact that the gauge theory we consider 
only contains the operator $Tr\phi \bar{\phi}$.

For $K <1/4$ the probe sees a completely smooth moduli space, indicating
that the one loop approximation is always valid. Those solutions 
should then be dual to weakly coupled vacua. Indeed we see from 
equations~(\ref{locusI}), (\ref{locusII}) that the VEVs are distributed on a circle with
radius $z_0>\Lambda$. In the large $N$ limit, all instantonic corrections
are suppressed by factors $(\Lambda/z_0)^N\ll 1$. The probe thus sees
the one loop results for $z>z_0$ and a constant coupling 
at scales below that set by the VEV distribution.

For $K=1/4$, $z=\Lambda$ ($u=0$ in the natural coordinates) is a
singularity for the probe action where it becomes tensionless ($\tau=0$). On
the field theory side, the gauge coupling $g_{YM}$ diverges. One is then
tempted to associate the corresponding supergravity solution to a 
strongly
coupled vacuum of the gauge theory. Every vacuum where the moduli
$u_i$ in the SW curve are distributed on a circle with radius
$r\le \Lambda$ would reproduce the result seen by the probe\footnote{The case $r=\Lambda$ is particularly intriguing because it corresponds
to an Argyres-Douglas point \cite{ad}, where the ${\cal N}=2$ theory becomes conformally invariant.}. 
The singularity in $\tau$ is interpreted
as the point on the moduli space where the perturbative approximation breaks
down and instanton corrections come into play. The region  at which the probe
becomes tensionless is usually called enhan\c con \cite{jpp} 
and it is identified with a
quantum distribution of VEVs, where the constituent branes have expanded to
form a shell. We will discuss the issue of singularities 
in more details in Section 4.5. 
Here we only want to stress that for this particular solution, the singular
region is actually point-like in the original coordinates, making the
interpretation as the enhan\c con less clear.

For solution $B$ the probe computation goes on as before, but the
identification of the appropriate ${\cal N}=2$ coordinates is more
difficult. The choice of coordinates $w = z + b^2/z$ brings 
the effective action in the form~(\ref{f1}) predicted 
by $\N=2$ supersymmetry. In solution $B$ the $U(1)_R$ symmetry 
is spontaneously broken and both the operators
${\mbox Tr} \phi\bar\phi$ and ${\mbox Tr} \phi^2$ have a VEV.
For $K= (1-b^4)/4$, the probe tension is given by
\begin{equation}
\tau(w)={iN\over \pi}\left ( {\rm arcosh}({w\over 2b}) 
+ {\mbox const}\right),
\label{tau}
\end{equation}
corresponding to the linear distribution of VEVs 
$\mu (a)=N/(\pi\sqrt{4b^2- a^2})$. 
Again one can interpret it as a strong
coupling vacuum.  
Curiously, this distribution of VEVs is of the same type as the one appearing 
for the \ntwo  moduli space region where all types of monopoles 
become massless \cite{ds}, which is the relevant one for 
the \ntwo $\ra$ \none $\,$ breaking. The situation for $K< (1-b^4)/4$ is similar to the analogous one in solution $A$, with two loci 
meeting along an ellipsis and a probe that sees a smooth moduli space.

For further works on the subject we refer to \cite{tutti,listaN2wrapped}.

\subsection{A supergravity dual of $\N=2^*$} 

Pure ${\cal N}=2$ SYM in four dimensions can also be obtained as a deformation
of ${\cal N}=4$ SYM with an equal mass for two of the chiral multiplets
$$
\delta {\cal L} = \int d\theta^2 m Tr(\Phi_1^2 + \Phi_2^2) = \sum_{i=1,2}
\left(m \lambda_i \lambda_i + |m|^2 |\phi_i|^2\right).
$$
At energies below the mass scale the theory should flow to pure ${\cal N}=2$ SYM.
Strictly speaking, to obtain pure SYM we should be able to decouple the massive
modes while keeping the low energy scale $\Lambda=me^{-8\pi^2/2Ng_{YM}^2}$ 
fixed. This requires a fine
tuning of the UV parameters 
which is outside the validity range of supergravity.
Thus, via gauge/gravity duality, we can only study theories with the 
massless content of pure ${\cal N}=2$ SYM but with additional massive 
states: ${\cal N}=2^*$ theories. Notice that the effective
Lagrangian will depend on the holomorphic
quantity $m$.

The general method for finding solutions corresponding to deformations of
$\N=4$ was discussed in Section 2. The solution can be found by using five 
dimensional gauged supergravity and then lifting  to ten dimensions.
In our case we are interested in the two fields 
\begin{eqnarray}
\alpha & \rightarrow &  \sum_{i=1}^4 Tr(\phi_i\phi_i) 
- 2\sum_{i=5}^6 Tr(\phi_i\phi_i),\nonumber\\   
\chi  & \rightarrow & Tr(\lambda_1\lambda_1+ \lambda_2\lambda_2) + h.c.
\end{eqnarray}
The ansatz for the five-dimensional solutions is of the form
\begin{eqnarray}
ds_5^2 &=& dy^2 + e^{2 Y(y)}dx_{\mu}dx^{\mu},\nonumber\\
\alpha &=& \alpha(y), \qquad \chi = \chi(y),\nonumber
\end{eqnarray}
with the boundary condition that the solution tends in the UV to $AdS_5$
($\alpha,\chi \rightarrow 0$, $Y \rightarrow y/R$). 
 In terms of ten dimensional fields, $\alpha$
 corresponds to the first KK mode
of the complex two form with both indices on $S^5$, and $\chi$ to the linear
combination of the internal part of the metric and of the four form potential.

The five dimensional solution was found in \cite{naked,pw,sfetsos}.
The presence of RR fields makes the lift to ten dimensions more complicated
than for the wrapped brane case \cite{pw}. In addition to the metric, the ten
dimensional Einstein-frame 
solution contains the self-dual five-form $F_{(5)}={\cal F} 
+ \star {\cal F}$, a complex
combination of the NSNS and RR two-forms\footnote{Since the 2-form is not
  needed in what follows, we refer the reader to \cite{pw} for its
  explicit expression.}, the dilaton and the axion \cite{pw}  
\begin{eqnarray}\label{pwsol}
ds^2  &=& 
\frac{(c X_1 X_2)^{1/4}}{\rho ^3} 
\left\{\frac{k^2\rho^6}{c^2-1}(dx_\mu)^2 + \frac{R^2}{\rho ^6 (c^2
    -1)^2}dc^2 + \right.\nonumber \\
&+& \left.   R^2
\left[ \frac{d\theta^2}{c} + \frac{\sin^2\theta}{X_2}d\phi^2 
+ \rho^6\cos^2\theta\left(
   \frac{\sigma_3^2}{cX_2} +
    \frac{\sigma_1^2+\sigma_2^2}{X_1} \right) \right] \right\},\nonumber\\
{\cal F} & =&
4 dx^0 \wedge dx^1 \wedge dx^2 \wedge dx^3 \wedge dw(r,\theta),\nonumber\\
\tau &=&  \frac{\tau_0 - \bar\tau_0 B}{1-B},
\end{eqnarray}
where the radial coordinate $y$ has been traded for 
$c=\cosh(2\chi)$ using the $\chi$ equation of motion\footnote{The
equations relevant for the change of coordinates are $\frac{d\chi}{dy} 
= - \rho ^4 \sinh(2 \chi)/2R,\, e^{Y}=k\rho^2/\sinh (2\chi)$.}. 

It is important to notice that, in these new variables, the boundary
corresponds to $c=1$ and the IR region to $c\rightarrow\infty$. 
As usual $R^2=\alpha'\sqrt{g_{YM}^2N}$, $\tau_0=\theta_s/2\pi + i/g_s$ 
is the asymptotic value of the complex dilaton and $\sigma_i$ are the 
left invariant one forms parameterizing the three-sphere. 
All the other functions in (\ref{pwsol}) are determined in terms of 
the five-dimensional solution \cite{pw}
\begin{eqnarray}
\label{pwdef}
&X_1 =  \cos^2\theta+c\rho^6\sin^2\theta, 
&B = e^{2i\phi} \frac{\sqrt{cX_1} - \sqrt{X_2}}{\sqrt{cX_1} 
+ \sqrt{X_2}}, \nonumber\\
&X_2 = c\cos^2\theta+\rho^6\sin^2\theta,
&w(r,\theta) = \frac{k^4\rho^6X_1}{4 \gs(c^2-1)^2} \nonumber\\
&\rho^6 = e^{6 \alpha}= c+(c^2-1)\left[\gamma+
  \frac{1}{2}\log\left(
{c-1\over c+1}\right)\right] \,.
\end{eqnarray}

The solution contains two parameters $k$ and $\gamma$.  
The parameter $k$ can
be identified with the mass perturbation $m$, while $\gamma$ 
parameterizes a family of
different solutions that should represent different flows to the IR 
${\cal N}=2$ theory.

As for the wrapped brane case, the solutions have a naked  singularity 
in the IR
and the value of $\gamma$ distinguishes among bad and
good ones. For $\gamma <0$ the metric and the dilaton become singular for 
$\rho \rightarrow 0$, which corresponds to a finite value $c=c_0$, 
while for  $\gamma=0$ the singularity is on
a ring at $\rho\rightarrow 0$, $c \rightarrow \infty$ and $\theta=\pi/2$. 
In these two cases the singularity has a physical interpretation. 
On the contrary the 
$\gamma > 0$ solutions
turn out to be unphysical.
Before discussing the probe results, notice that near the boundary
the solutions are asymptotic to $AdS_5 \times S^5$ for every value of  
$\gamma$, as expected.

The background geometry is generated by a stack of flat D3-branes, 
so that it is natural to use as a 
probe another D3-brane moving in the transverse directions 
 \cite{bpp,michela}. 
As discussed in Section \ref{nduewrapp}, one has to expand the probe
action for small velocities of the probe in the transverse directions, 
thus
obtaining a potential and a kinetic term for the transverse scalars 
plus the
usual $F^2$ and $F \tilde{F}$ terms for the gauge fields. 
We find again
that there are two loci where the potential vanishes: the plane ($c, \phi$),
$\theta = \pi/2$ (I) and the region where $\rho = 0$ (II), which
corresponds to a fixed $c_0$ and can be parameterized by $(\theta,\phi)$.  
The second region exists only for $\gamma <0$.
For the $\gamma <0$ solutions the two loci join
to give a completely smooth moduli space \cite{bpp}.
For $\gamma =0$, the gauge coupling is
\begin{equation}\label{tausug}
\tau (z) =  \frac{i}{g_s}
\left(\frac{z^2}{z^2-k^2 R^2}\right)^{1/2} 
+ \frac{\theta_{s}}{2\pi}, 
\end{equation}
where we define the complex coordinate, 
$z = kR (c \cos\phi - i \sin\phi)/ \sqrt{c^2-1}$,
in such a way that the scalar and gauge kinetic term have the standard 
${\cal N}=2$ form (see eq. (\ref{f1})).
The function $\tau(z)$  is singular for $z=\pm kR$ and  
has a branch cut on the segment $ -kR \le z \le kR$. In the original
coordinates this corresponds to $c=\infty$. 
For $\gamma =0$ the probe tension $\tau(z)$, expressed in the original
coordinates,  vanishes at the singularities.
This is the enhan\c con locus. We can determine the corresponding
brane distribution by comparing our result for the gauge
coupling to field theory expectations.
We need the one loop coupling constant of  ${\cal N}=2$ SYM with 
massive matter fields in the adjoint \cite{dhokphong}
\begin{equation}
\tau(z)= {i\over g_s} + \frac{\theta_{s}}{2\pi} + 
{i\over 2\pi }\sum_i \log \left (\frac{(z-a_i)^2}{(z-a_i)^2-m^2}\right )\,.\label{taud0}
\end{equation}
It will be sufficient to consider values of the moduli 
larger than the mass deformation 
\begin{equation}
\tau(z)\sim {i\over g_s} + \frac{\theta_{s}}{2\pi} + 
{i\over 2\pi }\sum_i {m^2\over
(z-a_i)^2 }\,.\label{taud}
\end{equation}
As in the previous section, by equating the continuum 
limit of eq. (\ref{taud})
to the supergravity result (\ref{tausug}) we obtain again a linear 
distribution for the VEVs \cite{bpp}
\begin{equation}
\mu(a) = \frac{2}{m^2 g_s}\sqrt{a_0^2 - a^2}\,,
\end{equation}
with $a_0^2 = m^2 g_s N/\pi$. In the supergravity limit, 
the size of the VEV distribution is much larger than the adjoint mass.
This justifies a posteriori the use of the one-loop approximation
in quantum field theory.

\subsection{$\N=2$ SYM from fractional branes}

In all the previous examples, we were forced to investigate particular 
points in moduli
space we could not select. The introduction of operators driving the theory to
different vacua
often induces other severe singularities. 
This is a general characteristic of models obtained with
the dimensional reduction to gauged supergravity. 
We now show that the use of fractional
branes allows, in principle, to study a generic point in moduli space.
The moduli indeed appear as free parameters in the solution.

The supergravity solutions corresponding to ${\cal N}=2$ fractional branes 
have been extensively discussed in~\cite{kn,bertolini,polch2,pg,torino,billo}. Here we
review the solution for our favorite example, the orbifold
singularity $\mathbb{R}^4/\mathbb{Z}_2$. As we saw in
Section~\ref{fractional}, with
$n_1$ fractional and $n_2$ anti-fractional branes at the orbifold
singularity $\mathbb{R}^4/\mathbb{Z}_2$ we can realize the theory $U(n_1)\times U(n_2)$.
As usual, the $U(1)$ factors are not described by the holographic dual.
The branes live
at $x_6=x_7=x_8=x_9=0$ and are arbitrarily distributed in the $(x_4,x_5)$  
plane. It is convenient to introduce the complex variable  $z=x_4+ix_5$ and to denote the positions of the fractional and anti-fractional branes by  
$a_i^{(1)},a_i^{(2)}$, respectively. 
In the gauge theory these correspond to  
VEVs of the Cartan 
values of the adjoint scalars parameterizing the generic vacuum.

Following~\cite{kn} we define
\begin{equation} 
\gamma = 2\pi\frac{(\tau_1-\tau_2)}{2} =  2\pi (c + \tau (b-\frac{1}{2})).  
\label{3} 
\end{equation} 
Notice that $\gamma (x_0,x_1,x_2,x_3,z)$ is a six-dimensional field living at the 
fixed plane $x_6=x_7=x_8=x_9=0$.  

Fractional and anti-fractional branes are sources for the RR fields $C_{(4)}$ and $C_{(4)}^T\sim c$.
Since they are oppositely charged under the twisted field $\gamma$
we can immediately write the linearized result~\cite{kn} 
\begin{equation} 
\gamma (z)=\gamma^{(0)}+2i\left (\sum_{i=1}^{n_1}\log (z-a_i^{(1)})
- \sum_{i=1}^{n_2}\log (z-a_i^{(2)})\right ),
\label{5} 
\end{equation} 
where the logarithms appear because the 3-brane is an extended source of real codimension two
for the localized six-dimensional field $\gamma$. Remarkably, $\gamma$ does not receive
any further correction. In fact, the supergravity equations only require 
$\gamma$ to be
holomorphic. Every holomorphic $\gamma$,  
combined with a black D3-brane ansatz \cite{pg,bertolini} 
\begin{eqnarray} 
ds^2&=&Z^{-1/2}dx_{\mu}dx^{\mu}+Z^{1/2}ds^2_K,\nonumber\\ 
F_5&=&dC_{(4)}+\star dC_{(4)},\qquad C_{(4)}={1\over Z}dx^0\wedge dx^1\wedge dx^2  
\wedge dx^3,
\label{6} 
\end{eqnarray} 
is a solution of Type IIB equations of motion provided that 
\begin{equation} 
- \Box_K Z=\rho_{D3}(x) + {\rm const}~
|\partial\gamma(z)|^2\delta^{(4)}(x_6,x_7,x_8,x_9).
\label{7} 
\end{equation} 
Here $\rho(x)$ is an arbitrary density of physical D3-branes
\cite{pg,bertolini}. The general solution of this equation is 
\begin{eqnarray} 
Z (x_T,z) &=& \sum_{i=1}^{n_1}{b^{(0)}\over (x_T^2+|z-a_i^{(1)}|^2)^2} 
+ \sum_{i=1}^{n_2}{1-b^{(0)}\over (x_T^2+|z-a_i^{(2)}|^2)^2}\nonumber\\
& & + \, {\rm const} \int d^2 w 
{|\partial\gamma (w)|^2\over (x_T^2+|z-w|^2)^2}. 
\label{8} 
\end{eqnarray} 
We see that, by taking $\gamma$ as in eq.~(\ref{5}), we obtain a solution
depending  on $n_1+n_2$ parameters representing the moduli of the $\N=2$ gauge theory.

The logarithmic behavior in~(\ref{5}) reproduces 
the one-loop beta function of the ${\cal N}=2$ gauge theory~\cite{kn}.
To see this more clearly, we can  introduce a probe in the system. 
Just send in  an extra fractional brane
represented  as a D5-brane wrapped on the vanishing cycle and 
positioned at $z$.  
One can immediately see that all factors of the warp function $Z$ cancel 
in the Born-Infeld action.
Indeed, the very same argument that led to eq.~(\ref{eee2}), tells us that the effective coupling constant on the probe is
\begin{equation}
\tau(z)=\frac{\gamma (z)}{2\pi}+\frac{\tau}{2}=
\tau^{(0)}+ \frac{i}{\pi}\left (\sum_{i=1}^{n_1}\log (z-a_i^{(1)})-
\sum_{i=1}^{n_2}\log (z-a_i^{(2)})\right ). 
\label{probe}
\end{equation}

This result obviously agrees with the 1-loop coupling constant of the $U(1)$ factor
as predicted by gauge theory\footnote{As far as the 1-loop result 
is concerned, the anti-fractional gauge group can be considered as inert.
We are left with an $SU(N)$ theory with $2n_2$ fundamental hypermultiplets with masses 
$m_i\equiv a_i^{(2)}$. The one loop result for an $SU(N)$ theory with $N_f$ fundamentals
hypermultiplets reads \cite{dhokphong}
$\tau(z)= \tau^{(0)}+ \frac{i}{\pi}\left (\sum_{i=1}^{N}\log (z-a_i)- \frac{1}{2}
\sum_{i=1}^{N_f}\log (z-m_i)\right )$.}.

The solution~(\ref{8}) 
presents various kinds of singularity.  
The supergravity background only reproduces the 1-loop result in the gauge
theory and  presents a singularity in the IR region, where the physics 
becomes non-perturbative. We can consider, for example, the case of
pure $SU(N)$ gauge group, $n_1=N$ and $n_2=0$, and a typical strongly coupled
vacuum, $a_i^{(1)}=0$. Equation~(\ref{probe}) becomes 
$\tau(z)=\frac{i}{\pi}\log z/\Lambda$. The supergravity solution
has an IR singularity at $z=0$.
The probe, on the other hand, becomes tensionless at the scale $\Lambda$
before reaching the singularity. From eq.~(\ref{3}) we see that, 
at the same scale, the space-time fields $b$ and $c$ vanish. 
In string theory new massless fields (in this case
tensionless strings) are expected.
The singularity is thus surrounded by a spherical shell, impermeable
to the probe, which is characterized by new space-time states
becoming massless. All these phenomena are usually associated with 
the enhan\c con mechanism, that will be discussed in Section \ref{theissue},
and suggests a possible resolution of the IR singularity.
In a general model, we may expect other
singularities at the positions of the constituent  
branes and a break down of the supergravity approximation 
near the  orbifold fixed planes. 

A general discussion of the interpretation of the supergravity
solution can be found in \cite{polch2,a,russo}.
More general solutions for systems of fractional branes at 
orbifold singularities
with $\N=2$ or $\N=1$ supersymmetry can be found in \cite{listatorin}. 
Models with fundamental matter fields can be obtained by adding D7-branes and the
corresponding solution, which involves a non-trivial holomorphic dilaton,
were discussed in \cite{pg,torino}.

\subsection{The issue of singularity}\label{theissue}

In all previous examples, we have seen that the supergravity solution only captures
the 1-loop result in the gauge theory and it is plagued by IR singularities. 
It seems widely believed that the resolution of singularities in the $\N=2$ models
is obtained with the mechanism known as the enhan\c con \cite{jpp}, 
where the constituent branes
reach an equilibrium configuration by forming shells.
Such a behavior is suggested by the SW solution of $\N=2$ gauge theories.
In the M theory approach to solving $\N=2$ gauge
theories, the SW curve actually describes what a system of branes looks like
after quantum corrections are taken into account. 
Consider, for example, pure $SU(N)$ gauge theory. In a generic strongly coupled vacuum,
the 1-loop result for a probe is valid until the scale $\Lambda$ where, extremely suddenly at large $N$,
instantonic corrections start to dominate the physics. What happens below 
the scale $\Lambda$
is accurately described by the SW curve $y^2=P(x)^2-\Lambda^{2N}, P(x)=\prod (x-u_i)$.
The importance of the curve for our purposes
is that, in the M theory description \cite{wittenM}, 
the moduli $u_i$ describe the positions of the constituent branes.
To exemplify the general situation, let us consider a circular
distribution of VEVs $|u_i|=r$, as that encountered in Section 4.2.
The curve reads $y^2=(x^N-r^N)^2-\Lambda^{2N}$. 
We can try to engineer this system  by forcing $N$ branes on a circle
of radius $r$.  
At large $N$, we can get a hint of what the quantum distribution of branes
looks like by taking a symmetric section of the curve at $y=0$
(the set of branch points of the curve)
\begin{equation}
(x^{N}-r^N)^2-\Lambda^{2N}=0.
\label{enh}
\end{equation}
In a strongly coupled vacuum $r<\Lambda$, the solutions of this equation
are distributed on a circle of radius $\Lambda$, since 
all powers of $(r/\Lambda)^N$ are negligible for large $N$.
We see that the distribution of 
branes, at the quantum level, expand to form a spherical shell of radius 
$\Lambda$, that we will call  enhan\c con. One can also see
that a probe cannot move beyond $x=\Lambda$\footnote{A probe can be indeed introduced considering the $SU(N+1)$ theory
in the vacuum $(u_1-z/N,....,u_N-z/N,z)$. Eq.~(\ref{enh}) is replaced,
for large $N$, by $(x^{N}-r^N)^2(x-z)^2-\Lambda^{2N+2}=0$. We see that, for $z>\Lambda$
we have a pair of branch-points
 $x\sim z$, corresponding to the probe moving outside the enhan\c con.
For $z<\Lambda$, instead, the $2N+2$ solutions of the equation are distributed
on the circle $x\sim\Lambda$. The probe cannot enter the enhan\c con and instead
dissolves in the spherical shell.}.
A similar analysis can be repeated for all points in moduli space
showing that this phenomenon occurs for all strongly coupled vacua, with an enhan\c con that
may change shape, and even degenerate (into a segment, for example) in
particular situations. In weakly
coupled vacua, on the other hand, one can check that a probe 
can move freely everywhere. The reader can easily check it in the
case of a circular distribution of radius $r>\Lambda$.
The case $r=\Lambda$ is special since it corresponds to an 
Argyres-Douglas conformal fixed point \cite{ad}.
These results are quite general for ${\cal N}=2$ theories:
using the much more complicated curve 
discussed in Appendix \ref{swcurve}, one can also check for instance
that the same phenomena occur for $SU(N+M)\times SU(N)$ groups 
\cite{russo}.

We would like to use this information from  quantum field theory to learn 
about the holographic dual. We identify
solutions where a probe can move freely everywhere with weakly
coupled vacua. We already made this identification in the previous Sections.
On the other hand, we identify solutions where the probe encounters
a barrier with strongly coupled vacua. The SW curve then suggests
that the constituent branes form shells.
We may expect, using the Gauss law, that the space-time field becomes 
constant inside the shell thus resolving the
singularity. The supergravity result is then an accurate description of 
the physics only outside the enhan\c con.
The name enhan\c con is used in the literature
associated to the following features:
\begin{itemize}
\item{It is the natural boundary for the motion of a probe. At the enhan\c con
the probe stops being a BPS object or stops being an elementary object. It
is supposed to dissolve in the enhan\c con.}
\item{It is a gravitational shell where gravity stops being attractive. It is a shell
around a repulson singularity.}
\item{It is a locus where extra massless string states become important.}
\item{It is a locus where there is an enhanced symmetry in space-time; from this the name!}
\end{itemize}
Actually, all the previous four features were realized only in the original
example with D2 and D6 branes on $K3$ in Type IIA \cite{jpp}. Since then,
some of these features are randomly realized in $\N=2$ models discussed in the literature.
The enhan\c con picture seems a good description for the fractional brane system and a slightly
less good description for the wrapped brane system, as discussed
in Section 4.2.
In the latter case, one can also study the system using world-sheet 
methods. A world-sheet $CFT$ describing a T-dual of the 
supergravity solution $A$ described in Section 4.2 has been written in 
\cite{horikapustin} and seems a good description for the weakly 
coupled vacua $K<1/4$, but not for $K=1/4$ where it becomes singular.


\section{Supergravity duals of $\N=1$ gauge theories}

In the first part of this Section we review the basic features of 
four dimensional $\N=1$  gauge theories that will be used 
in the following. 
We will then focus on two string-supergravity duals obtained with 
wrapped and fractional
branes. These models are known as the Maldacena-Nu$\tilde{\mbox n}$ez 
(MN) \cite{mn2} and Klebanov-Strassler (KS) \cite{ks} solutions,
respectively. The basic difference with respect to the corresponding $\N=2$ 
solutions is
that they are completely regular. Both theories exhibit confinement and spontaneous breaking 
of the chiral symmetry. We will also briefly describe 
the Polchinski-Strassler (PS) solution \cite{ps}, describing the 
$\N=1^*$ theory,
obtained as a deformation of the $\N=4$ $CFT$.    
 We conclude with some comments about $\N=0$ models obtained as soft 
breaking of $\N=1$.

\subsection{Some remarks on $\N=1$ SYM}

In this Section we review some of the basic features of
$\N=1$ supersymmetric gauge theories, focusing on the aspects that are
relevant for the comprehension of string duals.
For a general  and more complete discussion about $\N=1$  gauge theories 
we refer the reader to good reviews such as \cite{peskin,intseib}. In 
recent years, several string
models generalizing pure $\N=1$ SYM have appeared 
in the literature, from MCQD \cite{MQCD} 
to the many $AdS$-inspired models. Here we 
review pure $\N=1$ SYM and discuss how this results
can be adapted to its generalisation encountered in string duals. 

It is widely believed that pure $\N=1$ SYM confines and has a mass gap. 
The characteristic scale of the theory $\Lambda$ 
is set by the tension of the color flux tubes, or briefly QCD strings. 
They are not BPS objects and the value of their tensions cannot be fixed 
in terms of central charges or symmetries. Strings connecting 
external sources in different representations of the gauge group
are, in general, different physical objects. 
They  are classified by the center of the gauge group. 
In a confining $\N=1$ $SU(N)$ SYM theory, 
we can define $N-1$ different types of QCD strings,
since there are exactly $N-1$ representations of the 
gauge group that are not screened by gluons.
A $k$-string, $k=1,...,N-1$, connects external sources 
in the $k$-fold antisymmetric 
representation of $SU(N)$. It is then interesting to ask what is the
ratio of the tensions for $k$-strings. In many stringy-inspired
models one can derive the sine formula
\begin{equation}    
{T_k\over T_{k^\prime}}={\sin k \pi /N\over \sin k^\prime \pi /N}.    
\label{sine} \end{equation} 
This formula, or mild modifications of it, is valid 
in a variety of toy models exhibiting confinement,  
from softly broken ${\cal N}=2$ SYM 
\cite{ds} to MQCD \cite{hsz}. As we will see, it is also 
realized in the MN solution (and, with a small correction, 
in the KS model). It is certainly not an universal formula.
There are many quantum field theory counterexamples showing that it
can have corrections \cite{counterexamples}.
It would be quite interesting to understand if this formula 
is valid in pure YM theories. 
Unfortunately, since the QCD strings are not BPS, 
there is no known method of 
performing a rigorous computation in $\N=1$ SYM. Interestingly, 
the sine formula has been supported by recent lattice 
computations for pure non supersymmetric YM \cite{pisani}.

Another common feature of ${\cal N}=1$ theories is
spontaneous breaking of chiral symmetry.
The ${\cal N}=1$ Lagrangian can be written in superfield notation as 
\begin{equation}
L = -\frac{i}{16\pi}\int d^2 \theta \tau W_{\alpha}^2 +{\rm h.c.}
\label{eq111}
\end{equation}
There is a classical $U(1)_R$ symmetry, rotating the gaugino, which
is broken to a discrete $\mathbb{Z}_{2N}$ subgroup by instantons. 
This theory has $N$ vacua associated with the spontaneous breaking
of the $\mathbb{Z}_{2N}$ symmetry to $\mathbb{Z}_2$ by gaugino condensation,
\begin{equation}
<\lambda \lambda >\sim N\Lambda ^3 e^{2\pi in/N}\label{gaugino},
\end{equation}
where $\Lambda$ is the physical, RG invariant mass scale, 
and may be written in terms of the bare coupling 
$\tau$ at some UV scale as $\Lambda = \Lambda_{UV} e^{2 \pi i \tau / 3N}$.
The integer $n=0,..,N-1$ in (\ref{gaugino}) labels the
different vacua. 
The gaugino condensate is an operator with protected dimension three, 
since it is part of a chiral multiplet.
All information about  the vacuum can be conveniently      
described by a non-perturbative holomorphic superpotential    
\begin{equation}     
W = N^2 \Lambda^3  e^{2 \pi i n/ N}.
\label{n1w}   
\end{equation} 
Indeed from eq.~(\ref{eq111}) we see that the vacuum expectation value of the
superfield $W_{\alpha}^2$, whose lowest component is the bilinear $\lambda \lambda$,
can be obtained by differentiating the effective superpotential with respect to $\tau$.
The result~(\ref{gaugino}) then follows from eq.~(\ref{n1w}).
In presence of a spontaneous breaking of the $\mathbb{Z}_{2N}$ symmetry, we expect the existence
of domain walls (classical field configurations 
of codimension one) separating different vacua \cite{shif,MQCD}. 
The domain walls 
in $\N=1$ gauge theories are BPS saturated and 
their tension is determined by a central charge
of the supersymmetry algebra \cite{shif,MQCD}, in terms of holomorphic data. 
The tension of a domain wall connecting the vacua $i$ and $j$ 
is determined by the difference of the superpotential
\begin{equation}
T_{DW}\sim |W(i)-W(j)|,
\label{sup}
\end{equation}
which for pure $\N=1$ explicitly reads
\begin{equation}
T_{DW}\sim N|(\lambda\lambda)_i-(\lambda\lambda)_j|\sim N^2\Lambda^3 
\sin{{(i-j)\pi \over N}}.
\end{equation}
In the large $N$ limit the tension is then linear in $N$. By analogy
with D-branes, it was conjectured that the QCD strings can end on 
${\cal N}=1$ domain walls\cite{MQCD}. This typically happens in all stringy-inspired
generalization of pure ${\cal N}=1$ SYM.

The properties that are constrained by holomorphicity and symmetries are also 
valid in many generalization of the pure $\N=1$ SYM. 
In theories with spontaneous breaking of the $\mathbb{Z}_{2N}$ symmetry, 
we may expect  
$N$ vacua, a vacuum superpotential determining the condensates
 and domain walls, whose tensions are fixed 
by eq.~(\ref{sup}). This happens indeed in
MQCD, and in the MN and KS solutions. In pure $\N=1$ SYM, the 
characteristic scales of chiral symmetry breaking and of
QCD strings are fixed in terms of a single scale: 
$T_{DW}\sim N \Lambda^3$ and $T_s\sim\Lambda^2$.
In more general models, $T_{DW}$ and $T_s$ can be distinct.
While $T_s$ is not protected and computable only in the
semiclassical approximation, $T_{DW}$ is BPS-protected.
Its explicit value may nevertheless depend on the extra parameters 
in the theory, as it happens in MQCD, for example \cite{MQCD}, or 
in the MN and KS
solutions, as we will see.
Notice that some of the previous results are not applicable to the PS model
\cite{ps}, which describes $\N=1^*$, 
because the chiral symmetry is not visible in the supergravity approximation.

\subsection{${\cal N}=1$ SYM from wrapped five-branes}\label{none}

In this Section we will review the solution corresponding to 
${\cal N}=1$  SYM that can be constructed with wrapped 
five-branes \cite{mn2}.
The set up is similar to the ${\cal N}=2$ case of Section \ref{nduewrapp}.
In order to have a four dimensional world-volume theory,
we wrap $N$ Type IIB NS5-branes  on a two-cycle.
This will be a gauge field theory, since at low energies the LST on 
the world-volume of flat NS5-branes
reduces to six dimensional SYM.
The difference with the ${\cal N}=2$ case comes from the ambient geometry for the two-sphere.
In order to preserve only four supercharges the manifold in which the two-sphere is embedded must be a Calabi-Yau threefold.
We refer to Section 3.2 for conventions about charges and a detailed discussion
of the twist. Recall that, with the transverse group of symmetries of a 5-brane
written as $SO(4)\sim SU(2)^+\times SU(2)^-$, the abelian field responsible for the
twist is now $U(1)^+\subset SU(2)^+$. 
It was shown in Section 3.2 that with this choice of twist the brane field content at low energies is simply an ${\cal N}=1$ vector multiplet. 
$U(1)^+$ appears as the surviving $U(1)_R$ symmetry
of the ${\cal N}=1$ theory.

The supergravity solution will be a deformation of the linear dilaton
 background, dual to the LST.
As in the ${\cal N}=2$ case, the solution can be found using
seven dimensional gauged supergravity. We can consistently truncate
this theory  to the sector invariant under $SU(2)^-$.
Only one scalar field, the dilaton
$\phi=5\lambda$ ($\lambda_i=\lambda$, $i=1,2,3$ in
 eq. (\ref{scalars})), 
survives this truncation.
In addition to the dilaton, we expect that at least 
the metric warp factors and the $U(1)_R$ abelian field 
should be turned on.
A supersymmetric solution with these fields exists but it is singular. 
It turns out
that a regular solution can be found by turning on also
the non Abelian part of the $SU(2)^+$ gauge connection.
As we will show below, the new field is dual to the gaugino condensate.
It is remarkable that the space-time field de-singularizing the solution 
is associated with the non-trivial 
IR dynamic of the ${\cal N}=1$ SYM theory.

We are interested in solutions of the form
\begin{eqnarray}
ds_7^2 &=& e^{2f} (dx_4^2 + N\alpha' d\rho^2) 
           +  e^{2g} (d\theta^2 + \sin^2 \theta
d\phi^2) , \nonumber\\
A &=& \frac{1}{2} \left[ \sigma^3 \cos \theta d \phi +
{a\over 2} (\sigma^1 +i\sigma^2) (d \theta  -i\sin \theta d \phi) + \mbox{c.c} 
\right],\label{nonefields} \\ 
T_{ij}& =& e^{2\lambda} \delta_{ij},
\nonumber
\end{eqnarray}
where all the fields $f$, $g$, $\lambda$  and $a$ only depend on $\rho$.
The BPS equations for supersymmetric solutions are considerably more complicated than in
the ${\cal N}=2$ case. They can be found in Appendix \ref{unasol}.
Alternatively, we can write an effective action allowing for a
superpotential and solve the corresponding 
first order equations as we did in Section 4.2.

Substituting the ansatz (\ref{nonefields}) in the seven dimensional 
supergravity Lagrangian
and integrating over $S^2$, we
obtain the following one dimensional effective Lagrangian 
\begin{equation}
{\cal L}= {3\over16}e^{4Y}[16(Y')^{2}- 2(h')^{2} -\frac{1}{2}e^{-2h}|a'|^2+2e^{-2h}-\frac{1}{4}e^{-4h}(|a|^2-1)^2+4],
\label{lagmn2}
\end{equation}
where $h=f-g$, $4Y= 2h -2\phi +\log(16/3)$. Analogously to what we did 
for the $\N=2$ case, we set by hand $f=-2\lambda$.
The associated superpotential is
\cite{pandotse,gubsevol,tutti}
\begin{equation}
W=-{3\over8} e^{-2h}\sqrt{(1+4e^{2h})^2 +2(-1 + 4e^{2h})|a|^2 +|a|^4}.
\end{equation}
The supersymmetric solution was first found in \cite{chamsevolkov} and
subsequently reinterpreted in the context of gauge/gravity duals in
\cite{mn2}. It reads
\begin{equation}
e^{2h}=\rho \coth{2\rho}-{\rho^2 \over \sinh^2{2\rho}}-{1\over4},\qquad
a={2\rho \over \sinh{2\rho}},\qquad
e^{2\phi}={2e^h\over \sinh{2\rho}}.
\end{equation}
The ten-dimensional solution is obtained using formulae~(\ref{cvlift}) 
for the lift and 
. We write the solution using 
the Euler angles on the three-sphere
\begin{eqnarray}
g &=& e^{ i \psi \sigma^3 \over 2 } e^{ i \tilde \theta \sigma^1 \over 2 } 
e^{ i \tilde\phi \sigma^3 \over 2}, 
\nonumber \\
 { i \over 2} w^a \sigma^a &=& dg g^{-1}, 
\nonumber \\
w^1 + i w^2 &=& e^{ - i \psi } ( d \tilde \theta + i \sin \tilde \theta d \tilde\phi),
\nonumber \\
w^3 &=& d \psi + cos\tilde \theta  d\tilde\phi,
\end{eqnarray}
with  $\psi \in [0, 4\pi]$.

In the string frame, the 10$d$ solution for the wrapped NS5-branes is 
($A={1\over 2}A^a \sigma^a$) \cite{mn2}
\begin{eqnarray}\label{mn2metricNS}
ds^2_{str} &=& dx_4^2 +N \alpha'  [
d \rho^2 + e^{ 2 h(\rho)} ( d \theta^2 + \sin^2  \theta d\phi^2 ) + {1 \over 4} \sum_a
( w^a - A^a)^2],\nonumber\\
H^{NS}_{(3)} &=& {N\alpha' \over 4} [ - (w^1 -A^1)\wedge (w^2 - A^2) \wedge ( w^3-A^3)  + 
\sum_a F^a \wedge (w^a -A^a)], \nonumber\\
e^{2 \Phi_{NS}} &=& e^{ 2 \Phi_{NS,0}}{2e^{h(\rho)} \over  \sinh 2\rho}.
\end{eqnarray}
This is the Maldacena-Nu$\tilde{\mbox n}$ez solution. 
The metric is completely
regular in the IR, where it is of the form $\mathbb{R}^7\times S^3$. Indeed, $a\rightarrow 1$ in the IR and $A$ is a pure gauge which can be reabsorbed by a coordinate transformation on $S^3$. Moreover, since 
$e^{2 h} \rightarrow \rho^2$, the original $S^2$ is now contractible and combines with $\rho$ to give $\mathbb{R}^3$.
The (squared) radius of the three-sphere
is of order $N\alpha^\prime$ and the supergravity 
approximation is valid when $ N\gg 1$. 
The string coupling 
is vanishing for large $\rho$ and reaches its maximum value,
$e^{\Phi_{NS,0}}$, at $\rho=0$. For $e^{\Phi_{NS,0}} \ll 1$ 
the string coupling 
is everywhere small and all loop corrections are suppressed. 
However from the gauge theory point of view, we would like to decouple 
the KK modes in order to get pure SYM. As we will see in detail below, 
the ratio between the scales of the QCD strings and the KK modes is 
of order $e^{-\Phi_{NS,0}}N$ so that to decouple to scales we need
to send $e^{\Phi_{NS,0}}\rightarrow \infty$ \cite{mn2,nastase,aharony}. 
This forces us 
to use the S-dual D5 solution (see eq.~(\ref{eq2}))
\begin{eqnarray}\label{mn2metric}
ds^2_{str} &=& e^\Phi \left[ dx_4^2 +N \alpha'  [
d \rho^2 + e^{ 2 h(\rho)} ( d \theta^2 + \sin^2  \theta d\phi^2 ) + {1 \over 4} \sum_a ( w^a - A^a)^2] \right], \nonumber\\
F_{(3)} &=& \frac{N\alpha'}{4} [ -  (w^1 -A^1)\wedge (w^2 - A^2) \wedge ( w^3-A^3)  +  \sum_a F^a \wedge (w^a -A^a)],
\nonumber\\
e^{2 \Phi} &=& e^{ 2 \Phi_0}{\sinh 2\rho\over 2e^{h(\rho)}}.
\end{eqnarray}
Notice that for the D5 solution the (squared) radius of the IR 
three-sphere is $e^{\Phi_0}N\alpha'$ and 
the smallest value of the string coupling is 
$e^{\Phi_0}=e^{-\Phi_{NS,0}}$, reached
for $\rho =0$. The string coupling grows with $\rho$ and eventually will diverge in the UV.
We would also like to stress that the solution in the IR is very similar 
to the Klebanov-Strassler one \cite{ks}, which will be discussed 
in Section \ref{kleba}, since both models involve an IR geometry 
that corresponds to a deformed conifold.

The rest of this section is devoted to the analysis of the properties of the MN solution. In this discussion we will always use the 
D5 solution.

We first show how confinement is realized in the MN model. 
The natural candidate for a QCD string is a fundamental string.
Using a standard argument in $AdS/CFT$, we can see 
that there is confinement and compute the QCD string tension.
Confinement is expected because the space-time components of the metric 
at $\rho=0$ are  non-vanishing. 
The value of a Wilson loop, indeed, can be computed using
a fundamental string coming from infinity, with endpoints on the boundary at $\rho=\infty$. 
The string will minimize its energy
by reaching $\rho=0$ where  the metric components $\sqrt{g_{xx}g_{tt}}\,$ have a minimum. 
All the relevant contribution to the energy between two external sources 
is then due to a string sitting at $\rho=0$ and stretched in the $x$ 
direction. The estimate for the string tension is
then easily obtained
\cite{mn2,loeson}
\begin{equation}
T_s={e^{\Phi_0}\over 2 \pi \ap}.\label{ten}
\end{equation}
As reminded in Section 5.1, in a confining ${\cal N}=1$ $SU(N)$ SYM theory, 
we can define $N-1$ different type of strings connecting different external sources.
It has been shown in \cite{klebstring} that the ratio of the tensions in the MN
solution follows the sine formula
\begin{equation}    
{T_k\over T_{k^\prime}}={\sin k \pi /N\over \sin k^\prime \pi /N} .   
\label{sine2} \end{equation} 
This formula can be obtained by considering the IR metric $\mathbb{R}^7\times S^3$. 
The QCD string is described by a bound state of $k$ fundamental strings that
minimize its energy by expanding in a D3-brane with $k$ units of flux 
wrapping an $S^2$ inside $S^3$. The tension of the brane is balanced by the
space-time three-form and a stable configuration is obtained for a specific, $k$-dependent,
 $S^2$ inside $S^3$. Formula~(\ref{sine2}) then follows from a ratio of volumes 
\cite{klebstring}.
The ratio of $k$-tensions is a genuine, non BPS, prediction of the MN solution.

Let us now discuss the spontaneous breaking of the chiral symmetry.
From our general discussion in Section 5.1 we expect 
various phenomena:
\begin{itemize}
\item{The anomaly. In quantum field theory, $U(1)_R$ is anomalous 
and broken to $\mathbb{Z}_{2N}$ by
instantonic effects. These non-perturbative effects in the field theory
are already captured at the supergravity level. 
The existence of an anomaly can be detected with 
an UV computation in quantum field theory and therefore
should be already visible in the UV region of the solution.
In the MN solution the $U(1)_R$ symmetry acts as a shift
of the angle $\psi$. The UV form of the metric is invariant under
such shift, but this is not the case for the RR two-form 
$C_{(2)}\sim -\frac{N\alpha'}{2}\psi \sin\theta d \theta \wedge d \phi$.
In particular, the flux $\frac{1}{2\pi\alpha'}\int_{S^2} C_{(2)}$
varies by $-N\delta\psi$ under a shift of $\psi$. Since, as we discussed in Section 3.3,
the flux is periodic with period $2\pi$, the only allowed
transformations are those with $\delta\psi=\frac{2\pi n}{N}$ 
\cite{anomaly}:
the R-symmetry is then broken to the $\mathbb{Z}_{2N}$
subgroup $\psi\rightarrow\psi+2\pi k/N$ ($\psi$ has period $4\pi$).
This is a purely supergravity result. Nevertheless,
we can explicitly see the role of instantons in the anomaly
by considering an instantonic probe.  
In the D5 solution, instantons are identified  
with an Euclidean  D1-brane wrapping the same $S^2$ as the D5-branes.
An important point is that the $S^2$ where the branes 
can be wrapped in a stable way is not the original
$S^2$ parameterized by $(\theta,\phi)$, but it is mixed with an $S^2$ 
contained in the transverse $S^3$ \footnote{"For a detailed
discussion of the geometry of these cycles and their  relation to the 
anomaly see \cite{bertolinil}.}.
Indeed, since in the UV (where $a$ is vanishing) $d\psi$ only
appears in the combination $w^3-A^3=d\psi+\cos\tilde \theta d\tilde\phi-\cos\theta d\phi$, a stable D1-brane can  
live at a fixed $\psi$ only when it wraps
the sphere $\theta=\tilde\theta,\phi=\tilde\phi$. 
The action for a D1-brane should reproduce the coupling constant and
theta angle of the gauge theory,
\begin{equation}
-\frac{1}{2\pi \alpha' } \int_{S^2} e^{-\Phi}\sqrt{G} +\frac{i}{2\pi \alpha' } \int_{S^2} C_{(2)} = - \frac{8\pi^2}{g^2_{YM}} + i \theta_{YM}.
\label{coup}
\end{equation}
In particular, the theta angle is 
$\theta_{YM} \sim - N\psi$.
The anomaly of $U(1)_R$ and its breaking to $\mathbb{Z}_{2N}$ 
are evident in the shift of the theta angle.
Only those transformations that shift $\theta_{YM}$ by a multiple
of $2\pi$ remain good symmetries.} 
\item{The $N$ vacua. The spontaneous breaking $\mathbb{Z}_{2N}\rightarrow\mathbb{Z}_2$
is manifest in eq.~(\ref{mn2metric}).
$\mathbb{Z}_{2N}$ indeed is a good symmetry only in the UV.
It is broken to the $\mathbb{Z}_{2}$ 
symmetry $\psi \rightarrow \psi + 2\pi$ 
by the explicit form of the supergravity solution, due to
the presence of a non-zero $a$.
In general, different vacua of the theory
correspond to different regular solutions 
with the same asymptotic behavior. 
In our case, one can show that precisely $N$ solutions are nonsingular, 
corresponding to the $N$ vacua of ${\cal N}=1$ SYM \cite{mn2}.
The  vacua are permuted by the elements of $\mathbb{Z}_{2N}$,
which multiply $a$ by a phase. 
The $N$ regular solutions are then as in eq.~(\ref{mn2metric})
with $a$ replaced by $a= e^{2\pi n i/N}{2\rho \over \sinh{2\rho}}$,
$n=0,...,N-1$.
}

\item{The gaugino condensate. 
We expect that each vacuum is associated with a non-zero gaugino 
condensate. We can show this \cite{noi} by using 
the $AdS/CFT$ philosophy of Section 2.3.
Recall that we can determine if a given operator has a VEV
by looking at the asymptotic UV behavior of the dual supergravity
field. One solve the asymptotic 
second order equations of motion for the field, 
and associates a non-normalizable solution
with a deformation by the dual operator and a normalizable one
with a VEV. In our case, the background is not asymptotically $AdS$
but we can still try to see what this philosophy suggests.
The natural candidate for the field dual to the gaugino condensate
is $a$, since it has the right $U(1)_R$ charge and its phase
distinguishes among the various vacua.
From the reduction on $S^2$ of the 5-brane coupling
\begin{equation}
A^\mu_{ij}\bar{\Psi}\gamma^\mu\Gamma^{ij}\Psi,\label{apsipsi}
\end{equation}
one can indeed show that $a$ couples to the fermionic bilinear
$a \bar{\lambda^c}\lambda$, 
corresponding to the gaugino condensate \cite{noi}. 
The asymptotic solutions of 
the second order equations for $a$ can be derived from 
the Lagrangian (\ref{lagmn2}) 
\cite{gubsevol} 
\begin{equation}    
a \sim { Y\over \sqrt{2\rho}} +2 C \rho e^{-2\rho}.\label{asy}
\end{equation}
In the BPS MN solution $Y=0$ and $C\ne 0$, so the AdS/CFT dictionary
adapted to our case suggests that $a$ behaves as 
appropriate for a condensate. The quantity $C$ determines the 
value of the condensate in a given 
vacuum. We can also strengthen this interpretation by studying the
UV radial dependence of the field $a$. For this we have to choose an
energy/radius relation. As recalled in Section 2.1, 
this relation is ambiguous for D5-branes \cite{pp}.
We can choose to determine the relation by looking at the
coupling constant behavior\footnote{The following identification can be motivated by the fact that,
below the compactification scale, the one-loop $\beta$-function is fixed by
the chiral anomaly and can be extrapolated from the weak 
coupling result.}. An estimate for the UV behavior 
of the gauge coupling follows from our general discussion in Section \ref{genobs}
\begin{equation}\label{raggen}
{1\over g_{YM}^2}= \frac{1}{2(2\pi)^3\alpha^\prime}\int_{S^2} e^{-\Phi}\sqrt{G}\ra{N\over4\pi^2}\rho.
\end{equation}
This result has been obtained by using a stack of D5-branes, but 
can be equivalently
derived by using the action for an instanton, eq.~(\ref{coup}).
If we try to enforce the one-loop gauge theory result
${1\over g_{YM}^2}\sim {3N\over 8\pi^2}\log\mu $, 
we obtain the asymptotic radius/energy relation
$\rho\sim{3\over2}\log\mu$. 
From $a\sim \rho e^{-2\rho}$ we see that $a$ scales as $1/\mu^3$, 
as appropriate for a
protected dimension three operator \cite{noi}.}
\item{Domain walls. In a theory with spontaneous breaking of $\mathbb{Z}_{2N}$ and multiple vacua, we expect the
existence of domain walls.
In the string solution, they correspond to D5-branes wrapped on $S^3$, located at $\rho=0$ in order to minimize the energy.
One can estimate the tension $T_{DW}$ of the domain wall from the fact that the metric in the IR is approximately of the form $\mathbb{R}^{7}\times S^3$ and the radius of the three-sphere goes as $\sqrt{e^{\Phi_0}N\alpha'}$. We have 
\begin{equation}
T_{DW}\sim \frac{1}{\alpha^{'3}}\int_{S^3}e^{-\Phi}\sqrt{G}=\frac{e^{2\Phi_0}N^{3/2}}{\alpha^{'3/2}}.
\end{equation}
Since a fundamental string can end on a D5-brane and
the QCD string is a fundamental string, we see that a QCD string can end
on a domain wall.}
\end{itemize}

We can also estimate in supergravity the masses of glueballs and Kaluza-Klein states.
These can be determined by studying the equations of motion for 
supergravity fields in the background (\ref{mn2metric}).
The masses of the lightest glueballs are given by the lower value of the gravitational redshift.
The order of magnitude for the KK masses can be deduced from the 
inverse radius of the three-sphere.
Both of them are proportional to $1/\sqrt{\alpha' N}$.
We see from formula (\ref{ten}) 
that the ratio $T_s/m_{KK}^2$ is of order $Ne^{\Phi_0}$.
In order to decouple the KK and gauge theory scales
we need $Ne^{\Phi_0}\ll 1$, a condition that requires large
curvatures in the IR and cannot be obtained
in the supergravity approximation.
Nevertheless, as we have seen, the theory
described by the supergravity solution exhibits confinement and chiral symmetry
breaking and shares many properties of its cousin, pure ${\cal N}=1$ SYM.
We then have a family of string backgrounds dual to gauge theories
that are a one-parameter generalization of ordinary SYM. 
The extra parameter
can be identified with $Ne^{\Phi_0}$. Notice that the 
scales determined by the string and domain-wall tensions are not equal
and explicitly depend on the extra parameter. 
In this family, pure SYM corresponds to a  strongly coupled 
string background.

We end this Section with another quantitative property that
is not determined by the symmetries of the problem.
As noticed in \cite{tutti}, by using the full background and making some 
assumptions, one can make more precise
predictions on the behavior of the beta function.
Since there is no moduli space in ${\cal N}=1$, there
is no intrinsic prescription for computing the behavior of 
the gauge coupling with the scale;
both definitions of coupling and scale are ambiguous. 
In order to fix the radius/energy relation,
the authors in \cite{tutti} proposed to enforce the 
equation $a = {\Lambda^3\over \mu^3}$
that defines the gaugino condensate. They also proposed to 
define the coupling constant 
as that seen by a stack of D5-branes for all values of $\rho$.
Using (\ref{raggen}) with $S^2$ being the two sphere at
$(\theta=\tilde\theta,\phi=\tilde\phi)$\footnote{It is the same cycle we introduced to discuss instantons in the MN background.} \cite{merlatti}, one  
obtains the same result as in the UV limit (\ref{raggen}),
up to exponentially suppressed terms\footnote{Some mistakes in \cite{tutti} were eventually
corrected in \cite{merlatti}.}.
These two pieces of information uniquely determine
 the $\beta$-function \cite{tutti,merlatti}
\begin{equation}
\beta=-\frac{3Ng_{YM}^3}{16\pi^2}(1-\frac{Ng_{YM}^2}{8\pi^2})^{-1}.
\label{NSVZtor}\end{equation}
This formula coincides with the NSVZ $\beta-$function \cite{NSVZ}.
In field theory it gives the full perturbative result and it is not corrected by instanton contributions. 
In supergravity it is exact up to exponential terms, which can be 
interpreted as fractional
instantonic corrections.  The non-trivial content of this formula is the 
analyticity in $g_{YM}$ and the
one and two-loop coefficients, that are the only scheme-independent objects. It is not clear why this result is captured by 
 the supergravity approximation
which only describes a cousin of the ordinary pure $SU(N)$ gauge theory.

For other works on the MN solution we refer to \cite{listaMN}.

\subsection{${\cal N}=1$ SYM from fractional D-branes on the conifold}\label{kleba}

As discussed in Sections 1.5 and \ref{fractional}, 
$N$ physical and $M$ fractional D3-branes  placed at the apex of a conifold
realize on their world-volume a four-dimensional ${\cal N}=1$ supersymmetric 
gauge theory with gauge group $SU(N+M)\times SU(N)$. There is also an
$SU(2)\times SU(2)\times U(1)_R$ global symmetry inherited from 
the isometries of $T^{1,1}$. The gauge theory is coupled 
to bi-fundamental chiral multiplets $A$ and $B$,
 interacting through the superpotential $W$ given in eq.~(\ref{s10}).
$A$ and $B$ transform in the ({\boldmath $N+M$},{\boldmath $\bar{N}$}) 
 and in the ({\boldmath $\overline{N+M}$},{\boldmath $N$})
representation of the gauge group, respectively, and are a doublet of one of 
the global $SU(2)$'s each.
In this Section we will describe the corresponding Type IIB
solution.
On the supergravity  side, we expect to find a metric of warped form
\begin{equation}
ds^2_{10} =   h^{-1/2}(\tau)   dx^2_4 +  h^{1/2}(\tau) ds_6^2 \ ,
\label{warp}\end{equation}
with non-trivial $F_{(3)}$ and $F_{(5)}$ RR-fields induced by the D5 and D3 sources.
Since, as discussed in Section \ref{fractional}, the integral 
$\int_{S^2}B_{(2)}$ determines the difference of the gauge 
couplings, we also expect a non-zero $H_{(3)}$ reflecting the
running of the couplings in the non-conformal theory. 
A supersymmetric solution with this minimal set of fields 
and internal metric given by the conifold one, was found in \cite{kt1},
but it has a naked singularity in the IR. 
In \cite{ks}, a  regular solution was found  by considering 
 a  deformed conifold instead of the original singular one. 
We will see that the deformation of the conifold corresponds 
to the requirement that 
the supergravity background knows of the gaugino condensation 
in the dual 
field theory. This situation is then similar to that
occurring in the MN solution.

In terms of complex geometry, the deformation of the singular conifold
$\sum w_a^2 =0$ is described by the equation in $\mathbb{C}^4$
\begin{equation}
\sum w_a^2 =\varepsilon^2. \label{def}
\end{equation}
The deformation
consists in blowing-up an $S^3$ at the apex of the conifold, so to obtain
a smooth manifold. The deformed conifold metric can be written as 
\begin{equation}\label{metricd}
ds_6^2 = \frac{\varepsilon^{4/3}}{2} K(\tau)\left[\frac{(d\tau^2 + g_5^2) }{3 K^3(\tau)}
+ \cosh^2\frac{\tau}{2}\, (g_3^2 +g_4^2)
+ \sinh^2 \frac{\tau}{2} \, (g_1^2 + g_2^2) \right],
\end{equation}
where $K(\tau)=  (2^{1/3}\sinh \tau)^{-1}(\sinh (2\tau) - 2\tau)^{1/3}$ and the $g_i$ are as in (\ref{fbasis}).

The regular solution is known as the Klebanov Strassler solution \cite{ks}.
It consists of a metric of the form (\ref{warp}), with $ds_6$ as 
in (\ref{metricd}), warp factor given by 
\begin{equation}\label{KSh}
h(\tau) = (g_s M\alpha')^2 2^{2/3} \varepsilon^{-8/3}\int_\tau^\infty d x {x\coth x-1\over \sinh^2 x} (\sinh 2x - 2x)^{1/3}, 
\end{equation}
and antisymmetric fields 
\begin{eqnarray}
B_{(2)} &=& {g_s M \alpha'\over 2} [f(\tau) g^1\wedge g^2
+  k(\tau) g^3\wedge g^4 ]\ ,\nonumber \\
F_{(3)} &=& {M\alpha'\over 2} \left[(1- F) g^5\wedge g^3\wedge g^4
+  F g^5\wedge g^1\wedge g^2 + F' d\tau\wedge
(g^1\wedge g^3 + g^2\wedge g^4) \right],\nonumber \\
 F_{(5)} &=& {\cal F}_{(5)} + \star {\cal F}_{(5)}, \nonumber \\
{\cal F}_{(5)} &=& B_{(2)}\wedge F_{(3)} = {g_s M^2 (\alpha')^2\over 4}[f (1-F) + k F] 
g^1\wedge g^2\wedge g^3\wedge g^4\wedge g^5.
\label{forms}
\end{eqnarray}
The functions of $\tau$ appearing in the previous formulae read
\begin{eqnarray}
F(\tau )&=& \frac{\sinh \tau -\tau}{2\sinh\tau},\nonumber\\  
f(\tau )&=& \frac{\tau\coth\tau - 1}{2\sinh\tau}(\cosh\tau -1),\nonumber\\  
k(\tau )& =& \frac{\tau\coth\tau - 1}{2\sinh\tau}(\cosh\tau+1).\label{ooo}
\end{eqnarray}
The complex dilaton of Type IIB is constant and this allows for a small 
string coupling everywhere.

Let us examine the  asymptotic behavior of the solution.
For large values of $\tau$ (which correspond to the UV limit of the dual gauge theory) it is convenient to introduce the radial coordinate $r\sim\varepsilon^{2/3} e^{\tau/3}$. The metric thus reduces to
\begin{equation}
ds_{10}^2\rightarrow  h^{-1/2}(r)dx_4^2 + h^{1/2}(r)(dr^2 
+ r^2ds_{T^{1,1}}^2),
\label{KTh}
\end{equation}
with $r_s\sim\varepsilon^{2/3}$ and $h(r)= {81(\alpha'g_sM)^2\over 8r^4}\log(r/r_s)$.
It can be viewed, in some sense, as a logarithmic deformation 
of $AdS_5\times T^{1,1}$. This was the solution first found in
\cite{kt1}. 
If we would allow $r$ to range in $[0, \infty)$ it would be singular for $r=r_s$. 
In this limit, the RR and NSNS forms  reduce to 
\begin{eqnarray}\label{uvf}
F_{(3)} &\ra& \frac{M\alpha'}{4} g_5\wedge(g_1\wedge g_2 + g_3\wedge g_4) , \quad
B_{(2)} \ra \frac{3g_s M\alpha'}{4}\log(r/r_s)(\ g_1\wedge g_2+ \  g_3\wedge g_4),\nonumber \\
{\cal F}_{(5)} &\ra& \frac{3g_sM^2(\alpha')^2}{8}\log(r/r_s) g_1\wedge g_2\wedge g_3\wedge g_4\wedge g_5.
\end{eqnarray}
For small $\tau$ the metric instead approximates to 
\begin{eqnarray}\label{KSIR}
ds_{10}^2 \rightarrow{ \varepsilon^{4/3}\over  b g_s M\alpha'} 
dx_{\mu} dx^{\mu} 
+ c(g_s M\alpha')
\left[{1\over 2}(d\tau^2 +g_5^2)+ (g_3^2 + g_4^2)+{1\over 4}\tau^2(g_1^2 + g_2^2)\right],
\end{eqnarray}
where $b$ and $c$ are numerical constants.
From the definition of the forms $g_i$ (eq.(\ref{fbasis})), it is easy to see that the angular part  splits in a non-vanishing $S^3$ and a shrinking $S^2$ fibered over it, 
just as in the MN case. 
The curvature is controlled by the value of $g_sM$, and it is small 
when this parameter is large. The antisymmetric fields $B_{(2)},F_{(5)}$ go to zero in the limit, while $F_3\ra (M\alpha'/2) g_5\wedge g_3\wedge g_4$ is supported only by the non-vanishing $S^3$. 

Let us now see how the KS solution encodes 
the properties of the dual gauge theory. 
Since there exist many good reviews in the literature, we will
just sketch the basic results referring the reader to \cite{ks,kreview}
for more details. 
It is believed that the $SU(N+M)\times SU(N)$ theory exhibits a series of Seiberg dualities
until it eventually reduces in the deep IR to pure $SU(M)$. At each step of 
the cascade, the group
is $SU(N+M-kM)\times SU(N-kM)$. The strongly coupled factor $SU(N+M-kM)$ undergoes 
a Seiberg duality to $SU(N-M-kM)$, while the other factor remains inert\footnote{We refer 
to \cite{peskin,intseib} for a detailed discussion of Seiberg duality.
We simply mention that this duality occur for a $SU(N)$ theory with $N_f>N+1$ flavors 
of quark chiral superfields 
$A_i$, $\bar A_i$, $i\, =1,...,N_f$, in the $\bf{N}$, $\bf{\overline N}$ representations. 
In this case the theory is dual to another ${\cal N} = 1$ SYM with $SU(N_f-N)$ gauge group,
$N_f$ flavors 
$C_i$, $\bar{C}_i$, 
and an extra gauge singlet chiral superfield $N^{ij}$ interacting by the superpotential 
$W=CN\bar{C}$.}. 
As a result, $k$ is increased by one unit.
 In the KS solution,
this can be seen from the UV limit of the RR five-form field strength (\ref{uvf}) which can be rewritten as
\begin{equation}
\mathcal{F}_{(5)}\sim N_{eff}(r)vol(T^{1,1}), \qquad  N_{eff}(r) = N + {3\over 2\pi}g_sM^2\log(r/r_0).
\end{equation}
We introduced a convenient reference scale $r_0$ defined so that the effective D3-charge $N_{eff}(r_0)=N$. 
The logarithmic decreasing of $N_{eff}$ with the 
radius was interpreted in \cite{kt1} as a decreasing 
in the rank of the dual gauge theory group as the theory flows to the IR.
At the UV scale $r=r_0$, $N_{eff}=N$ and the dual field theory has $SU(N+M)\times SU(N)$ as 
gauge group. At $r_k= r_0 {\rm exp}(-2\pi k/3g_sM)$, with $k$ integer, the dual gauge group 
is $SU(N-kM+M)\times SU(N-kM)$. If $N=kM$, we thus find that after $k$ cascade 
steps the gauge group flows to  $SU(M)$. 
The  UV completion of the theory is somewhat peculiar. The
inverse cascade never stops. In a sense, the UV limit is a $SU(\infty)\times SU(\infty)$
gauge theory.

The metric in eq. (\ref{KTh}) can be used to
study the UV properties of the $SU(N+M)\times SU(N)$ gauge theory
when $M\ll N$. 
Indeed, the curvature, which is determined by $Ng_s$
at the reference scale $r_0$, decreases for larger values of $r$.
Moreover, if $Mg_s$ is sufficiently small the cascade steps will be 
well separated. In these conditions, the
singular metric~(\ref{KTh}), which is a logarithmic deformation of $AdS_5\times T^{1,1}$, 
will give a convenient description of the almost conformal theory
$SU(N+M)\times SU(N)$. As shown in Section 3.3,  the gauge couplings are
 related to some of the supergravity moduli
\begin{equation}
{1\over g_{1}^2} + {1\over g_{2}^2}= {1\over 4\pi g_{s}};\qquad
{1\over g_{1}^2} - {1\over g_{2}^2}= {1\over 4\pi^2 g_{s}}\left(\frac{1}{2\pi\alpha'}\int_{S^2}B_{(2)}-\pi \right) .
\label{KScouplings}
\end{equation}
In order to use this formula, we must identify the cycle $S^2$ 
where the D5 branes are wrapped. As discussed in Section 1.5,
we can identify the $S^2$ 
with $\theta_1=\theta_2,\phi_1=-\phi_2$.
In the large $r$ limit, we thus find that
 the sum of the gauge couplings is constant while (see  (\ref{uvf}))  the difference runs as 
\begin{equation}\label{betadiff}
{4\pi^2\over g_1^2} - {4\pi^2\over g_2^2}= 3 M \log (r/r_s)=3 M \log(\mu/\Lambda).
\end{equation} 
The last equality in the above equation requires a specific choice of how to 
relate the radial coordinate to the energy scale of the field theory.
Also for the KS solution there are ambiguities in determining the radius/energy relation\footnote{The different methods for computing the radius/energy relation give different results \cite{kreview}. However, if we are only interested in the leading logarithmic UV behavior all methods agree. The comments on the holographic dual of the gaugino condensate we will give in the following also confirm the above identification.}.
We use the same relation as for the conformal  $AdS_5\times T^{1,1}$ solution, $r/r_s =\mu/\Lambda$, $\Lambda$ being the IR scale. This is obtained by 
considering the energy of a string stretched in the background. 
One can show that eq. (\ref{betadiff}) reproduces, up to orders $M/N$, 
the UV gauge theory 
result  obtained  from the exact NSVZ beta function \cite{NSVZ} 
for ${\cal N}=1$ gauge theories. Using formula~(\ref{betaex}) we can indeed write
\begin{eqnarray}
\frac{4\pi^2}{g_1^2}&=&\frac{1}{2}(3(N+M)-N(6-2\Delta_A-2\Delta_B))
\ln(\mu/\Lambda),\nonumber\\ 
 \frac{4\pi^2}{g_2^2}&=&\frac{1}{2}(3N-(N+M)(6-2\Delta_A-2\Delta_B))
\ln(\mu/\Lambda).\label{bf}
\end{eqnarray}
At leading order in $M/N$, $\Delta_A+\Delta_B=3/2$, which is the result for the conformal
case. The difference of the two equations in (\ref{bf}) then reproduces the supergravity result.
We refer the reader interested in a more detailed comparison to 
\cite{ks,kreview}.

The IR region of the KS solution should describe a pure $SU(M)$ SYM. 
We notice however that, in the supergravity approximation, $g_sM\rightarrow\infty$  and so the cascade steps are not well separated and the 
additional massive fields of the original theory $SU(N+M)\times SU(M)$ 
are not decoupled. As usual, we can get a pure SYM theory only beyond the supergravity regime. The supergravity solution is
dual to a four-dimensional gauge theory with a 
large number of massive matter fields. As its cousin, pure SYM,
the theory confines and presents the standard pattern of chiral symmetry 
breaking.  
The analysis is similar to that for the MN solution.
The $U(1)_R$ symmetry of the theory corresponds to 
shifts of the 
transverse angular variable $\psi$. The transformation law for the RR
field $C_{(2)}$ (or the analysis of an instantonic probe) shows that
$U(1)_R$  is anomalous  and broken 
to $\mathbb{Z}_{2M}$ \cite{anomaly,kreview}.  
The spontaneous symmetry breaking $\mathbb{Z}_{2M}\rightarrow \mathbb{Z}_2$ is
manifest in the full KS solution. 
The IR expression (\ref{KSIR}), which depends on $\psi$ through $\cos \psi$ and $\sin \psi$, 
has in fact only a $\mathbb{Z}_2$ invariance under $\psi \rightarrow \psi + 2\pi$. 
The breaking is also evident in eq.~(\ref{def})
that is not invariant under arbitrary phase shifts of the $w_a$, but only under
$w_a\rightarrow - w_a$.
The IR limit of the KS background also shows that the dual field theory 
confines. This is due to the fact that the warp factor 
approaches a constant value $h\sim (g_sM\alpha')^2\varepsilon^{-8/3}$ when $\tau\rightarrow0$. The tension for confining strings $\alpha'T_s\sim h^{-1/2}$ is thus of the order $\varepsilon^{4/3}/g_sM\alpha'$, and the glueball masses scale as $\varepsilon^{2/3}/g_sM\alpha'$. The deformation parameter thus gives the fundamental scale of the dual field theory.
The KS background also allows to extract information about other 
field theory features like  baryons, domain walls, etc. (we refer the interested reader 
to \cite{loeson,kreview}).

In the final part of this Section we derive the effective Lagrangian for the KS
solution. This will fill a gap in the previous discussion and show how
the KS solution can be derived from a set of first order equations. 
This will also allow us to give  the map 
between at least some of the fields 
appearing in the KS solution and the gauge invariant operators in the 
 dual gauge theory. 
In particular we will identify the holographic duals of the gaugino bilinears, and check that their behavior is consistent with the existence of a condensate in the dual field theory.
To this purpose it
is convenient to write the ansatz for the solution in the 
form \cite{pandotse}\footnote{Since we want to use the methods of Section \ref{nonconfI}, we
introduce a radial coordinate $u$ analogous to that of Section \ref{nonconfI}.}
\begin{eqnarray}
ds^2&=&2^{1/2}3^{3/4}\left[ e^{-5q(u)}(du^2+e^{2Y(u)}dx_{\mu}dx^{\mu})
+ds^2_{int}\right],\nonumber\\
ds^2_{int}&=&e^{3q(u)}\left[\frac{e^{-8p(u)}}{9}g_5^2+
\frac{e^{2p(u)+y(u)}}{6}(g_1^2+g_2^2)
+\frac{e^{2p(u)-y(u)}}{6}(g_3^2+g_4^2)\right],\nonumber\\
B_{(2)}&=&-(\tilde f(u)\ g_1\wedge g_2+\tilde k(u)\  g_3\wedge g_4), \quad  \Phi=\Phi(u),\nonumber\\
F_{(3)}&=& 2P g_5\wedge g_3\wedge g_4+d[\tilde F(u)(g_1\wedge g_3+
g_2\wedge g_4)],\nonumber\\
F_{(5)} &=& {\cal F}_{(5)} + \star {\cal F}_{(5)},\,\, {\cal F}_{(5)} = -L(u)g_1\wedge g_2\wedge g_3\wedge g_4\wedge g_5.
\label{KTans}\end{eqnarray}
This is a quite general ansatz since it includes the conformal 
case $AdS_5 \times T^{1,1}$ as well as the singular and regular non-conformal 
solutions. In particular, the fields $y, \tilde f-\tilde k$ distinguish 
between the singular ($y= \tilde f-\tilde k=0$) and regular ($y, \tilde f-\tilde k\ne 0$) 
conifold geometries. For convenience, we have rescaled the functions
in the KS solution as follows: $\tilde f=-2g_sPf,\, \tilde k =-2g_sPk,\, \tilde F=
2PF$. 
The supergravity equations 
for $F_{(5)}$ set $L(u)= Q + (\tilde k-\tilde f)\tilde F +2P\tilde f$, where $Q$ and $P$ are 
constants related to the number of physical and fractional branes.
More precisely $P=\frac{M\alpha'}{4}$~; for $P=0$, $Q$ is proportional
to $N$, while for  $P\ne 0$ it can be reabsorbed in a redefinition of $\tilde f,\tilde k$.

Using the ansatz (\ref{KTans}) and integrating over the 10$d$ coordinates, 
the Type IIB Lagrangian reduces to 
the following effective action \cite{pandotse,bgz}
\begin{equation}
S=\int du e^{4Y}\left ( 3({\dot Y})^2-{1\over 2}G_{ab}{\dot\varphi}^a
{\dot\varphi}^b -V(\varphi )\right ),
\label{KTl}\end{equation}
supported by the constraint 
$3{\dot Y}^2-{1\over 2}G_{ab}{\dot\varphi}^a{\dot\varphi}^b +V(\varphi )=0$, with
\begin{eqnarray}
G_{ab}{\dot\varphi}^a{\dot\varphi}^b&=&15{\dot q}^2+10{\dot p}^2+ {{\dot y}^2\over 2}+
{{\dot\Phi}^2\over 4}\nonumber\\
 &+&
 e^{-\Phi-6q-4p}
(e^{-2y}{\sqrt{3}\dot {\tilde f}^2\over 2}+e^{2y}{\sqrt{3}\dot {\tilde k}^2\over 2})+
 \sqrt{3} e^{\Phi
-6q-4p}\dot{\tilde F}^2,\nonumber\\
V(\varphi)&=& e^{-8q} [e^{-12p}-6e^{-2p}\cosh{y}+{9\over 4}e^{8p}
(\sinh{y})^2 ]+ {9\sqrt{3}\over 8}e^{4p-14q}
e^{-\Phi}(\tilde f-\tilde k)^2\nonumber\\
&+& {9\sqrt{3}\over 4}e^{4p-14q+\Phi}
[e^{-2y}\tilde F^2+e^{2y}(2 P - \tilde F)^2]
+{27\over 2}e^{-20q}
L^2.
\label{potks}
\end{eqnarray}
For supersymmetric solutions, the second order equations of motion from eq. (\ref{KTl}) can 
be reduced to a set of first order ones, 
since the previous Lagrangian admits a superpotential (in the sense of 
 Section 2.3)
\begin{equation}
W=-3e^{4p-4q}\cosh y -2e^{-6p-4q}-3\sqrt{3}e^{-10q}\left (Q+\tilde F(\tilde k-\tilde f)+2P\tilde f\right ).
\label{superpo}
\end{equation} 
One can check that the KS solution satisfies the 
first order equations following 
from this superpotential (the KS radial coordinate 
is related to $u$ by $d\tau=3e^{-4q+4p}du$  and $e^{y}=\tanh{\tau/2}$). 
The dilaton does not appear in $W$ and so it is constant for 
all the possible solutions.

In absence of fractional branes, $P=0$, the potential $V$ in (\ref{potks}) 
has an ${\cal N}=1$ critical point, corresponding to the
conformal background  $AdS_5\times T^{1,1}$ generated by a stack of
physical D3-branes. If we choose to rescale $Q=-2/3\sqrt{3}$,
the critical point is at $q=p=y=\tilde F=0$ (with arbitrary $\Phi$ and
$\tilde k=\tilde f$). With this conventions the $AdS$ radius is $R=1$. 
If we consider $P$ (and so $M/N$) as a small
deformation of the conformal background, we can still relate 
the supergravity fields we turn on
to deformations or changes in the vacuum of the theory, using the rules
of the $AdS/CFT$ correspondence \cite{bgz}.
For $P=0$, the potential and superpotential around the critical point read
\begin{eqnarray}
V&\approx& -3 + 32 q^2 +12 p^2 -3 y^2 +21\xi_1^2 -3\xi_2^2, \nonumber \\
W&\approx& -3 + 4 q^2 -6 p^2 -3 y^2 +3\xi_1^2 -3\xi_2^2,
\end{eqnarray}
where all the scalars have been redefined in order to have
 diagonal and canonically
normalized kinetic terms.  The fields $p,q,y$ have only been rescaled while
the $\xi$ have been defined as 
$^4\sqrt{12} \tilde F = \xi_1-\xi_2$, $^4\sqrt{3}(\tilde f-\tilde k) ={\sqrt2} (\xi_1 +\xi_2)$; we also define 
$2s=^4\sqrt{3}(\tilde f+\tilde k)$.
For canonically normalized scalars, the quadratic terms in the expansion of the potential give the masses of the supergravity fields. 
As discussed in Section \ref{definition}, with the normalization chosen, 
the mass/dimension relation is $\Delta = 2 + \sqrt{4+m^2}$ and 
we thus find
that $q,p,y,\xi_1,\xi_2$ correspond to operators of dimension 
$\Delta= 8, 6, 3,  7, 3$ respectively. The fields $\Phi$ and $s$ do not 
appear in the superpotential: their mass squared is zero and thus they 
correspond to operators of dimension $\Delta=4$. 
Using the results in \cite{ferrara}, we can tentatively 
identify the fields 
with gauge theory  operators in the following multiplets \cite{bgz}
\begin{eqnarray}
&q,p\rightarrow {\mbox Tr} (W^2\bar W^2),\quad \,\,\,\, \Delta=8,6; \qquad &
\xi_1\rightarrow  
 {\mbox Tr}(A\bar A+B\bar B)W^2 ,\quad \Delta=7; \nonumber\\
&\xi_2\rightarrow {\mbox Tr} (W^2_{1}+W^2_{2}),\quad \quad  \Delta=3; \qquad 
& y\rightarrow {\mbox Tr} (W^2_{1}-W^2_{2}), \qquad \quad \Delta=3.
\label{op}
\end{eqnarray}
In particular we see that the fields $y, \xi_2$ can be read as holographic duals of the gaugino bilinears \cite{bgz}. The field $s$ is the massless field $\int_{S^2}B_{(2)}$  associated with a marginal direction in the $CFT$ \cite{kw1,kw2,kn}. The corresponding  operator is 
${\mbox Tr}(F^2_{1}-F^2_{2})$. Finally, the dilaton $\Phi$ 
corresponds to ${\mbox Tr}(F^2_{1}+F^2_{2})$.

When $P\ne 0$ a tadpole term for $s\sim \tilde f+\tilde k$ 
is introduced in the effective potential which makes the coupling 
constant run as 
in~(\ref{betadiff}). In the limit where the solution is a small 
deformation of the
conformal case, we can still reasonably use the identifications 
made above.
From the quadratic terms in the $W$ expansion one can read the 
leading asymptotic behavior of the fields near the critical point and, 
applying the rules of Section 2.3, tell whether they
correspond to a deformation ($\Delta -4$) or a choice of a 
different vacuum 
($- \Delta$). As the reader can see, $q$ and $\xi_1$ correspond to deformations, 
while the other fields are related to VEVs.
The fields $y,\xi_2$ can thus be related to field theory vacua with a non zero gaugino condensate \cite{bgz}. As a check one can explore their asymptotic behavior using the full KS solution. In the UV  the fields $\xi_2,y$ go 
like $\varepsilon^2/r^3$ and this is indeed appropriate for a protected dimension 3 operator \cite{loeson}.
As mentioned before, $\xi_2$ and $y$ are precisely the fields that control the
deformation of the conifold. Thus, their asymptotic behavior 
confirms the relation between the deformation of the geometry 
and the chiral
symmetry breaking on the field theory side.

For further works on the KS model we refer to \cite{listKS}.

\subsection{Supergravity duals of ${\cal N}=1^*$}\label{N1star}

In this Section we will briefly describe the known supergravity 
solutions
dual to ${\cal N}=1^*$, referring the reader to the original 
papers for the
details of the computations \cite{gppz3, ps, pwN1}. ${\cal N}=1^*$ 
theories
are obtained deforming  ${\cal N}=4$ with a supersymmetric mass 
term for the
three chiral superfields. The potential then reads (here we consider for
simplicity the case of equal masses, the generalization being straightforward)
\begin{equation}
\int d\theta^2 \left( 2\sqrt{2} \:
Tr(\Phi_1 [\Phi_2,\Phi_3]) + m \sum_{i=1}^3 (\Phi_i)^2 \right).
\label{pot1}\end{equation}
The theory possesses a very rich vacuum structure, parameterized by the $N$
dimensional, generally reducible, representations of $SU(2)$. These are indeed
the solutions of the F-term equations for supersymmetric vacua 
$[\Phi_i,\Phi_j] = - \frac{m}{\sqrt{2}} \epsilon_{ijk} \Phi_k$.
For a generic vacuum the matrices $\Phi$ will have a block diagonal structure,
where the blocks represent irreducible $SU(2)$ representations of different
dimension $n_i$ (including dimension 1) such that $\sum_i n_i = N$.
A detailed discussion of the classical and quantum properties of 
the various vacua can be found in \cite{ps,dw}. 
Here we only focus on the two
cases that have the simplest interpretation on the supergravity side.
One is the Higgs vacuum, corresponding to the $N$ dimensional irreducible
representation of $SU(2)$. In this case the gauge group is completely broken
and there is a mass gap already at the classical level. 
The other vacuum is characterized by zero VEVs for the scalar fields,
$<\Phi> =0$ (i.e. $N$ copies of the trivial
representation). $SU(N)$ is unbroken and the theory is expected to confine and
to have $N$ distinct vacua parameterized by the gaugino condensate 
$<\lambda \lambda >$.

Following the philosophy of Section \ref{nonconfI}, the supergravity duals of 
${\cal N}=1^*$ should be given by Type IIB solutions with non-zero profile
for the modes corresponding to the mass deformation (\ref{pot1}).
From the perspective of 5$d$ ${\cal N}=8$ supergravity, fermion bilinears are
dual to scalars in the $\bf{10}$ of $SU(4)$.
The supersymmetric mass term for the chiral multiplets, $m_{ij}$,
transforms as the $\bf{6}$ of $SU(3)\subset SU(4)$, and the corresponding
supergravity mode appears in the
decomposition of the $\bf{10}\rightarrow
\bf{1}+\bf{6} +\bf{3}$ of $SU(4)$ under 
$SU(3)\times U(1)$. The term $\bf{1}$ in this 
decomposition corresponds instead to the scalar $\sigma$ dual to the gaugino
condensate in ${\cal N}=1$ SYM. If we further require an $SO(3)$ symmetry,
by taking equal masses $m_{ij}=m \, \delta_{ij}$, one can show that 
the Lagrangian can be consistently truncated to the fields $m$
and $\sigma$. In units $R=1$ it reads \cite{gppz3}
\begin{eqnarray}
{\cal L} &=& \sqrt{-g}\left\{- {{\cal R}\over 4}
+ {1\over 2}(\partial m)^2+{1\over 2}(\partial \sigma)^2 + \right. \nonumber\\
& & \left. - \frac{3}{8} \left[\left(\cosh{{2m\over \sqrt{3}}}\right)^2
+4\cosh{{2m\over \sqrt{3}}}\cosh{2 \sigma}
-\left(\cosh{2 \sigma}\right)^2 +4 \right]\right\},
\label{bohh}
\end{eqnarray}
and admits the superpotential $W=-\frac{3}{2} \left(\cosh{\frac{2m}{\sqrt{3}}} + \cosh{2\sigma}\right)$. 
The explicit supersymmetric
solution of the 5$d$ equations of motion
was found in \cite{gppz3}, and consists of a family of solutions
depending on two independent parameters.
Using the rules discussed in Section 2.3 we can show that these
solutions correspond to vacua with a non-zero gaugino condensate.
Indeed, by expanding the superpotential around the UV fixed point
$W\sim -3 -m^2-3 \sigma^2$ 
we see that $m$ and $\sigma$ have the right asymptotic behavior 
to be identified with a  mass term for the matter superfields
and the gaugino condensate, respectively.
One can further check that 
the theory has a mass gap. It is then natural to identify these solutions
with the confining vacua of ${\cal N}=1^*$.
However, the gaugino condensate in the solution is a continuous 
parameter rather than a discrete one as expected from field theory. 
Moreover, the solutions have a naked singularity, which 
is still present in the ten dimensional lift \cite{pwN1}, thus 
making the physical
interpretation not very clear. Nevertheless, the five-dimensional 
solution, which is one of the few analytically known,
has proved to be useful for computing Green functions
along the RG flow \cite{ans,bianchi}. Interestingly, 
sensible results have been obtained 
despite the presence of a singularity.

The analysis of the five dimensional flow
 seems to suggest that the supergravity approximation is not 
sufficient to
describe duals of ${\cal N}=1^*$ theories and that a stringy mechanism is
required to resolve the singularity. 
One possibility is to introduce D-brane sources in the $AdS_5 \times S^5$
background: this is the idea behind the Polchinski Strassler 
solution \cite{ps}.  In this case, the sources are D3-branes 
polarized via Myers' effect \cite{myers} into five-branes with world-volume 
$\mathbb{R}^4 \times S^2$ ($S^2$ is an equator of $S^5$ and $\mathbb{R}^4$ 
is a slice of $AdS_5$ at fixed radius).

To see how this works consider first the Higgs vacuum. 
In the field theory we have scalar
VEVs in the $N$ dimensional irreducible representation of $SU(2)$ which 
are distributed on a non-commutative sphere
\begin{equation}
\sum_{i=1}^3 \left| \phi_i \right|^2 \sim m^2 N^2 I_{N}.
\label{fuzzy}
\end{equation}
On the string theory side, ${\cal N}=1^*$ is realized 
as the world volume theory on a set of $N$ D3-branes, 
the scalar fields being the transverse coordinates
of the branes ($x_i =  2\pi \alpha' \phi_i$). 
Equation (\ref{fuzzy}) then corresponds to a configuration of D3-branes non-commutatively expanded into an $S^2$. From the non-abelian
generalization of the CS action \cite{myers} 
\begin{equation}
S \sim \mu_3 \int C_{(4)} + \mu_3 (2 \pi \alpha')^2 \int F_{0123ijk}^{(7)} [x_i,x_j ] x_k,
\end{equation}
we see that the expanded D3-branes have an additional electric coupling to the
RR 6-form (equivalently a magnetic coupling to the RR 2-form, $C_{(2)}$) and are
therefore equivalent to a single D5-brane with world volume 
$\mathbb{R}^4 \times S^2$, $N$ units of D3-brane charge and 
zero net D5-brane charge. 
It is then possible to identify the Higgs vacuum with such a single D5-brane. 
Notice that this interpretation  also fits with the standard $AdS/CFT$ 
dictionary. Indeed the mass deformation $m \lambda \lambda$ 
corresponds in 10$d$ to the linear combination of the NSNS and RR two-forms,
which  are the fields to which the D5-brane is coupled. 
The full supergravity solution corresponding to the wrapped D5 in the $AdS_5
\times S_5$ background is not known. In \cite{ps}, the asymptotic solutions
near the boundary and near the D5 were given. The solution is stable due to
the balance between the 2-form potential and the energy of the non-commutative
expansion. As a result the D5-brane sits at a fixed radius $z \sim \alpha' m
N$. By analyzing fundamental strings in this background,
one can check that electric charges are screened, as appropriate for an
Higgs vacuum \cite{ps}.

The basic idea behind the derivation of the solution is that the D5-brane
can be seen as a small perturbation of the $AdS$ background. 
This is however not
the case for the confining vacuum which should correspond to $N$ coincident D5-branes. 
Instead one can use S-duality, since it can be proved that 
in  ${\cal N}=4$ SYM it maps the Higgs vacuum to one of the $N$ confining vacua. On the supergravity side
S-duality corresponds to the $SL(2,\mathbb{Z})$ symmetry of Type IIB. 
Hence each of the  confining vacua should be described by an NS5-brane.
By analyzing fundamental strings, one can check that, in this vacuum,
the electric charges confine. 
From the equation of motion for the 2-form field, one can also check the
existence of a subleading solution corresponding to a gaugino condensate 
in field theory, according to the prescription of Section \ref{nonconfI}.

We refer to \cite{ps} for more information about the rich physics
of this model and for a study of the dual solutions to the other vacua of
${\cal N}=1^*$.


\subsection{A ${\cal N}=0$ solution}\label{nzeros}

We will only marginally discuss four-dimensional 
non-supersymmetric theories in this review. Non supersymmetric theory can be
obtained introducing a finite temperature in a higher dimensional theory 
that possesses a holographic 
dual, or starting directly
with non-supersymmetric string theories, like Type 0. 
Alternatively,
we can also study non-supersymmetric deformations of four dimensional
gauge theories. 

Here we will only consider 
models that can be obtained as soft breakings of the solutions
discussed in this and the previous Sections.   
The general strategy is to consider solutions with the same fields as the supersymmetric
ones. Some of the fields discussed in the MN or KS solutions, or in the ${\cal N}=2$ 
solutions in Section 4, are dual to scalar or fermionic
bilinears and they can be used to introduce mass terms in the theories.
Known examples of supersymmetry  breaking in the literature are
a massive sector
of ${\cal N}=1$ \cite{aaroni2}, soft deformations of the MN solution \cite{gubsevol,bcz2,epz,apreda} and
of the KS one \cite{gubser2,apreda}.

For simplicity, we will only consider the MN solution.
We can introduce a gaugino mass term in the system, leaving us with pure YM
in the far IR. We will need a non-supersymmetric solution of the Lagrangian~(\ref{lagmn2}).
Of course, we cannot use the BPS equations any longer and the equations of 
motions are needed. As already mentioned, a solution with an asymptotic behavior
as in~(\ref{asy}) with $Y\ne 0$ represents a deformation of the MN solution with a mass
term for the gaugino.
The general  analytical solution of the equations of motion corresponding to 
the Lagrangian~(\ref{lagmn2})
is not known. 
The asymptotics for large and small $\rho$, together with a numerical interpolation between the two, which proves that the solutions actually exist, have been discussed in \cite{gubsevol}.
In the UV the solution is 
\begin{eqnarray}    
a &=& {Y\over \sqrt{2\rho}} (1 + {1-|Y|^2/2\over 2\rho}+...) + 2 C \rho e^{-2\rho}(1+{\gamma\over 2\rho}+...),\nonumber\\    
h &=& {1 \over 2} \log \rho   -{|Y|^2\over 8\rho^2}(1+...) +   
P \sqrt{2\rho} e^{-2\rho}(1 + { \alpha \over 2\rho}+...),\\    
\Phi &=& \Phi_0 + \rho - 1/4\log{\rho} + {5|Y|^2\over 64\rho^2}(1+...) - P' \sqrt{2\rho} e^{-2\rho}  ( 1 + {\beta \over 2\rho}+...),  \nonumber  
\label{asymp}    
\end{eqnarray}    
where dots stand for corrections in $1/\rho$.
Using the equations of motion and choosing a convenient parameterization
we find $P = P' = k Re(\bar{C} Y)$
where $k$ is a free parameter. The other parameters $\alpha,\beta,\gamma$
are uniquely determined as functions of $k$ and $Y$. 
We work in the D5-brane setup.

The striking point is that there exists a family of $regular$ 
solutions whose expansion in the IR is
\begin{eqnarray}  
a &=& 1 - b \rho^2 + ..., \nonumber\\    
e^h &=&\rho - ( \frac{b^2}{4} + \frac{1}{9}) \rho^3 + ...,\\   
\Phi& =& \Phi(0) + (\frac{b^2}{4}+\frac{1}{3})  \rho^2 + ...,\nonumber    
\end{eqnarray}
where $b\in (0,2/3]$. The full solutions can be found     
by numerically integrating the IR solutions to the UV and solving for     
the UV parameters as functions of $b$ and $\Phi(0)$. 
$\Phi(0)$ matches with $g^2_{YM}$ and $b$ with the gaugino mass term. 
The other UV parameters can be expressed in terms of these two.
The ${\cal N}=1$ solution corresponds to $b=2/3$ and, of course, $Y=0$.
The other values of the parameter $b$ correspond to non supersymmetric solutions.

In ${\cal N} = 1$ $SU(N)$ SYM, soft breaking terms may be introduced into supersymmetric theories by    
promoting the parameters of the theory to background superfields. 
A non-zero $F$-component for $\tau$     
introduces in the bare Lagrangian a gaugino mass. 
If the mass is small compared to    
$\Lambda$, we can treat the supersymmetry breaking term as a perturbation.
We see from formula~(\ref{n1w}) in Section 5.1 
that the vacuum energy is no longer zero but, at leading order 
in the mass, it is   
 given by (for $\theta_{YM}=0$ and $m_\lambda$ real)        
\begin{equation} \label{pot}    
\Delta V \sim  m_\lambda \Lambda^3 \cos \left[ { 2 \pi n \over N}     
\right].    
\end{equation}    
The degeneracy of the vacua is removed and there is a single unique vacuum ($n=0$).    
The differences in energies between the $N$ different 
vacua, after supersymmetry breaking, can be computed also
using the ${\cal N}=1$ supergravity solution \cite{epz}. 
The computation of the free-energy in the supergravity solutions 
was done, for somehow different purposes, in \cite{gubsevol}. 
Here we just quote the result: the difference in energy between
the non-supersymmetric solution and the reference BPS one is   
\begin{equation}    
\Delta I \sim e^{ 2 \Phi_0} P \sim
e^{ 2 \Phi_0} k Re(\bar{C} Y).\label{result}    
\end{equation}    
In the supersymmetric limit where $Y$,  the 
gaugino mass, is zero, the     
solutions with different phases of the condensate $C$ are degenerate.
We can compute from this formula the energy of the vacua 
when a gaugino mass term $Y=m_\lambda$ is introduced.
At leading order in $m_\lambda$, we write $C=\Lambda^3 e^{2\pi in/N}+
O(m_\lambda)$ and, as can be shown numerically, $k={\rm constant}+O(m_\lambda)$. It is important that $k$ is a $U(1)_R$ invariant quantity
not depending on the phase of $C$. The vacuum energy then reads
\begin{equation}   
E \sim Re ( \bar{C} Y) \sim Re ( m_\lambda \Lambda^3 e^{-2\pi in/N}),  
\end{equation}    
reproducing the field theory result (\ref{pot}).

It is interesting to notice that the deep IR    
form of the metric is exactly the same for all solutions, ${\cal N}=1$ supersymmetric    
and not.
Furthermore, all the features of the ${\cal N}=1$ solution which depend only on the far IR form of the metric, such as the Wilson loop and the string tension, are similarly realized 
in the non supersymmetric case.
The presence of the mass gap can be taken as an argument for the classical stability of this 
background: even if not supersymmetric, the gap should prevent any mode to become tachyonic 
if the mass deformation is small enough.
This way of reasoning is due to the authors of \cite{aaroni2}.

A similar analysis was performed in \cite{gubser2,apreda} in the
case of the KS solution.


\vskip .2in

\noindent

\begin{center}

\textbf{Acknowledgments}

\end{center}

\vskip .1in

\noindent

We would like to thank R. Apreda, N. Evans, O. Gobbi, K. Peeters,  A. Sagnotti
and  M. Zamaklar for useful comments and discussions. 
We would also like to thank all our collaborators in the papers cited
in this review. 
F. B. is partially
supported by INFN. 
M. P. is supported by an European Commission Marie Curie Postdoctoral
Fellowship under contract number HPMF-CT-2001-01277, and partially by
the EU contract HPRN-CT-2000-00122.  A. Z. is partially
supported by INFN and MURST under contract 2001-025492, and by 
the European Commission TMR
program HPRN-CT-2000-00131, wherein he is associated to the
University of Padova.
This review is partially based on F. B. and A. L. C.  PhD thesis.

\appendix

\section{SW curves for $\N=2$ systems of branes}\label{swcurve}
Here we list the SW curves associated with the $\N=2$ theories
that are discussed in this review. 
For the majority of $\N=2$ models 
obtained with fractional and wrapped branes the curve can be 
determined using the M-theory lifting of the Hanany-Witten set-up
for NS and D4-branes \cite{wittenM}:

\begin{description} 
\item{(i)} The curve for $SU(N)$ with $N_f$ flavors is given by the genus-$(N-1)$ 
hyperelliptic Riemann surface
\begin{equation}
y^2=P(x)^2-\Lambda^{2N-N_f}P_f(x),\label{curvesun2}
\end{equation}
where the polynomials $P(x)$ and $P_f(x)$ are expressed in terms of the moduli and hypermultiplet
masses
\begin{equation}
P(x) = \prod _{i=1} ^N (x-u_i),
\qquad \qquad
P_f(k) = \prod _{\alpha =1} ^{N_f}(x-m_\alpha)\, .
\end{equation}
This theory can be realized with $N$ D4-branes stretched between two NS-branes and $N_f$ 
semi-infinite D4-branes. The zeros of the polynomials $P(x)$ and $P_f(x)$ 
represent the positions of the finite  and semi-infinite D4-branes, respectively.
\item{(ii)} The curve for $\prod_{i=1}^k SU(N_i)$ with bi-fundamental 
hypermultiplets in the 
({\boldmath $N_i$},{\boldmath $N_{i+1}$}) and fundamental hypermultiplets
 for $SU(N_1)$ and $SU(N_k)$ is given by
the polynomial in $t$ and $x$   
\begin{equation}
P_{k+1}(x) t^{k+1}+P_{k}(x) t^{k}+... +P_{1}(x) t + P_{0}(x)=0,
\label{curveproduct}
\end{equation}
where $P_s(x)=\prod_{i=1}^{N_s} (x-u^s_i), s=1,...,k$ are degree-$N_s$ polynomials 
containing the information about the moduli for the s$^{th}$-group and $P_{0,k+1}
=\prod_i (x-m^{0,k+1}_i)$
are polynomials containing the information about the masses of the fundamental hypermultiplets.
The curve is not hyperelliptic.
The system can be realized with $k+1$ NS-branes, $N_i$ D4-branes stretching between the
$i-(i+1)$ pair of NS-branes and two set of semi-infinite D4-branes. The zeros of the
polynomials $P_i(x)$ represent the positions of the D4-branes. 
All the dependence on the dynamically generated scales have been suppressed for simplicity.
The case $k=1$ with $P_0\equiv \Lambda^{2N-N_f}P_f,\, P_1\equiv 2P,\, P_2\equiv 1$, reduces
to the curve given above for $SU(N)$ with $N_f$ flavors with the redefinition $y=t+P(x)$.
\item{(iii)} The ${\cal N}=2$ $CFT$ associated to the affine $A_{k-1}$ Dynkin diagram is 
the cyclic quiver $\prod_{i=1}^k SU(N)$
with bi-fundamentals in the ({\boldmath $N_i$},{\boldmath $N_{i+1}$}), 
$i=1,...,N$ where $i=N+1\equiv 1$. 
The theory is conformal. The SW curve is not hyperelliptic
and it is an $N$-sheeted covering of a torus with modular parameter
$\tau$ corresponding 
to the
diagonal coupling constant in the $CFT$ $\tau=\sum_i\tau_i$. The curve
can be written in 
terms
of a meromorphic section of an Higgs bundle on the torus (locally an
hermitian $N\times N$ 

matrix $\Phi$)
\begin{equation}
\det (x I - \Phi)|_{{\rm locally}}=\prod_{i=1}^N (x-u_i)=0.
\label{ellipticcurve}
\end{equation}
$\Phi$ is meromorphic on the torus with exactly
$k$ simple poles whose residues determine the hypermultiplet masses
\cite{wittenM}. 
The systems
can be realized with $N$ D4-branes wrapped on a circle in the presence
of $k$ NS-branes. 
The example
for the $A_1$ case, corresponding to the $\mathbb{R}^4/\mathbb{Z}_2$ singularity, was pictured in Section~\ref{fractional}.
The poles of $\Phi$ correspond to the positions of the NS-branes upon lifting to M-theory
where the circle combines with the M-theory circle in giving a torus\footnote{An explicit representation
of the meromorphic functions can be given in terms of theta functions; if $u$ is the standard
coordinate on the torus represented as a parallelogram in $\mathbb{C}$, the curve can be
represented in a form similar to the previous cases as an infinite polynomial in $t=e^{iu}$
with coefficients that are degree-$N$ polynomials in $x$. Only $k$ polynomials are independent and give
the position of the D4-branes.
The circle has been lifted to its universal covering
$\mathbb{R}$; there is accordingly a precise pattern of repetition in the $x$-polynomials.}.
The case $k=1$ corresponds to $SU(N)$ with a massless adjoint, i.e. the $N=4$ theory.
$SU(N)$ with a massive adjoint can be described with a suitable twist along the circle \cite{dw,wittenM}.
\end{description}

The latter case actually contains all the previous ones and all the models we are interested in.
Indeed the non-conformal 
cases $(i)$ and $(ii)$ can be obtained from case $(iii)$ 
by a combination of the following two operations.
Firstly, we can freeze some of the coupling constants $\tau_i$, thus obtaining non-cyclic models with
fundamental hypermultiplets as in $(i)$ or $(ii)$. 
Secondly, theories with generic group   $\prod_{i=1}^k SU(N_i)$
can be obtained by considering suitable corners in the moduli space of the original $CFT$.
$SU(N)\times SU(N+M)$ with two bi-fundamentals, for example, can be obtained from the $A_1$ theory
$SU(N+M)\times SU(N+M)$ at low energies by giving large VEVs to some of the moduli of the 
first group. We just saw that the $SU(N+M)\times SU(N+M)$ curve can be written as a 
 meromorphic function on the torus with two poles.
The curve depends on  two polynomials $P_i(x)$
representing the positions of the D4-branes. We can take, for example, 
$P_{2}=\prod_{i=1}^N (x-u^{(1)}_i)$ and $P_1(x)=\prod_{i=1}^N (x-u_i^{(2)})(x^M-Y^M)$,
with $Y$ much bigger than the surviving moduli $u^{(i)}$ for $SU(N)\times SU(N+M)$. The curve for
$SU(N)\times SU(N+M)$ is then obtained for $Y\rightarrow \infty$ with $\tau$ suitably scaled.

\section{Fermionic shifts for systems with wrapped branes}\label{fermishift}

Here we will review in details the seven dimensional gauged supergravity setup used in the text for the wrapped brane system.
We will derive the relevant BPS equations and solve them for the ${\cal N}=2$ and ${\cal N}=1$ duals.


\subsection{Supergravity equations}\label{supergrav}

The strategy to obtain supersymmetric solutions in the $SO(4)$ 7$d$ gauged sugra
 for the systems of wrapped five-branes, is setting to zero the supersymmetry variations, which gives first order equations, and then check if the second order equations of motion are satisfied by the solutions.
We will write the general formulae \cite{noi} for the
supersymmetry variations of fermions with only three diagonal scalars\footnote{This and the following ones turn out to be sufficiently general ansatz for the dual theories considered here and probably for some generalization.}
\begin{equation} \label{scalari}
V_I{}^i = {\rm diag}( e^{-\lambda_1},  e^{-\lambda_1},
e^{-\lambda_2},  e^{-\lambda_3}).
\end{equation}
We take the $SO(4)=SU(2)^+ \times SU(2)^-$ gauge fields of the form
\begin{eqnarray}
A & = & \alpha \,[\cos{\theta} \, d\phi \, \eta _{1}^{+} + a(\rho) \,d\theta \, \eta
_{2}^{+}+ b(\rho) \, \sin{\theta} \, d\phi \, \eta _{3}^{+}]+ \nonumber \\
&& \beta \, [\cos{\theta}
\, d\phi \, \eta _{1}^{-} + \tilde{a}(\rho) \, d\theta \, \eta _{2}^{-} +
\tilde{b}(\rho) \, \sin{\theta} \, d\phi \, \eta _{3}^{-}], \label{AA}
\end{eqnarray}
where $\alpha$, $\beta$ are constants and the $\eta$ matrices are the
generators  of the $SU(2)^{\pm}$ in the $SO(4)$ notation and take the form
\begin{equation}
\eta _{1}^{\pm}={1\over 2} \texttt{\footnotesize 
$\left( \begin{array}{cccc} 0 & 1 & 0 & 0
\\ -1 & 0 & 0 & 0 \\ 0 & 0 & 0 & \pm 1 \\ 0 & 0 & \mp 1 & 0 \end{array}
\right)$} \,\, 
\eta_{2}^{\pm}={1\over 2}  \texttt{\footnotesize $
\left( \begin{array}{cccc} 0 & 0 & \mp 1 & 0 \\ 0 & 0 & 0 & 1 \\
\pm 1 & 0 & 0 & 0 \\ 0 & -1 & 0 & 0
\end{array} \right) $}\,\,
\eta _{3}^{\pm}={1\over 2}  \texttt{\footnotesize $\left(\begin{array}{cccc} 0 & 0 & 0 & 1 \\
0 & 0 & \pm 1 & 0 \\ 0 & \mp 1 & 0 & 0 \\ -1 & 0 & 0 & 0 \end{array}
\right)$}
\label{eta}
\end{equation}
The field strength is normalized as $F=dA+2m[A,A]$.
The ansatz for the metric (in the Einstein frame) is
\begin{equation}
ds_{7}^2 = e^{2f}(dx_{4}^2 + d\rho^2) + e^{2g}(d\theta^2 + \sin^{2}{\theta} d\phi^2).
\label{metrica2}
\end{equation}
It has to be thought as the seven dimensional part of a ``warped'' linear dilaton metric with the two-sphere compactification.
We do not use different notations for curved and flat indices.
To  pass from the former to the latter one must multiply
$\gamma^\phi,\gamma^\theta,\gamma^{\rho,\chi}$ by the inverse vielbein
($\chi=0,1,2,3$ labels the four dimensional coordinates).
From (\ref{metrica2}) it follows that the non trivial components of the spin
connection are
\begin{equation}
\omega_{\chi}^{\chi\rho}= f', \quad \omega_{\theta}^{\theta\rho}=g'e^{g-f}, \quad \omega_{\phi}^{\phi\rho}=g'e^{g-f}\sin{\theta},  \quad \omega_{\phi}^{\phi\theta}=\cos{\theta},
\label{spinconn}
\end{equation}
where the prime denotes differentiation with respect to $\rho$.
The general form of the supersymmetry variations can be obtained as a singular limit of the
ones in \cite{minasian} and reads
\begin{eqnarray}
\delta\psi_\mu &=& \left[ {\cal D}_\mu+
{1\over4} \gamma_\mu\gamma^\nu V_i^{-1\,I} {\partial}_{\nu} V_I{}^i
+{1\over 4} \Gamma^{ij}F_{\mu\lambda}^{ij}\gamma^\lambda
\right] \epsilon,\label{psimu}\\
\delta(\Gamma^{\hat{i}}\lambda_{\hat{i}}) &=& \left[ {m \over 2}(T_{\hat{i}j}-{1\over 5} T \delta_{\hat{i}j})\Gamma^{\hat{i}j}+{1\over2}\gamma_\mu P^\mu_{\hat{i}j} \Gamma^{\hat{i}j}+{1\over 16}\gamma^{\mu\nu}(\Gamma^{\hat{i}}\Gamma^{kl}\Gamma^{\hat{i}}-{1\over 5}\Gamma^{kl})F_{\mu\nu}^{kl}
\right] \epsilon,\qquad \label{gammalambda}
\end{eqnarray}
with $F_{\mu\lambda}^{ij}=V_I{}^iV_J{}^jF_{\mu\lambda}^{IJ}$.
Notice that the index $\hat{i}$ is not summed over.

Now, let's impose $\partial_\theta\epsilon=\partial_\phi\epsilon=\partial_x\epsilon=0$ and concentrate our attention on the gravitino,
 for example on its $\theta$ component
\begin{eqnarray}
\delta\psi_\theta &=& \left[ {1\over2}(g+\lambda_1 +{\lambda_2 
+\lambda_3 \over 2})'e^{g-f}\gamma^{\theta\rho}+{1\over4}m
[(e^{\lambda_1-\lambda_2}+e^{\lambda_2-\lambda_1})(-\alpha \,a+\beta\, 
\tilde{a})\Gamma^{13} \right.  \nonumber \\
 & &\left. +(e^{\lambda_1-\lambda_3}+e^{\lambda_3-\lambda_1})
(\alpha \,a+\beta\, \tilde{a})\Gamma^{24}]+{1\over2}\Gamma^{12}
e^{-2\lambda_1}{1\over 2}[\alpha\, 
\sin{\theta}(2\,m\,\alpha\,a\,b-1) \right. \nonumber \\
& &  \left. +\beta\, \sin{\theta}(2\,m\,\beta\,\tilde{a}\,
\tilde{b}-1)]\gamma^\phi{e^{-g}\over \sin{\theta}}+\Gamma^{34}
e^{-\lambda_2-\lambda_3}{1\over 2}[\alpha\, \sin{\theta}
(2\,m\,\alpha\,a\,b-1) \right. \nonumber \\
& & \left. -\beta\, \sin{\theta}(2\,m\,\beta\,\tilde{a}\,\tilde{b}-1)]
\gamma^\phi{e^{-g}\over \sin{\theta}}+\Gamma^{14}e^{-\lambda_1-\lambda_3}
{1\over 2}[\alpha\, \cos{\theta}(b-2\,m\,\alpha\,a) \right. \nonumber \\
& & \left. +\beta\, \cos{\theta}(\tilde{b}-2\,m\,\beta\,\tilde{a})]\gamma^\phi{e^{-g}\over \sin{\theta}}+\Gamma^{23}e^{-\lambda_1-\lambda_2}{1\over 2}[\alpha\, \cos{\theta}(b-2\,m\,\alpha\,a)\right. \nonumber \\
& & \left. -\beta\, \cos{\theta}(\tilde{b}-2\,m\,\beta\,\tilde{a})]\gamma^\phi{e^{-g}\over \sin{\theta}}+\Gamma^{13}e^{-\lambda_1-\lambda_2}{1\over 2}[\alpha\, a'-\beta\,\tilde{a}']\gamma^\rho e^{-f}\right. \nonumber \\
& &\left. +\Gamma^{24}e^{-\lambda_1-\lambda_3}{1\over 2}[-\alpha\, a'-\beta\,\tilde{a}']\gamma^\rho e^{-f}
\right] \epsilon=0. \,\, \label{psiteta}
\end{eqnarray}
Since the spinor $\epsilon$ is charged under $SU(2)^+\times SU(2)^-$,
let's separate the two components letting $\epsilon=\epsilon^+ \oplus
\epsilon^-$.
Now take the following basis of sigma matrices for $\epsilon^{\pm}$,
\begin{equation}\label{basesigma}
\Gamma^{12}\pm \Gamma^{34}=2i\sigma_3^{\pm},\,\,\,\,\Gamma^{24}\pm \Gamma^{31}=2i\sigma_1^{\pm},\,\,\,\,\Gamma^{14}\pm
\Gamma^{23}=-2i\sigma_2^{\pm},
\end{equation} 
and the following basis
for the seven dimensional gamma matrices
\begin{equation}\label{basegamma}
\gamma^{\mu}=\gamma_{(4)}^{\mu}\otimes 1, \qquad \gamma^{\theta,\phi,\rho}=\gamma_{(4)}^5 \otimes
\sigma^{1,2,3}.
\end{equation}
In these notations the spinor $\epsilon$ can be seen as a $2 \times 2$ matrix, with each entry a five dimensional spinor
\begin{eqnarray}\label{epsilon}
\epsilon =\left( \begin{array}{cc}
p & q \\ iq^c & -ip^c \end{array} \right).
\end{eqnarray}
This form is dictated by the symplectic-Majorana condition to be imposed in seven dimensions.
The space-time $\s$'s act on the matrix from the left, while the (transposed) gauge ones from the right.

Let's concentrate only on $\epsilon^+$; $\epsilon^-$, which will be present in the ${\cal N}=2$ solutions, behaves in an analogous way.
There are some constraints which come from the dependence on
$\theta$ in the fermionic variation (\ref{psiteta}).
From the ${\cos\theta \over \sin\theta}$ terms one gets $b=2\,m\,\alpha\,a$ and $\tilde{b}=2\,m\,\beta\,\tilde{a}$.
The contribution in $\cos\theta$ gives instead the twist condition
\begin{equation}
\left[
\gamma^{\phi\theta}+m[(\alpha+\beta)+{1\over 2}(\alpha-\beta)(e^{\lambda_2-\lambda_3}+e^{\lambda_3-\lambda_2})]i\sigma_3^+\right]\epsilon^+=0,\label{Atwist0}
\end{equation}
which will set $2m\alpha=1$ in the $\N=1$ case and $2m\alpha=2m\beta=1$ in the $\N=2$ solutions.
The gauge field $A_{\mu}$ was taken in (\ref{AA}) to have a component, the $\cos{\theta}$ one, proportional to the sphere spin connection, and formula (\ref{Atwist0}) gives the complete twist condition.
After these relations are imposed, the remaining part of (\ref{psiteta}) gives the actual first order differential equation to be solved.

The $\psi_{\phi}$ component of the gravitino variation is very similar to the $\psi_{\theta}$ one and it ultimately has two contributions that have to vanish separately.
The $\cos{\theta}$ part gives again the twist condition, while the $\sin\theta$ one gives the very same equation of the $\psi_\theta$ variation once the twist is imposed.
In an analogous way one can deduce the equations following from the $\psi_{\chi}$ and $\psi_{\rho}$ components of the gravitino, as well as the ones coming from the gaugino variations \cite{noi}.

The full set of BPS equations then reads
\begin{eqnarray}\label{equazioni}
\delta\psi_{\chi} &\rightarrow& f'+x'
=0,\nonumber\\
\delta\psi_\rho &\rightarrow& \left[\partial_{\rho}+{1\over2}x'+{1\over2}e^{-h}\gamma^\theta
i\sigma_1^+(a'\cosh{z}+\tilde{a}'\sinh{z})\right]\epsilon^+=0,\nonumber \\
\delta\psi_\phi &\rightarrow& \left[ h'e^h+\gamma^{\rho\theta}
i\sigma_1^+(a\cosh{z}\cosh{y}+\tilde{a}\sinh{z}\sinh{y}) + \right. \nonumber \\
& & +{1\over2}\gamma^\theta 
i\sigma_1^+(a'\cosh{z}+\tilde{a}'\sinh{z})+ \nonumber \\
&&
\left.
-{1\over2}e^{-h}\gamma^\rho[(a^2-1)\cosh{y}-(\tilde{a}^2-1)\sinh{y}]\right]\epsilon^+=0,\nonumber \\
\delta\lambda_i &\rightarrow& \Biggl[e^{-y}\sinh{2z}-z'\gamma^\rho+\gamma^\theta
i\sigma_1^+e^{-h}(a\sinh{z}\cosh{y}+\tilde{a}\cosh{z}\sinh{y})+ \nonumber \\
&&+{1\over2}\gamma^{\rho\theta}
i\sigma_1^+e^{-h}(a'\sinh{z}+\tilde{a}'\cosh{z})\Biggr]\epsilon^+=0,\nonumber \\
&& \left[ {1\over5}(e^y+e^{-y}\cosh{2z})+{1\over10}e^{-2h}[(a^2-1)\cosh{y}-(\tilde{a}^2-1)\sinh{y}]\right. + \nonumber \\
&&\left. -x'\gamma^\rho-{1\over5}\gamma^{\rho\theta}
i\sigma_1^+e^{-h}(a'\cosh{z}+\tilde{a}'\sinh{z})\right]\epsilon^+=0,\nonumber \\
&& \Biggl[(e^y-e^{-y}\cosh{2z})-{1\over2}e^{-2h}[(a^2-1)\sinh{y}-(\tilde{a}^2-1)\cosh{y}] + \nonumber \\
&& -y'\gamma^\rho+2\gamma^{\theta}
i\sigma_1^+e^{-h}(a\cosh{z}\sinh{y}+\tilde{a}\sinh{z}\cosh{y})\Biggr]\epsilon^+=0,
\end{eqnarray}
where $x=\lambda_1 +{\lambda_2 +\lambda_3 \over 2}$, $y=\lambda_1 -{\lambda_2 +\lambda_3 \over 2}$ , $z={\lambda_2 -\lambda_3 \over 2}$ and $h=g-f$.

To proceed, the ansatz for the fields must be refined.
The way to do it depends primarily on how many supersymmetries are to be preserved.
In terms of the fields, one must embed the $U(1)_s$ spin connection on the sphere in the $SO(4)$ normal bundle.
There are essentially two ways to do it.
One can break $SO(4) \rightarrow U(1)_{(1)} \times U(1)_{(2)}$ and embed $U(1)_s$ in, say, $U(1)_{(1)}$; this will preserve ${\cal N}=2$ supersymmetry.
Or one can view $SO(4)=SU(2)^+ \times SU(2)^-$ and embed $U(1)_s$ in, say, $U(1)^+\subset SU(2)^+$; this leads to ${\cal N}=1$ theories.

\subsection{${\cal N}=2$ solutions}\label{duesol}
Let us begin with the first case, in which $U(1)_s \sim U(1)_{(1)}$.
The latter is the diagonal of the two $U(1)$ factors in $SU(2)^+ \times SU(2)^-$.
Moreover, in this case we want no condensate or mass term for the ${\cal N}=2$ fermions, so the relevant equations follow from (\ref{equazioni}) putting $a=\tilde{a}=0$.
They read
\begin{eqnarray}\label{first22}
f' &=& -({\lambda_1}' +\frac{{\lambda_2}'+ {\lambda_3}'}{2}),\nonumber \\
g' &=& -({\lambda_{1}}' +{{\lambda_{2}}'+{\lambda_3}'\over 2}) 
+{1\over 2}e^{f-2g-2\lambda_{1}},\nonumber \\
{\lambda_{2}}'+2{\lambda_3}'+2{\lambda_{1}}' &=& 
-e^{f+2\lambda_3},\nonumber \\
{2\lambda_{2}}'+{\lambda_3}'+2{\lambda_{1}}' &=& 
-e^{f+2\lambda_{2}},\nonumber \\
3{\lambda_{1}}'+{\lambda_{2}}'+{\lambda_3}' &=& -e^{f+2\lambda_{1}}
+{1\over 2}e^{f-2g-2\lambda_{1}}.
\end{eqnarray}
Note that we have linear differential equations.
The solutions then read
\begin{eqnarray}
f &=& -(\frac{\lambda_2+\lambda_3}{2}+\lambda_1) ,\nonumber \\
e^{2g-2f} &=& u , \nonumber \\
e^{\frac{\lambda_2+\lambda_3}{2}-\lambda_{1}} &=& \sqrt{{{e^{4u}+b^4}\over{e^{4u}-b^4}}-{1\over2u}+ {2Ke^{2u}\over{u(e^{4u}-b^4)}}}, \nonumber \\
e^{\frac{\lambda_2+\lambda_3}{2}+\lambda_{1}} &=& \left({e^{2u}\over{e^{4u}-b^4}}\right)^{1/5} \left[{{e^{4u}+b^4}\over{e^{4u}-b^4}}-{1\over2u}+ {2Ke^{2u}\over{u(e^{4u}-b^4)}}\right]^{-{1\over10}} ,\nonumber \\
e^{\lambda_2 -\lambda_3} &=& {{e^{2u}-b^2}\over{e^{2u}+b^2}},
\label{duscasol2}
\end{eqnarray}
with
\begin{equation}
{du\over d\rho}\equiv e^{\frac{\lambda_2+\lambda_3}{2}-\lambda_{1}}.
\label{unuova}
\end{equation}
These solutions reduce to the two-scalars ones when $\lambda_2=\lambda_3$, i.e. when $b=0$. 

\subsection{${\cal N}=1$ solutions}\label{unasol}
As explained in section \ref{none}, the ${\cal N}=1$ dual solution corresponds to the identification $U(1)_s \sim U(1)^+\subset SU(2)^+$.
This twist amounts then on taking only the $SU(2)^+$ part of the connection in
equations (\ref{equazioni}).
Together with a single scalar in the matrix $T_{ij}$, call it $\lambda_1$, this provides a consistent truncation of the gauged supergravity \cite{cve1}.
In general, care must be taken in that the spinor $\epsilon^+$ is a two component $SU(2)$ vector and not only $\sigma_3$ is present the equations, which will thus retain their nonlinear structure.
There is an exception to this statement, namely if one considers the $U(1)^+$
case $a=0$, which reduces immediately the equations (\ref{equazioni}) to
\begin{equation}
h'={1\over2}e^{-2h},\qquad \qquad
\lambda'=-{1\over 5}+{1\over 20}e^{-2h},\label{u11}
\end{equation}
where $\lambda_i = \lambda,$ with $i=1,2,3$. The solution is
\begin{equation}\label{badsingular}
e^{2h}=\rho, \qquad \qquad 5\lambda={1\over 4}\log{\rho} - \rho.
\end{equation}
This solution has a bad type singularity, so it's not interesting.
As argued in \cite{mn2}, the way around this problem is turning on the non Abelian part of the connection, i.e. $a$.
This is not only a technical trick to get a nonsingular solution, but it is the right physical answer to the problem, as $a$ is dual to the gaugino condensate of the ${\cal N}=1$ theory.

Now the equations are truly non-linear, and defining
\begin{equation}
A={1\over2}h'e^h, \quad B={1\over2}a, \quad C={1\over4}e^{-h}(a^2-1), \quad D=-{1\over4}a',
\end{equation}
the $\psi_\theta$ equation becomes
\begin{equation}
[\gamma^{\theta\rho}A+iB\sigma_1^+ +i\gamma^\phi C\sigma_3^+ +i\gamma^\rho D\sigma_1^+]\epsilon^+=0,
\label{pippo}\end{equation}
that can be rewritten as
\begin{equation}
i\sigma_1^+\gamma^{\rho\theta}\epsilon^+=(\Delta + \Pi \gamma^\rho)\epsilon^+,\label{ventitre}
\end{equation}
with
\begin{equation}
\Delta=-{AB-CD\over A^2-C^2}, \qquad \Pi=-{AD-BC\over A^2-C^2}.\label{delta1}
\end{equation}
Multiplying (\ref{ventitre}) by $i\sigma_1^+\gamma^{\rho\theta}$, one obtains the consistency relation
\begin{equation}
\Delta^2\,-\,\Pi^2=1. \label{eq1}
\end{equation}
The gaugino variation (only the second gaugino variation in (\ref{equazioni}) survives), in the notation
\begin{equation}
E={2\over5}+{1\over10}e^{-2h}(a^2-1), \quad F=-2\lambda', \quad G={1\over5}e^{-h}a',
\end{equation}
reads
\begin{equation}
[E+\gamma^\rho F+i\sigma_1^+\gamma^{\rho\theta} G]\epsilon^+=0,
\end{equation}
that again can be rewritten in the form (\ref{ventitre}) but now with
\begin{equation}
\Delta={4e^h+e^{-h}(a^2-1) \over 2a' }, \qquad \Pi=-{10e^h\lambda' \over a' }, \label{delta2}
\end{equation}
so that from consistency of (\ref{delta1}) with (\ref{delta2}) one obtains the other two equations
\begin{equation}
{AB-CD\over A^2-C^2}={E\over G}, \qquad {AD-BC\over A^2-C^2}={F\over G}.\label{eq22}
\end{equation}
Finally, there is the $\psi_\rho$ variation, giving the $\rho$ dependence of the spinor and another equation
\begin{equation}
[2+{1\over2}e^{-2h}(a^2-1)]\,\Delta-\,\partial_\rho\Pi=0. \label{eq4}
\end{equation}
One can verify that all these equations are solved by the functions in \cite{chamsevolkov,mn2}
(the seven dimensional dilaton in the text is $\phi=5\lambda$) 
\begin{equation}
e^{2h}=\rho \coth{2\rho}-{\rho^2 \over \sinh^2{2\rho}}-{1\over4},\qquad
a={2\rho \over \sinh{2\rho}},\qquad
e^{10\lambda}={2e^h\over \sinh{2\rho}}.
\end{equation}

As a final remark, note that the system of equations (\ref{equazioni}) should have the right ingredients to provide also a solution corresponding to the breaking ${\cal N}=2$ $\rightarrow {\cal N}=1$.
The ${\cal N}=2$ $\rightarrow$ ${\cal N}=1$ field theory contains the gaugino condensate, which is dual to $a$, and a mass term for the $\psi$ fermion, which is dual to $\tilde{a}$.
Then the full $SU(2)^+\times SU(2)^-$ group is needed.
Moreover, one can expect that both operators ${\mbox Tr}\phi\bar\phi$ and
${\mbox Tr}\phi^2$ , which are dual to $(\lambda_2 \pm \lambda_3)/2$, have a VEV in QFT.
All in all, on the supergravity side this should then correspond to a solution
where all the fields $(\lambda_1,\lambda_2,\lambda_3,a,\tilde a)$
are turned on.
The BPS equations for this case are nothing else than (\ref{equazioni}).
The solution is still lacking, due to the technical difficulty in solving the equations.
However, a simplified model with $\lambda_2=\lambda_3$
and only the $SU(2)^+$ group admits a solution \cite{noi}, despite the fact that the number of
equations is redundant.
The ten dimensional solution has a good type singularity and its UV normalizable behavior indicates that it corresponds to the attempt of giving a VEV to scalar fields.
Since these scalars are massive to begin with, one expects an instability in QFT that may explain the singular behavior of the supergravity solution.



\end{document}